\documentclass{aastex62}
\pdfoutput=1 
\usepackage{amsmath}
\usepackage{forest}
\usepackage{threeparttable}
\usepackage{enumitem}

\newcommand\coonemath{\operatorname{CO(1-0)}}
\newcommand\cotwomath{\operatorname{CO(2-1)}}
\newcommand\cothreemath{\operatorname{CO(3-2)}}
\newcommand\cofourmath{\operatorname{CO(4-3)}}
\newcommand\comath{\operatorname{CO}}
\mathchardef\mhyphen="2D

\shorttitle{ASPECS LP power spectrum analysis at 3~mm}
\shortauthors{Uzgil et al.}

\begin{document}

\title{The ALMA Spectroscopic Survey in the HUDF: Constraining cumulative CO emission at $1 \lesssim z \lesssim 4$ with power spectrum analysis of ASPECS LP data from 84 to 115~GHz}

\correspondingauthor{Bade D. Uzgil}
\email{buzgil@nrao.edu}

\author{Bade D. Uzgil}
\affil{National Radio Astronomy Observatory, Pete V. Domenici Array Science Center, P.O. Box 0, Socorro, NM 87801, USA}
\affil{Max-Planck-Institut f\"ur Astronomie, K\"onigstuhl 17, D-69117, Heidelberg, Germany}

\author{Chris Carilli}
\affil{National Radio Astronomy Observatory, Pete V. Domenici Array Science Center, P.O. Box 0, Socorro, NM 87801, USA}
\affil{Cavendish Laboratory, University of Cambridge, 19 J~J~Thomson Avenue, Cambridge CB3 0HE, UK}

\author{Adam Lidz}
\affiliation{Department of Physics \& Astronomy, University of Pennsylvania, 209 South 33rd Street, Philadelphia PA 19104, USA}

\author{Fabian Walter}
\affil{National Radio Astronomy Observatory, Pete V. Domenici Array Science Center, P.O. Box 0, Socorro, NM 87801, USA}
\affil{Max-Planck-Institut f\"ur Astronomie, K\"onigstuhl 17, D-69117, Heidelberg, Germany}

\author{Nithyanandan Thyagarajan}
\affil{National Radio Astronomy Observatory, Pete V. Domenici Array Science Center, P.O. Box 0, Socorro, NM 87801, USA}

\author{Roberto Decarli}
\affil{INAF-Osservatorio di Astrofisica e Scienza dello Spazio, via Gobetti 93/3, I-40129, Bologna, Italy}

\author{Manuel Aravena}
\affil{N\'ucleo de Astronom\'ia de la Facultad de Ingenier\'ia y Ciencias, Universidad Diego Portales, Av. Ej\'ercito Libertador 441, Santiago, Chile}

\author{Frank Bertoldi}
\affil{Argelander-Institut f\"ur Astronomie, Universit\"at Bonn, Auf dem H\"ugel 71, 53121 Bonn, Germany}

\author{Paulo C. Cortes}
\affil{Joint ALMA Observatory - ESO, Av. Alonso de C\'ordova, 3104, Santiago, Chile}
\affil{National Radio Astronomy Observatory, 520 Edgemont Rd, Charlottesville, VA, 22903, USA}

\author{Jorge Gonz\'alez-L\'opez}
\affil{N\'ucleo de Astronom\'ia de la Facultad de Ingenier\'ia y Ciencias, Universidad Diego Portales, Av. Ej\'ercito Libertador 441, Santiago, Chile}
\affil{Instituto de Astrof\'isica, Facultad de F\'isica, Pontificia Universidad Cat\'olica de Chile Av.  Vicu\~na Mackenna 4860, 782-0436 Macul, Santiago, Chile}

\author{Hanae Inami}
\affil{Univ. Lyon 1, ENS de Lyon, CNRS, Centre de Recherche Astrophysique de Lyon (CRAL) UMR5574, 69230 Saint-Genis-Laval, France}
\affil{Hiroshima Astrophysical Science Center, Hiroshima University, 1-3-1 Kagamiyama, Higashi-Hiroshima, Hiroshima 739-8526, Japan}

\author{Gerg\"o Popping}
\affil{Max-Planck-Institut f\"ur Astronomie, K\"onigstuhl 17, D-69117, Heidelberg, Germany}

\author{Dominik A. Riechers}
\affil{Department of Astronomy, Cornell University, Space Sciences Building, Ithaca, NY 14853, USA}
\affil{Max-Planck-Institut f\"ur Astronomie, K\"onigstuhl 17, D-69117 Heidelberg, Germany}
\affil{Humboldt Research Fellow}

\author{Paul Van der Werf}
\affil{Instituto de Astrof\'isica, Facultad de F\'isica, Pontificia Universidad Cat\'olica de Chile, Av.  Vicu\~na Mackenna 4860, 782-0436 Macul, Santiago, Chile}

\author{Jeff Wagg}
\affil{SKA Organization, Lower Withington Macclesfield, Cheshire SK11 9DL, UK}

\author{Axel Weiss}
\affil{Max-Planck-Institut f\"ur Radioastronomie, Auf dem H\"ugel 71, 53121 Bonn, Germany}

\begin{abstract}

We present a power spectrum analysis of the ALMA Spectroscopic Survey Large Program (ASPECS LP) data from 84 to 115~GHz. These data predominantly probe small-scale fluctuations ($k=10$--100~h~Mpc$^{-1}$) in the aggregate CO emission in galaxies at $1 \lesssim z \lesssim 4$. We place an integral constraint on CO luminosity functions (LFs) in this redshift range via a direct measurement of their second moments in the three-dimensional (3D) auto-power spectrum, finding a total CO shot noise power $P_{\comath, \comath}(k_{\cotwomath}) \leq 1.9\times10^2$~$\mu$K$^2$~(Mpc~h$^{-1}$)$^3$. This upper limit (3$\sigma$) is consistent with the observed ASPECS CO LFs in \citet{Decarli2019_3mm}, but rules out a large space in the range of $P_{\comath, \comath}(k_{\cotwomath})$ inferred from these LFs, which we attribute primarily to large uncertainties in the normalization $\Phi_*$ and knee $L_*$ of the Schechter-form CO LFs at $z > 2$. Also, through power spectrum analyses of ASPECS LP data with 415 positions from galaxies with available optical spectroscopic redshifts, we find that contributions to the observed mean CO intensity and shot noise power of MUSE galaxies are largely accounted for by ASPECS blind detections, though there are $\sim20$\% contributions to the CO(2-1) mean intensity due to sources previously undetected in the blind line search. Finally, we sum the fluxes from individual blind CO detections to yield a lower limit on the mean CO surface brightness at 99~GHz of $\langle T_{\comath} \rangle = 0.55\pm0.02$~$\mu$K, which we estimate represents 68--80\% of the total CO surface brightness at this frequency.

\end{abstract}

\section{Introduction}\label{sec:intro}

The formation of molecular clouds from atomic hydrogen gas, and their subsequent consumption as fuel for star formation, are important transitions linking the early stages in the lifecycle of the interstellar medium (ISM) to the evolution of galaxies.  Obtaining an unbiased and complete measure of the cold gas content and star formation activity of galaxies as functions of cosmic time provides insight into the underlying physical processes that regulate this evolution. With the cosmic star formation rate density (SFRD) well characterized out to $z\sim3$--4, and rest-frame UV observations setting constraints on the SFRD into the first billion years after the Big Bang (e.g., \citet{MD2014} for a review), there are ongoing efforts to complement this understanding with concurrent trends in the atomic \citep[HI; e.g.,][]{Neeleman2016} and molecular \citep[H$_2$; e.g.,][]{Decarli2019_3mm} gas history---particularly during the epoch of galaxy mass assembly at $z\sim2$, when cosmic star formation activity was approximately ten times higher than in the present epoch, and more than half of the stellar mass in the Universe was accumulated \citep{MD2014}.

While the cosmic HI gas density has been inferred from observations of damped Ly-$\alpha$ systems in quasar spectra at $z\lesssim5$ \citep{Wolfe2005}, a more direct method is to observe the HI gas in emission via the 21~cm hyperfine transition. 21~cm experiments have constrained the atomic gas density in cosmic volumes out to $z\sim0.8$ \citep{Switzer2013,Chang2010}, and extended surveys are underway (e.g., CHIME, Tianlai, HIRAX, BINGO, Ooty Wide Field Array) to push this redshift limit in a continuous range out to $z\sim3.5$. As the primary science goal of many of these experiments is to use HI as a tracer of large-scale structure in order to measure the imprint of baryon acoustic oscillations, these experiments utilize an observational technique known as line intensity mapping to survey large areas of sky ($\mathcal{O}(10^3$--10$^4$~deg$^2$)) with coarse angular resolution ($\mathcal{O}(10$~arcmin)) in a spectral line across a wide fractional bandwidth (30--60\%), resulting in 3D maps of spatially confused line emission throughout cosmological volumes. Rather than detecting, then, at high significance, emission from galaxies on an individual basis, line intensity mapping experiments measure the surface brightness fluctuations in the targeted spectral line, as well as any additional line or continuum emission contributing to the aggregate surface brightness at the observed frequencies, via the power spectrum. 

Owing to the large collecting areas and wide bandwidths available in existing facilities such as the Atacama Large Millimeter Array (ALMA), Karl G. Jansky Very Large Array (JVLA), and IRAM NOrthern Extended Millimeter Array (NOEMA), the cosmic evolution of molecular gas density has already been measured out to $z\sim4$---well into the epochs of galaxy mass assembly and peak cosmic star formation history---with various surveys targeting different rotational $J$ transitions of the CO molecule as an H$_2$ gas tracer \citep{Pavesi2018, Walter2016_survey, Walter2014}. Unlike the HI intensity mapping experiments, the CO surveys performed blind spectral scans, or so-called molecular deep fields, to build a census of galaxies' gas content by detecting emission from individual CO-emitting sources that are brighter than the survey's flux limit. Given the relatively small fields of view and longer baselines of the telescopes employed in these efforts, the molecular deep fields are characterized by survey areas ($\mathcal{O}(10^0$--10$^1$~arcmin$^2$)) and angular resolutions ($\mathcal{O}(10^0)$~arcsec) well-suited for observing individual galaxies. 

The ALMA Spectroscopic Survey Large Program (ASPECS LP) in the \emph{Hubble} Ultra Deep Field (HUDF) is the latest example of a blind spectral scan that has resulted---with 68 hours of total telescope time to scan the full ALMA Band 3 from 84 to 115~GHz---in the tightest blind constraints to date on the evolution of CO luminosity functions, which directly translate to measurements on the cosmic molecular gas density \citep{Decarli2019_3mm} (hereafter, D19), over $\sim12$~Gyr of the Universe's history, revealing the levels of accumulation and consumption of molecular gas in galaxies from $z\sim4$ to the present day. ASPECS LP targeted a $\sim4.6$~arcmin$^2$ field in a region of HUDF containing the deepest near-infrared (near-IR) photometric data on the sky \citep{Illingworth2013, Koekemoer2013}, and $\sim$1,500 spectroscopic redshifts for rest-frame optically/UV-selected galaxies \citep{Inami2017}, which facilitated the confirmation and redshift-identification of blindly detected line candidates, and further enabled the characterization of physical properties such as molecular gas mass, stellar mass, AGN fraction, metallicity, IR luminosity, and star formation rate (SFR) for all secure detections in the field, and for hundreds of fainter sources, as well. Papers from the ASPECS team discuss key results from the ASPECS LP scan in ALMA Band 3, including the observed CO luminosity functions (D19), which also contains a detailed description of the ASPECS survey and ancillary datasets, blind searches for spectral line and continuum detections \citep{GL2019_3mm} (GL19), MUSE-based CO identifications and demographics of the ASPECS CO sample from spectral energy distribution (SED) modeling \citep{Boogaard2019_3mm} (B19), theoretical perspectives on the cosmic molecular gas density evolution \citep{Popping2019}, ISM properties \citep{Aravena2019_3mm}, and stacking analysis with MUSE galaxies in UDF (Inami et al., in prep).

In this paper, we consider the ASPECS LP Band 3 data in the context of a power spectrum analysis. Although we adopt the power spectrum approach used in line intensity mapping experiments, our dataset is inherently distinct from those produced by aforementioned line intensity mapping experiments, given the marked differences in sky coverage and angular resolution. Our overarching goals, however, to (1) probe flux from galaxies below the survey's sensitivity threshold for individual line detections and (2) improve constraints on the observed cumulative emission---from CO, in our case---by measuring surface brightness fluctuations within the survey volume, is akin to the objectives common throughout the line intensity mapping experimental landscape \citep[e.g.,][for a review]{Kovetz2017}, including experiments with goals of mapping the CO intensity field at $2 < z < 3$ to determine the cosmic molecular gas density \citep{Li2016, Keating2016}. Furthermore, the parallel analysis by the ASPECS team to extract individual CO detections, along with the rich multi-wavelength datasets available in HUDF, provide valuable information to aid in the interpretation of the power spectrum results, and enable an exploration of the complementarities between the two approaches. 

The organization of this work is as follows. In Section~\ref{sec:pspec_context}, we place ASPECS in the context of a power spectrum analysis, identifying, e.g., relevant scales that the survey covers in Fourier space. In Section~\ref{sec:data_methods}, we describe the Band 3 data, as well as details regarding our approach to measuring the power spectrum. Our results on lower limits from blindly detected sources, the 3-dimensional (3D) CO autopower spectrum, and statistical analysis of the CO fluctuation data including information from galaxy catalogs are presented in Section~\ref{sec:results}. In Section~\ref{sec:lim_vs_gs}, we discuss briefly our findings in the framework of a comparison between the power spectrum analysis and the blind line search in recovering the true CO power, and comment on the capability of current facilities to measure the CO power at high redshift. Finally, we summarize our findings in Section~\ref{sec:conclusion}.
 
Throughout this work we adopt a cosmological model with $\Omega_{\mathrm{M}}=0.7$, $\Omega_{\Lambda}=0.3$, $\Omega_{k}=0$, and $h = H_0/100 = 0.70$. 

\section{ASPECS in the context of a power spectrum analysis} \label{sec:pspec_context}
\subsection{Mapping survey dimensions from real-space to Fourier-space}

The original goal of the ASPECS Large Program (ASPECS LP)---``to reach a sensitivity such that the predicted `knee' of the CO luminosity function could be reached at $z\sim2$'' \citep{Walter2016_survey}---was a key driver of the chosen survey parameters,\footnote{See \citet{Walter2016_survey} for more on rationale behind the opted survey design.} including total observing time (and, hence, RMS sensitivity per beam per channel), spectral resolution $\Delta\nu_{chn}$, survey bandwidth $\Delta\nu_{BW}$, array configuration (or synthesized beam size $\Delta\theta_{b}$), and survey width $\Delta\theta_{S}$. We do not---and, in some cases, cannot---alter these experimental parameters for the purposes of the power spectrum analysis, except when redefining $\Delta\nu_{chn}$ and $\Delta\theta_{S}$, to be explained in more detail below. Upon adopting a target redshift for the observations, $\Delta\nu_{chn}$, $\Delta\nu_{BW}$, $\Delta\theta_{b}$, and  $\Delta\theta_{S}$ can be translated to physical co-moving length scales via the standard cosmological relations, and, thus, set the range of distances where statistical correlations between galaxies can be probed. Given that 11 of the 16 secure, blindly detected sources in GL19 with known redshifts in ASPECS correspond to CO(2-1) emitters, we have adopted a target redshift $z_{cen,\cotwomath}=1.315$ to represent the redshift of CO(2-1) emission observed at bandcenter, $\nu_{cen}=99.572$~GHz. We discuss the issue of redshift ambiguities in the CO line emission in Section~\ref{sec:line_ambiguity}.

\subsubsection{Real-space dimensions} \label{sec:realspacedims}

The largest physical scale, then, accessible in real-space in the line-of-sight dimension, $r_{\parallel,max}$, is determined by the survey's frequency coverage, $\Delta\nu_{BW}$:
\begin{equation}
r_{\parallel,max} = \frac{c}{H_0}\int_{z_{min}}^{z_{max}} \frac{\mathrm{d}z}{\sqrt{\Omega_{\mathrm{M}} (1+z)^3 + \Omega_{\Lambda}}} = \chi(z_{max}) - \chi(z_{min}),
\label{eq:rpar_max}
\end{equation}
where $\chi(z)$ is the co-moving line-of-sight distance to redshift $z$. In the above expression, $z_{min}$ and $z_{max}$ correspond to the minimum and maximum redshifts observed at, respectively, the highest and lowest frequencies, $\nu_{max}=114.750$~GHz and $\nu_{min}=84.278$~GHz, of the survey bandwidth, so that $z_{min} = \nu_{rest, \cotwomath}/\nu_{max}-1= 1.009 $, $z_{max} = \nu_{rest, \cotwomath}/\nu_{min}-1=1.735$, and $r_{\parallel,max}=1054.8$~Mpc~$h^{-1}$.

Similarly, the channel resolution establishes the smallest physical scale probed in the line-of-sight, $r_{\parallel,min}$:
\begin{equation}
r_{\parallel,min} = 
\chi(z_{chn,i+1})-\chi(z_{chn,i}).
\label{eq:rpar_min}
\end{equation}
Here, $z_{chn,i}$ and $z_{chn,i+1}$ correspond to redshifts of the $i$'th and $i$'th$+1$ channels at observed frequencies $\nu_i$ and $\nu_{i+1} = \nu_i+\Delta\nu_{chn}$, so that $z_{chn,i} = \nu_{rest, \cotwomath}/\nu_i -1$ and $z_{chn,i+1} =  \nu_{rest, \cotwomath}/(\nu_i+\Delta\nu_{chn}) -1$. Because $\Delta\nu_{chn}$ is a constant across the band, the physical separation between channels increases gradually with redshift. In practice, then, to facilitate computing the Fourier transform, we do not use Eq.~\ref{eq:rpar_min} when converting channel widths from frequency to physical distance. Rather, we define channel separations of equal width in space, dividing the total line-of-sight distance, $r_{\parallel,max}$, by the number of channels, $N_{chn}=196$, in the band, yielding $r_{\parallel,min} = 5.38$~Mpc~$h^{-1}$. This value is equal to the channel width at $\nu_{cen}$, and is a reasonable substitute for the true $r_{\parallel,min}$ per channel, due to the modest 12.6\% relative change in $r_{\parallel,min}$ from either band edge to bandcenter. Note that the choice of $N_{chn}=196$ reflects the fact that we have imaged the ASPECS data cube while re-binning the native ALMA channel resolution by a factor of 40 (i.e., $\Delta\nu_{chn}=\Delta\nu_{40chn}$), compared to the factor of 2 re-binning ($\Delta\nu_{chn}=\Delta\nu_{2chn}$) used when imaging the data cube for purposes of CO line searches, etc. The coarser spectral resolution $\Delta\nu_{40chn}=0.156$~GHz, with a velocity width $\Delta v_{40chn}\sim470$~km~s$^{-1}$ at $\nu_{cen}$, ensures that most CO emission is spectrally unresolved throughout the data cube: the median full-width half maximum (FWHM) of the Gaussian line flux profiles of the blindly detected lines in ASPECS is 355~km~s$^{-1}$, and the full range of observed FWHMs span 40.0~km~s$^{-1}$ to 617~km~s$^{-1}$ (GL19). We favor the larger channel width to avoid significant contributions to the power spectrum from emission lines with $\mathrm{FWHM} \gg \Delta v_{chn}$, since we do not attempt to characterize the effect of falsely elongating the observed flux density of these lines from the $\sim$kpc-scales of localized emission within the CO-bright galaxy to the $\sim$Mpc-scales when converting channel widths to cosmological line-of-sight distances.

In the transverse, or on-sky, dimensions the largest and smallest physical scales accessible to probe CO fluctuations in real-space,  $r_{\perp, max}$ and $r_{\perp, min}$, are determined by the survey width and synthesized beam size:
\begin{align}
r_{\perp, max} &= D_{\mathrm{A,co}}(z_{cen,\cotwomath}) \Delta\theta_S \label{eq:rperp_max} \\
r_{\perp, min} &= D_{\mathrm{A,co}}(z_{cen,\cotwomath}) \Delta\theta_b, \label{eq:rperp_min}
\end{align}
where the units of $\Delta\theta_{S}$ and $\Delta\theta_{b}$ are in radians, and $D_{\mathrm{A,co}}(z)$, the co-moving angular diameter distance at redshift $z$, is equal to $\chi(z)$ for $\Omega_{k}=0$.

As the antenna primary beam size grows with observed wavelength, $\Delta\theta_{b}$ and $\Delta\theta_{S}$ gradually increase with redshift across the survey bandwidth. As with $r_{\parallel, min}$, we adopt fixed values for each quantity calculated at $\nu_{cen}$, where change is modest to either band edge. At this frequency, the synthesized beam---an ellipse described by the FWHMs of its major and minor axes, $\Delta\theta_{b,maj}$ and $\Delta\theta_{b,min}$, respectively---for the full dataset is $\Delta\theta_{b} = \Delta\theta_{b,maj} \times \Delta\theta_{b,min} = 1.80 \ \mathrm{arcsec} \times 1.48 \ \mathrm{arcsec}$, corresponding to co-moving transverse distances 0.0250~Mpc~$h^{-1} \times 0.0205$~Mpc~$h^{-1}$. (For reference, $\Delta\theta_b=2.11~\mathrm{arcsec} \times 1.67~\mathrm{arcsec}$ at $\nu_{min}$, and $\Delta\theta_b=1.51~\mathrm{arcsec} \times 1.32~\mathrm{arcsec}$ at $\nu_{max}$.) Expressing the beam area as $A_b = \Delta\theta_{b,maj} \Delta\theta_{b,min}\pi/\left(4\ln2\right)$, we let $r_{\perp, min}=\sqrt{A_b}=0.0240$~Mpc~$h^{-1}$.  Note that the data cube for an interferometric image is gridded with rectangular cell sizes $\Delta\theta_{cell}$ a factor of a few times smaller than $\Delta\theta_{b}$, chosen such that $\Delta\theta_{cell}$ represents Nyquist sampling of the longest baseline visibility data \citep{TCP1999}. Thus, the smallest transverse dimension present in the dataset is actually $\Delta\theta_{cell}=0.36$~arcsec ($=0.005$~Mpc~$h^{-1}$ at $\nu_{cen}$), though there is no information on CO fluctuations contained within physical scales smaller than $\Delta\theta_{b}$.

\begin{figure}
\center
\begin{tabular}{c c c}
\includegraphics[width=0.33\textwidth]{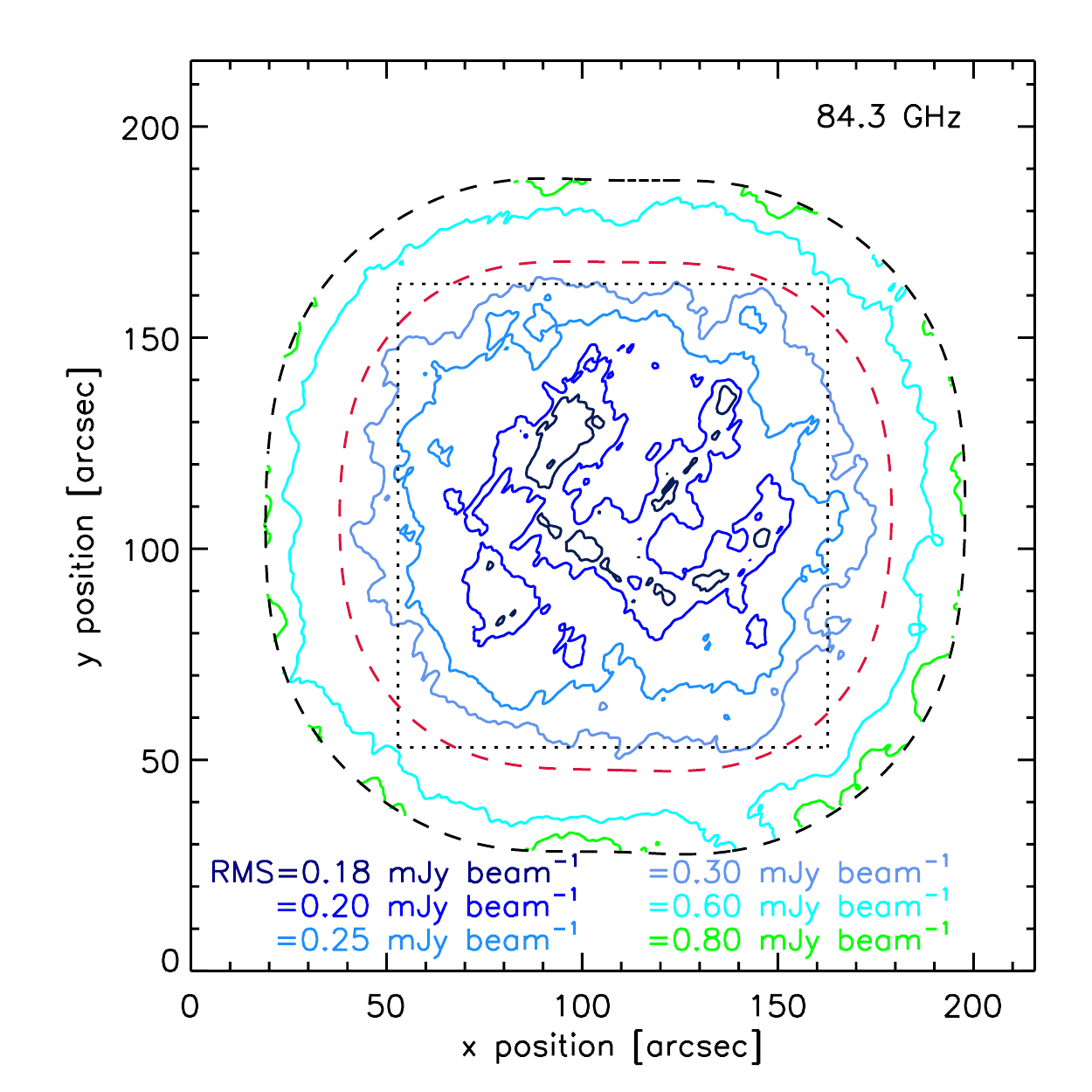} &
\hspace{-0.5cm}
\includegraphics[width=0.33\textwidth]{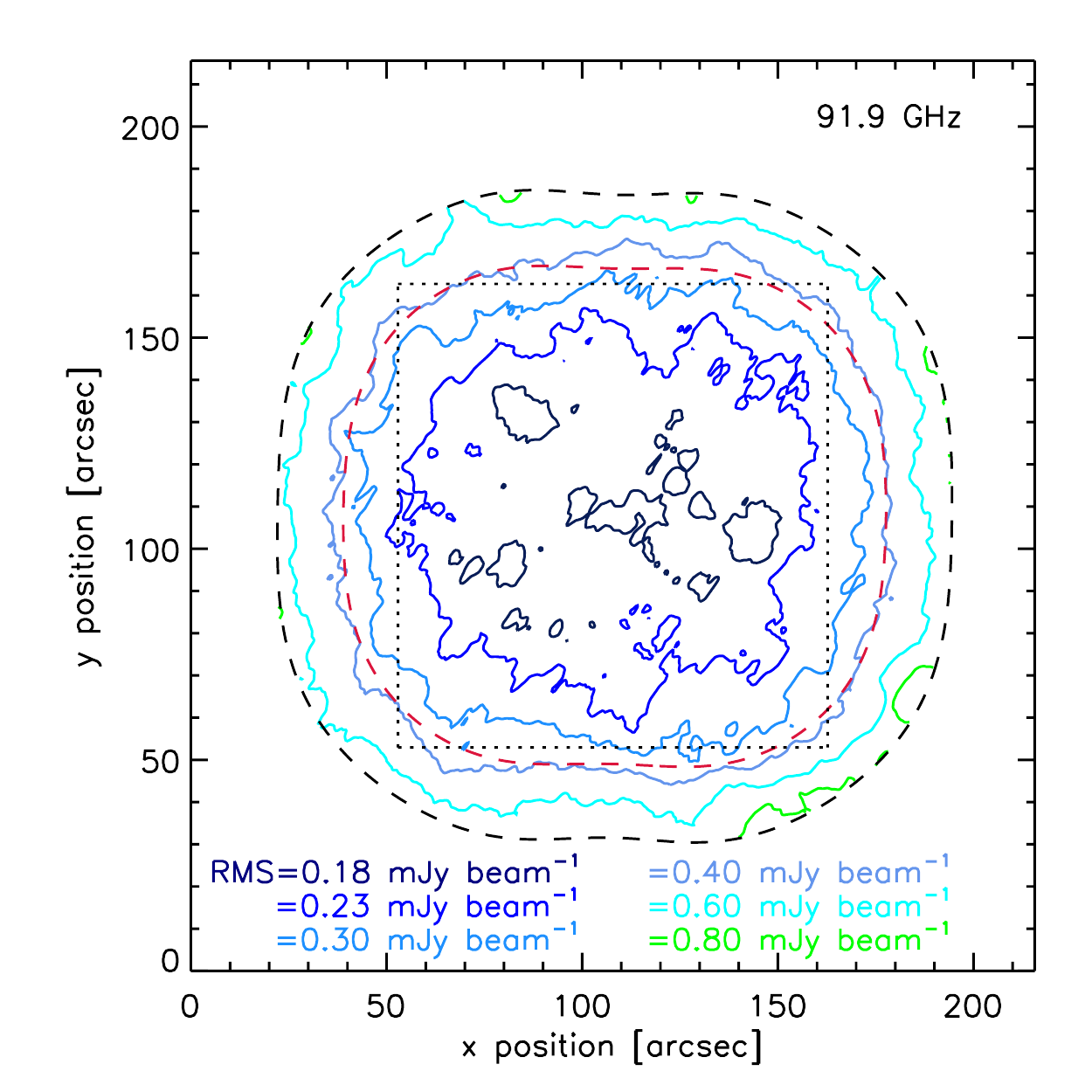} &
\hspace{-0.5cm}
\includegraphics[width=0.33\textwidth]{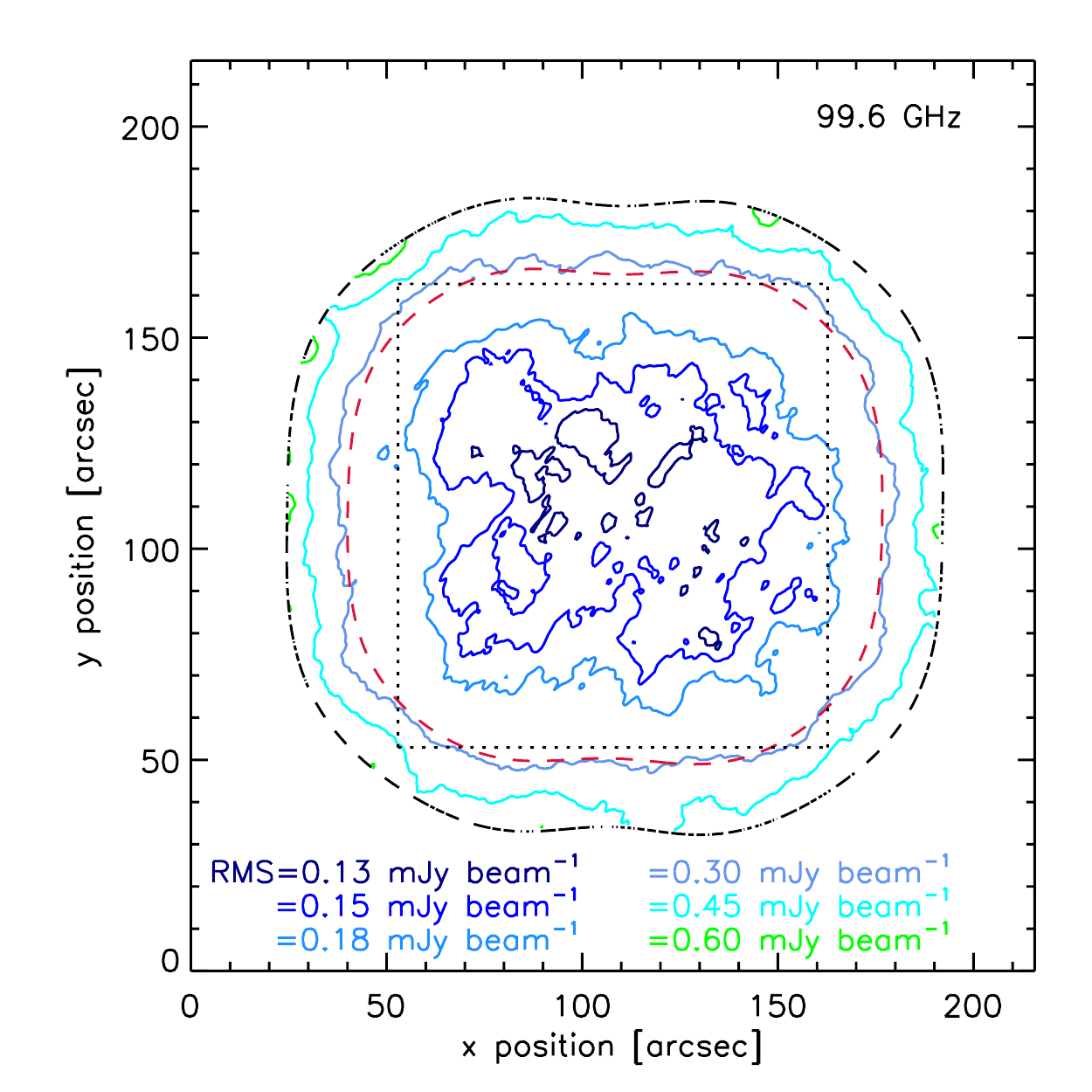} \\
\hspace{-0.5cm}
\includegraphics[width=0.33\textwidth]{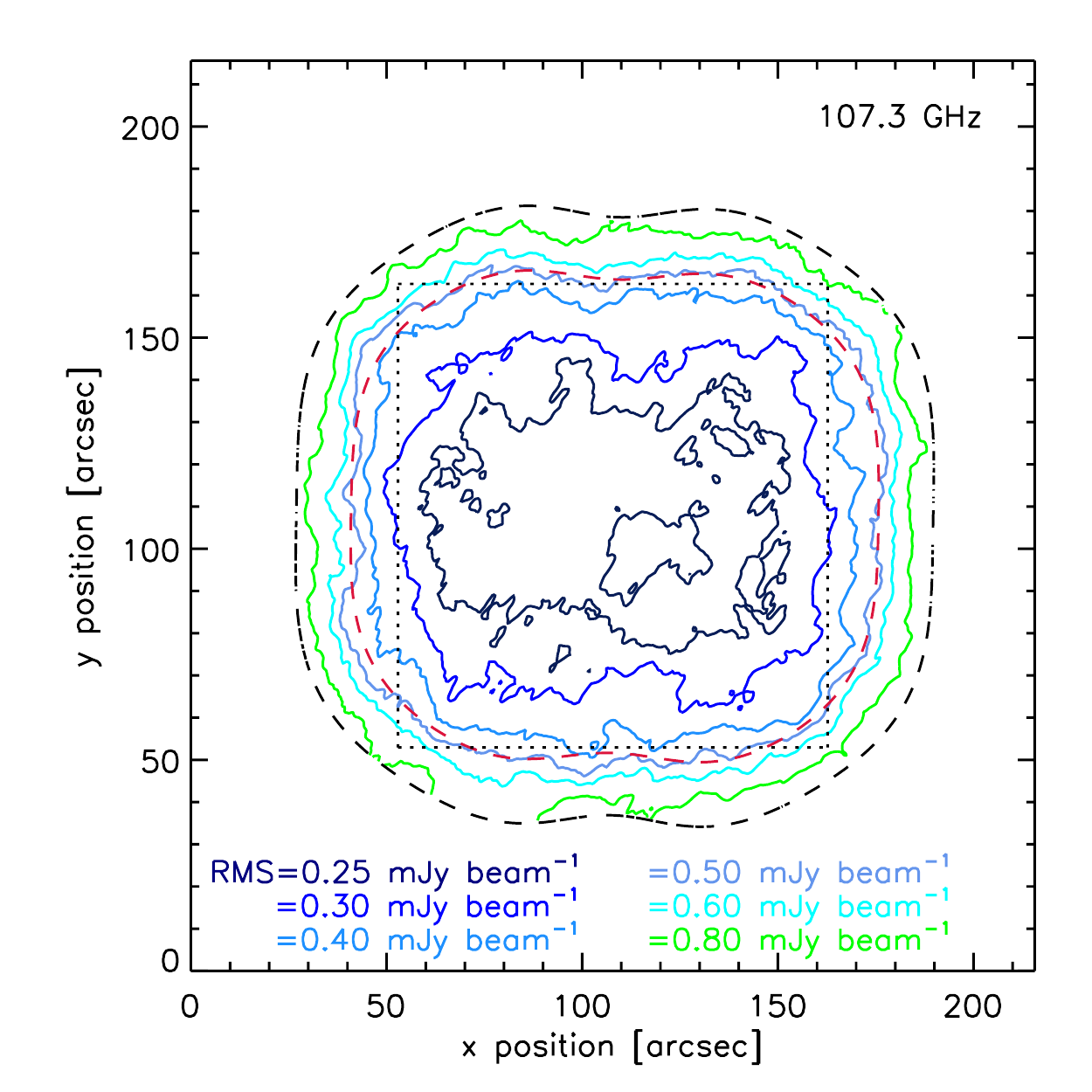} &
\includegraphics[width=0.33\textwidth]{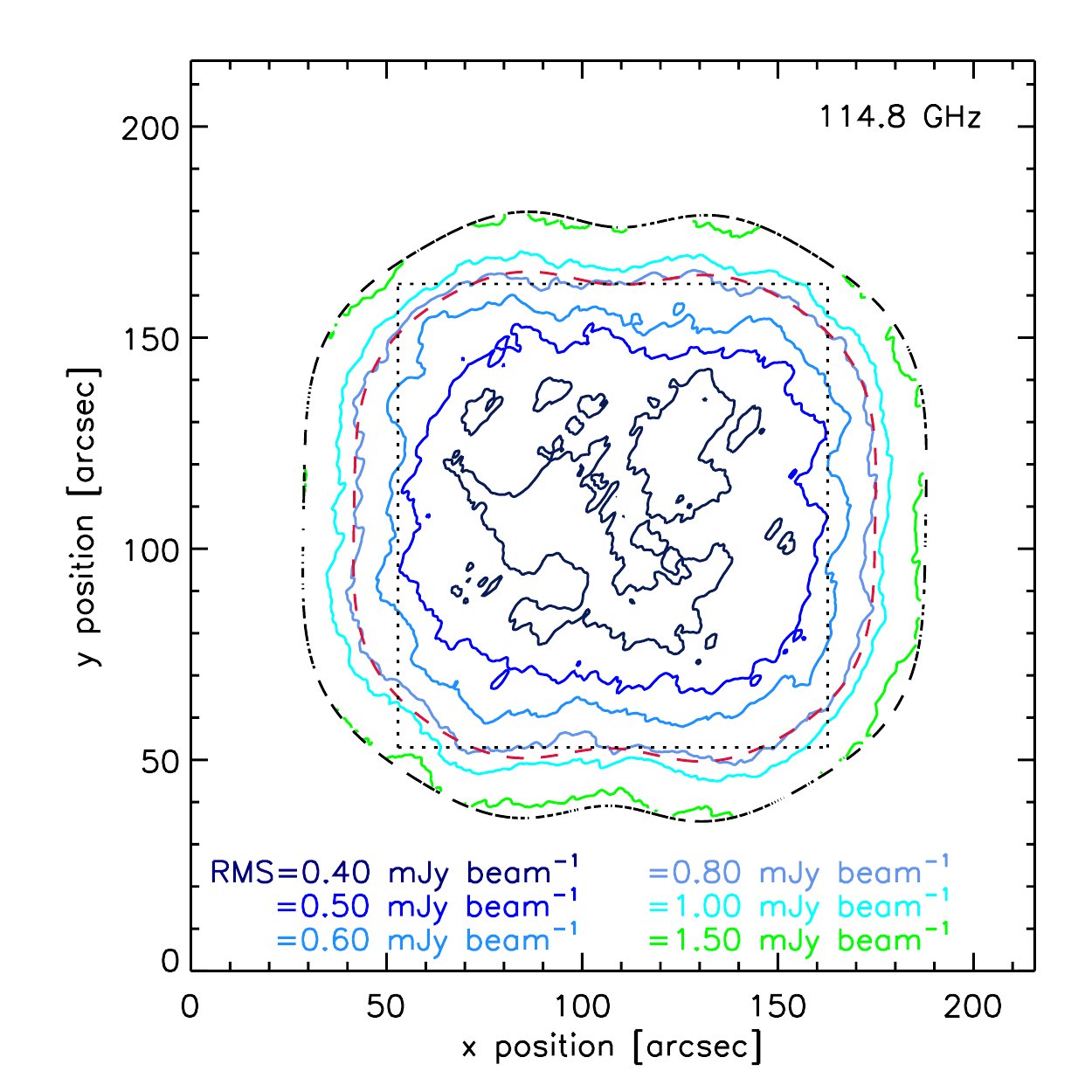} 
\end{tabular}
\caption{Noise maps at $\nu_{min}=84.3$~GHz, 91.9~GHz, $\nu_{cen}=99.6$~GHz, 107.3~GHz, and $\nu_{max}=114.8$~GHz. Solid contours indicate curves of constant RMS noise (in units of mJy~beam$^{-1}$). Notable changes in the respective RMS at different frequencies are due to overlap of frequency bands in the observations (top right panel) and decreasing atmospheric transmission at higher frequencies (bottom row) (GL19). The black and red dashed contours show, respectively, the mosaic primary beam cutoff and half-power point. The square (black, dotted) boundary used to define survey area, $\Delta\theta_S^2$, in the power spectrum analysis is overlayed on each map.}
\label{fig:noise_maps}
\end{figure}

At $\nu_{cen}$, the full width of the survey spans roughly $\Delta\theta_{S,tot}=2.83$~arcmin at a primary beam response cutoff of 20\%. The sensitivity profile of the mosaic primary beam implies, however, that the antenna response drops to 50\% at $\Delta\theta_{S,\mathrm{HPBW}}\approx2.15$~arcmin. Beyond this threshold, the RMS noise statistics deteriorate rapidly, as indicated by the noise map\footnote{Noise maps were generated by calculating the RMS, in units of mJy~beam$^{-1}$, for each pixel, using all surrounding data within a 70~pixel $\times$ 70~pixel box.} in Figure~\ref{fig:noise_maps} (lefthand panel, middle row for $\nu_{cen}$). Not only is the overall RMS noise higher for the survey field past the half-power point of the mosaic primary beam, but the spatial variation of the RMS is also significantly greater in this region, compared to the central $\sim4$~arcmin$^2$, e.g., where the RMS remains mostly between 0.13~mJy~beam$^{-1}$ per channel and 0.18~mJy~beam$^{-1}$ per channel, gradually reaching 0.3~mJy~beam$^{-1}$ per channel at the half-power point;\footnote{Here, the quoted RMS values refer to channel widths with $\Delta\nu_{chn}=\Delta\nu_{2chn}$, or, mJy~beam$^{-1}$ per 7.81~MHz channel.} beyond the half-power point, the RMS increases from 0.3~mJy~beam$^{-1}$ to 0.8~mJy~beam$^{-1}$ at the outermost edge of the mosaic, defined by the primary beam cutoff at 20\% antenna response.  We note, as well, that results from the blind search for individual CO emitters in ASPECS data (GL19) suggest lower fidelity (i.e., higher probability of false identification) of line candidates in the survey volume corresponding to $<50\%$ antenna response, so other studies (e.g., CO luminosity function measurements presented in D19 within the ASPECS collaboration have excluded data that lie outside $\Delta\theta_{S,\mathrm{HPBW}}$. Thus, we limit our analysis to a square region (shown as the black, dotted square boundary in Figure~\ref{fig:noise_maps}) with area $\Delta\theta_S^2 = \left(1.84~\mathrm{arcmin}\right)^2= \left(1.53~\mathrm{Mpc}~h^{-1}\right)^2$, chosen to lie within the 50\% power threshold at all observed frequencies. For reference, this region encompasses roughly 85\% of the volume contained within $\Delta\theta_{S,\mathrm{HPBW}}$, and 55\% of the volume within $\Delta\theta_{S,tot}$.

\subsubsection{Fourier-space dimensions}

Since we will be characterizing the CO fluctuation field by its power spectrum, we must relate the relevant physical scales probed in real-space (Equations~\ref{eq:rpar_max}--\ref{eq:rperp_min}) to the wavevectors with magnitude $k=\sqrt{k_{\parallel}^2+k_{\perp}^2}$ that are accessible to ASPECS in Fourier space:
\begin{gather}
k_{\parallel,min} = \frac{2\pi}{r_{\parallel,max}}~\mathrm{and}~k_{\perp,min} = \frac{2\pi}{r_{\perp,max}} \\
k_{\parallel,max} = \frac{2\pi}{2r_{\parallel,min}} ~\mathrm{and}~k_{\perp,max} = \frac{2\pi}{2r_{\perp,min}},
\end{gather}
where $k_{\parallel,min}$ and $k_{\perp,min}$ represent the lowest $k$ modes available in the line-of-sight and transverse dimensions, respectively; note that these fundamental modes map to the largest scales accessible in real space, with frequencies spanning a single oscillation across $\Delta\nu_{BW}$ and $\Delta\theta_S$. The highest $k$ modes probed by the survey, $k_{\parallel,max}$ and $k_{\perp,max}$, correspond to Nyquist frequencies, $k_{\parallel,\mathrm{Nyq}}\propto1/(2r_{\parallel,min})$ and $k_{\perp,\mathrm{Nyq}}\propto1/(2r_{\perp,min})$, mapping these modes to the smallest physical scales in the survey.

Table~\ref{tab:coord_mapping} summarizes the ASPECS survey parameters adopted for the power spectrum analysis, and their mappings to physical dimensions in real- and $k$-space. In Figure~\ref{fig:pco21_model_kscales}, we indicate the location of $k_{\parallel}$ and $k_{\perp}$ values for ASPECS relative to the predicted total CO(2-1) power---including both clustering and shot noise contributions---at $z=1$ from \citet{Sun2018}.  The ranges of transverse and line-of-sight $k$ modes have important implications to be considered when computing the power spectrum. 

For example, the large bandwidth and relatively narrow survey area of ASPECS dictate that the CO brightness fluctuations on physical scales larger than the survey width, $r_{\perp,max}=1.53$~Mpc~$h^{-1}$ (i.e., for $k < k_{\perp,min}=4.107~h$~Mpc$^{-1}$), will be probed exclusively by $k_{\parallel}$ modes. Thus, the power spectrum measured at $k<4.107$~$h$~Mpc$^{-1}$ will be an inherently one-dimensional (1D) measurement, dominated by power from the shorter wavelength, high-$k_{\perp}$ modes projected into the line-of-sight.  These physical scales are important, however, for extracting information about large-scale clustering. Based on models \citep[e.g.,][]{Sun2018, Pullen2013} for the total CO power spectrum at the redshift range relevant to this study, we expect any power from galaxy clustering between dark matter halos, $P_{\mathrm{CO,CO}}^{clust}(k,z)$, to dominate the total CO power spectrum, $P_{\mathrm{CO,CO}}^{tot}(k,z)$, up to $k \lesssim 1$~$h$~Mpc$^{-1}$ compared to contributions from small-scale clustering of galaxies that share a common host dark matter halo or shot noise power, $P_{\mathrm{CO,CO}}^{shot}$ (cf. Figure~\ref{fig:pco21_model_kscales}). If the CO surface brightness, $\langle T_{\mathrm{CO}} \rangle$, fluctuations trace the large-scale clustering of galaxies with some mean bias factor, $\langle b_{\mathrm{CO}}\rangle$, that offsets CO-emitting galaxies from the underlying dark matter distribution, i.e., if 
\begin{equation}
\label{eq:pclust}
P_{\mathrm{CO,CO}}^{clust}(k) = \langle T_{\mathrm{CO}}(z)\rangle^2 \langle b_{\mathrm{CO}}(z)\rangle^2 P_{\mathrm{m,m}}(k,z),
\end{equation}
where $P_{\mathrm{m,m}}(k,z)$ is the linear matter power spectrum (appropriate for $k<0.1$~h~Mpc$^{-1}$), then the low-$k$ component of the power spectrum is, in principle, useful for constraining the aggregate CO emission within a given cosmological volume. Note that the units of $P_{\mathrm{CO,CO}}^{clust}(k,z)$ in Equation~\ref{eq:pclust} are in $\mu$K$^2$~(Mpc h$^{-1}$)$^{3}$, for $\langle T_{\mathrm{CO}}(z)\rangle$ in $\mu$K, $P_{\mathrm{m,m}}(k,z)$ in (Mpc~h$^{-1}$)$^3$, and a dimensionless $b_{\mathrm{CO}}(z)$.

\begin{table*}
\centering
\caption{Mapping ASPECS survey parameters to real- and Fourier-space dimensions}
\begin{tabular}{l c}
\hline \hline
Survey bandwidth, $\Delta\nu_{BW}$ & 84.278--114.750~GHz \\
Channel resolution, $\Delta\nu_{chn}$ & 0.156~GHz \\
Survey width, $\Delta\theta_S$ & $1.84~\mathrm{arcmin}$ \\
Beam size, $\Delta\theta_b$ & $1.80 \ \mathrm{arcsec} \times 1.48 \ \mathrm{arcsec}$ \\
\hline
Central redshift, $z_{cen,\cotwomath}$ & 1.315 \\
$r_{\parallel, min} \le r_{\parallel} \le r_{\parallel, max}$ &  $5.38$~Mpc~$h^{-1} < r_{\parallel} < 1054.8$~Mpc~$h^{-1}$\\
$r_{\perp, min} \le r_{\perp} \le r_{\perp, max}$ &  $0.0240$~Mpc~$h^{-1} < r_{\perp} < 1.53$~Mpc~$h^{-1}$\\
$k_{\parallel, min} \le k_{\parallel} \le k_{\parallel, max}$ & $0.00596$~$h$~Mpc$^{-1} < k_{\parallel} < 0.584$~$h$~Mpc$^{-1}$\\
$k_{\perp, min} \le k_{\perp} \le k_{\perp, max}$ & $4.107$~$h$~Mpc$^{-1} < k_{\perp} < 130.900$~$h$~Mpc$^{-1}$\\
\hline
\label{tab:coord_mapping}
\end{tabular}
\end{table*}

For physical scales smaller than $r_{\perp,max}=1.53$~Mpc~h$^{-1}$ (i.e., for $k \ge k_{\perp,min}=4.107$~$h$~Mpc$^{-1}$), it is clear from Table~\ref{tab:coord_mapping} that $k$ can have contributions from both $k_{\perp}$ and $k_{\parallel}$ as long as $k=\sqrt{k_{\parallel}^2+k_{\perp}^2}$ is within the range $k=4.107$--130.900~$h$~Mpc$^{-1}$---we are measuring a full, three-dimensional (3D) power spectrum in this regime---though the $k_{\perp}$ modes, with wavelengths on order of $\Delta\theta_b$ up to $\Delta\theta_S$, provide most of the information on the power at these scales. At $z\sim1$, we expect any power from galaxy-galaxy clustering at $k\gtrsim4$~$h$~Mpc$^{-1}$ to be buried under a Poissonian shot noise component, $P_{\comath,\comath}^{shot}(k)$, dominated by bright CO emitters in the survey volume. By restricting our analysis to $k\gtrsim10$~h~Mpc$^{-1}$, where the true power spectrum is expected to be flat, the power spectrum measurement is unaffected by the highly anisotropic ASPECS survey window function.

\subsection{Inherent challenges to the auto-power spectrum measurement}

One of the intrinsic benefits of the auto-power spectrum measurement is its sensitivity to intensity fluctuations from all sources, faint and bright, contained within the survey volume. The inclusion in the power spectrum analysis of all flux densities present in the data cube also presents specific challenges to the interpretation of the measured power. For the purpose of the ASPECS power spectrum analysis, a primary concern is the redshift ambiguity of CO emission within the observed survey bandwidth. We also discuss briefly the effects of possible contribution from continuum emission.

\subsubsection{Redshift ambiguity of CO emission} \label{sec:line_ambiguity}

As with blindly detected, individual line candidates, where the redshift of a line candidate without spectroscopically or photometrically confirmed counterparts can be ambiguous, the true redshifts of sources contributing to the intensity fluctuations contained within the ASPECS survey volume are unknown. However, in defining a real- and Fourier space grid to perform our power spectrum calculations, we have assumed a specific target redshift $z_{cen,\cotwomath}=1.315$---corresponding to the central redshift of CO(2-1) in the ASPECS bandwidth---for the emission. While CO(2-1) is expected to dominate the mean surface brightness at 99~GHz, based on the number of blind detections in ASPECS relative to other line transitions, for example, it is not the only source of spectral line emission present in the survey volume. Figure~2 of D19 illustrates the number of different spectral lines that are, in principle, observable within the ASPECS frequency coverage, emitted from galaxies within the local Universe (such as CO(1-0) at $z < 0.37$) and high-redshift (such as any CO transition from $J>2$ at $z>2$). Thus, the measured auto-power spectrum of the ASPECS data should be interpreted as a sum of the power from surface brightness fluctuations from all relevant CO transitions:
\begin{align}
\label{eq:pspec_sum}
P_{\comath, \comath}(k) &= P_{\coonemath, \coonemath}(k_{\coonemath}) \nonumber \\
& \qquad {} + P_{\cotwomath,\cotwomath}(k_{\cotwomath}) \nonumber \\
& \qquad {} + P_{\cothreemath,\cothreemath}(k_{\cothreemath}) \nonumber \\
& \qquad {} + P_{\cofourmath,\cofourmath}(k_{\cofourmath}) \nonumber + \dots, \\
\end{align}
where we truncate the sum, in practice, to include contributions from CO(1-0), CO(2-1), CO(3-2), and CO(4-3), because the ASPECS survey has only resulted in CO blind detections up to $J=4$ (GL19). Each term in the righthand side of the above equation is expressed as a function of wavenumber $k_{\mathrm{CO(J\mhyphen(J-1))}}$, corresponding to the Fourier space defined at the emitted redshift of the respective $J$ transition at bandcenter. For reference, at $\nu_{cen}=99.572$~GHz, CO(1-0), CO(3-2), and CO(4-3) can be emitted from $z = 0.157, 2.470$ and 3.629, respectively. The $k$ appearing on the lefthand side of Equation~\ref{eq:pspec_sum} is intentionally ambiguous; ultimately, we would like to write $P_{\comath, \comath}(k)$ as $P_{\comath, \comath}(k_{\cotwomath})$. We can convert $k_{\mathrm{CO(J\mhyphen(J-1))}}$ to $k_{\cotwomath}$, defined at $z_{cen,\cotwomath}$, using so-called ``distortion" factors (as given in, e.g., \citet{LidzTaylor2016}),
\begin{equation}
\alpha_{\perp}(z_{cen,\mathrm{CO(J\mhyphen(J-1))}}) = \frac{k_{\perp,\cotwomath}}{k_{\perp,\mathrm{CO(J\mhyphen(J-1))}}} =  \frac{ D_{\mathrm{A,co}}(z_{cen,\mathrm{CO(J\mhyphen(J-1))}})}{D_{\mathrm{A,co}}(z_{cen,\cotwomath})}
\label{eq:alpha_perp}
\end{equation}
and
\begin{equation}
\label{eq:alpha_par}
\alpha_{\parallel}(z_{cen,\mathrm{CO(J\mhyphen(J-1))}}) = \frac{k_{\parallel,\cotwomath}}{k_{\parallel,\mathrm{CO(J\mhyphen(J-1))}}} = \frac{H(z_{cen,\cotwomath})}{H(z_{cen,\mathrm{CO(J\mhyphen(J-1))}})} \frac{\left(1+z_{cen,\mathrm{CO(J\mhyphen(J-1))}}\right)}{\left(1+z_{cen,\cotwomath}\right)}
\end{equation}
that relate transverse and line-of-sight co-moving distances, respectively, between the true emitted redshift of intensity fluctuations, $z_{cen,\mathrm{CO(J\mhyphen(J-1))}}$, and the adopted redshift $z_{cen,\cotwomath}$. In Equation~\ref{eq:alpha_par},  the expression $H(z)$ refers to the Hubble parameter at redshift $z$. Finally, combining Equations~\ref{eq:pspec_sum}--\ref{eq:alpha_par}, we obtain the total CO power in terms of $k_{\cotwomath}$:
\begin{align}
P_{\comath,\comath}(k_{\cotwomath}) &= P_{\cotwomath,\cotwomath}(k_{\cotwomath}) \nonumber \\
& \qquad {} + \sum_{J=1,3,4} \left[ \frac{1}{\alpha_{\perp}(z_{cen,\mathrm{CO(J\mhyphen(J-1))}})^2 \alpha_{\parallel}(z_{cen,\mathrm{CO(J\mhyphen(J-1))}})} \right. \nonumber \\
& \qquad {} \left. \times \ P_{\mathrm{CO(J\mhyphen(J-1)),\mathrm{CO(J\mhyphen(J-1))}}}\left(\frac{k_{\perp,\cotwomath}}{\alpha_{\perp}(z_{cen,\mathrm{CO(J\mhyphen(J-1))}})}, \frac{k_{\parallel,\cotwomath}}{\alpha_{\parallel}(z_{cen,\mathrm{CO(J\mhyphen(J-1))}})}\right) \right]
\label{eq:pspec_sum_kco21}
\end{align}
The multiplicative pre-factor $1/(\alpha_{\perp}^2 \alpha_{\parallel})$ in the second term on the righthand side of the above equation represents the ratio of volume probed by the survey in CO(2-1) relative to the other $J$ transitions. The ratio $1/(\alpha_{\perp}^2 \alpha_{\parallel})> 1$ for $J=1$, reflecting the fact that the volume probed by CO(1-0) is less than the volume probed by CO(2-1), while the opposite is true for the $J=3$ and $J=4$ transitions, where $\frac{1}{\alpha_{\perp}^2 \alpha_{\parallel}} < 1$. Contributions to the total measured power from different CO transitions are correspondingly magnified or demagnified when projected into the CO(2-1) frame (Equation~\ref{eq:pspec_sum_kco21}).

Note that Equation~\ref{eq:pspec_sum_kco21} is only valid for large separations in redshift between the sources of CO emission; if there is overlap between redshift ranges of the different transitions in the ASPECS survey volume, then Equation~~\ref{eq:pspec_sum_kco21} will contain cross-terms that represent the cross-power spectrum between the CO transitions that overlap in redshift. For the ASPECS spectral coverage, we point out that there is a small overlap in redshift for the CO(3-2) and CO(4-3) in the survey at $z=3.011$--3.107 (cf. Table~1 in D19), but cross-terms here will be negligible given that the mean redshifts $z_{cen,\cothreemath}=2.470$ and $z_{cen,\cofourmath}=3.629$ are widely separated. 

\subsubsection{Continuum emission} \label{sec:continuum}

A search for continuum emission in the ASPECS LP Band 3 data was presented in GL19. This study identified 6 continuum sources, with the brightest emission on the order $\sim10$~$\mu$Jy, indicating that the continuum level in each channel of the 3~mm cube is negligible for our purposes. To ensure, however, that our power spectrum measurements reflect power from spectral line (CO) fluctuations only, and do not contain contributions from continuum, we perform continuum subtraction on the cube with a linear baseline fit, described in Section~\ref{sec:data_methods}. This continuum-subtracted cube is used for all power spectrum and cross-power spectrum analyses.

\section{Data and Methods} \label{sec:data_methods}
\begin{table}[b]
\centering
\caption{Blind CO detections in ASPECS LP 3~mm survey.}
\begin{tabular}{c c c c c}
\hline \hline
ID & Line & $\nu_{obs}$ & Flux & Redshift \\
     &         &    [GHz]        & [Jy~km~s$^{-1}$] & \\
(1) & (2) &    (3)                &    (4)               & (5) \\     
\hline
ASPECS-LP-3mm.01$^{a}$ & CO(3-2) & 97.584 & $1.02\pm0.04$ & 2.543 \\
ASPECS-LP-3mm.02$^{a}$ & CO(2-1) & 99.513 & $0.47\pm0.04$ & 1.317 \\
ASPECS-LP-3mm.03$^{a}$ & CO(3-2) & 100.131 & $0.41\pm0.04$ & 2.454 \\
ASPECS-LP-3mm.04$^{a}$ & CO(2-1) & 95.501 & $0.89\pm0.07$ & 1.414 \\
ASPECS-LP-3mm.05$^{a}$ & CO(2-1) & 90.393 & $0.66\pm0.06$ & 1.550 \\
ASPECS-LP-3mm.06$^{a}$ & CO(2-1) & 110.038 & $0.48\pm0.06$ & 1.095 \\
ASPECS-LP-3mm.07$^{a}$ & CO(3-2) & 93.558 & $0.76\pm0.09$ & 2.696 \\
ASPECS-LP-3mm.08$^{a}$ & CO(2-1) & 96.778 & $0.16\pm0.03$ & 1.382 \\
ASPECS-LP-3mm.09 & CO(3-2) & 93.517 & $0.40\pm0.04$ & 2.698 \\
ASPECS-LP-3mm.10 & CO(2-1) & 113.192 & $0.59\pm0.07$ & 1.037 \\
ASPECS-LP-3mm.11$^{a}$ & CO(2-1) & 109.966 & $0.16\pm0.03$ & 1.096 \\
ASPECS-LP-3mm.12 & CO(3-2) & 96.757 & $0.14\pm0.02$ & 2.574 \\
ASPECS-LP-3mm.13 & CO(4-3) & 100.209 & $0.13\pm0.02$ & 3.601 \\   
ASPECS-LP-3mm.14 & CO(2-1) & 109.877 & $0.35\pm0.05$  & 1.098 \\
ASPECS-LP-3mm.15 & CO(2-1) & 109.971 & $0.21\pm0.03$ & 1.096 \\
ASPECS-LP-3mm.16 & CO(2-1) & 100.503 & $0.08\pm0.01$ & 1.294 \\
\hline
\multicolumn{5}{p{.6\textwidth}}{$^{a}$Source is classified as extended in GL19.} \\
\multicolumn{5}{p{.6\textwidth}}{---\emph{Notes:} (1) Catalog ID.  (2) Identified line transition. (3) Observed frequency at line center. (4) Integrated line flux from Table~6 of GL19. (5) Redshift of observed CO transition.} 
\end{tabular}
\label{tab:COdets}
\end{table}
The ASPECS LP survey consisted of two blind frequency scans at 1.2~mm and 3~mm in the \emph{Hubble} Ultra Deep Field \citep[HUDF; ][]{Beckwith2006}. The methods and subsequent power spectrum analysis presented here utilize the 3~mm observations, which consist of a 17-pointing mosaic over $\sim4.7$~arcmin$^2$ at five frequency tunings spanning the full extent of ALMA Band 3. As described in D19 (see their Section 2.2 for more detail) of this series, the resulting visibility data were imaged using the CASA task \textsf{tclean}---with natural weighting applied in the \emph{uv}-plane and frequency rebinning over 2 of the native 3.91~MHz spectral resolution elements---to produce an image cube with mean RMS $=\langle\sigma_{N,\mathrm{2chn}} \rangle = 1.96\times10^{-4}$~Jy~beam$^{-1}$ per channel, across all  3,935 channels in the cube. Here, the RMS per channel of the data cube has been inferred by computing the 
RMS in the central 70 by 70 pixels for each channel map; given the lack of known sources in this 70 by 70 pixel-wide skewer through the data cube, the RMS in this region is expected to be a valid representation of the noise in the cube. (Recall that Figure~\ref{fig:noise_maps} shows the spatial variation of the RMS noise at a number of representative frequencies.) At $\nu_{cen}$, for example, this mean RMS translates to a mean surface brightness sensitivity in units of Jy~sr$^{-1}$ and, via the Rayleigh-Jeans Law, a mean brightness temperature: $\langle\sigma_{N,\mathrm{2chn}} \rangle = 2.76\times10^{6}$~Jy~sr$^{-1} = 9.07\times10^3$~$\mu$K. In the context of a power spectrum analysis, this mean RMS can give rise to a spectrally featureless (i.e., ``white") noise power, $P_N$:
\begin{equation}
P_N = \langle\sigma_{N,\mathrm{2chn}} \rangle^2 V_{vox,\mathrm{2chn}},
\label{eq:pnoise}
\end{equation}
where $V_{vox,\mathrm{2chn}}$ refers to the voxel volume defined by the beam area and channel width. At $z_{cen}$, $V_{vox, \mathrm{2chn}} = \Delta\theta_b^2 \Delta\nu_{2chn} = (0.024~\mathrm{Mpc \ h^{-1}})^2 \times (0.27~\mathrm{Mpc \ h^{-1}}) = 1.57\times10^{-4}$~(Mpc~h$^{-1}$)$^3$, and implies $P_N = 1.29\times10^4$~$\mu$K$^2$~(Mpc~h$^{-1}$)$^3$. This image cube, referred to in this work as $T_{0,\mathrm{2chn}}$, has served as the principal data product exploited in a variety of analysis efforts by the ASPECS team, including the identification of blindly detected, individual line emitters. The catalog of reliable---specifically, where the probability that the line is due to noise has been determined in GL19 to be less than 10\%---blind detections is reproduced in Table~\ref{tab:COdets}. 

As already discussed in Section~\ref{sec:realspacedims}, the 7.81~MHz channel width is too fine a spectral resolution for the purposes of the power spectrum analysis. Thus, unless otherwise noted, we have imaged with a frequency re-binning over 40 native spectral resolution elements to obtain an image cube $T_{0, \mathrm{40chn}}$ (or $T_0$, hereafter, for brevity) characterized by a lower mean RMS, $\langle \sigma_{N,\mathrm{40chn}} \rangle = 4.55\times 10^{-5}$~Jy~beam$^{-1}$~channel$^{-1}$ $\approx \langle \sigma_{N,\mathrm{2chn}} \rangle / \sqrt{20}$, as depicted in the bottom righthand panel of Figure~\ref{fig:timesplit_properties}. Note that the noise power remains unchanged as the larger channel width counteracts the change in $\langle\sigma_{N,\mathrm{40chn}} \rangle^2$ (cf. Equation~\ref{eq:pnoise}). 

After imaging and applying a primary beam correction to correct  flux densities for the effect of the mosaic sensitivity pattern, we estimate and subtract any possible continuum emission by running CASA task \textsf{imcontsub} in the full cube to ensure that all surface brightness fluctuations in $T_0$ are due to spectral emission. The continuum was approximated using a linear baseline fit across all channels to prevent introducing artificial spectral structure. We inspected RMS levels and spectra at random positions in the cube before and after continuum subtraction, finding negligible (less than 0.1--1\%) change in both quantities, confirming our expectations based on GL19 (cf. \ref{sec:continuum}). 

For the purposes of the power spectrum analysis, however, we do not work directly with $T_0$ or $T_{0, \mathrm{2chn}}$. That is, we do not assess the level of astrophysical signal in the 3~mm dataset by taking the (auto-)power spectrum of $T_0$, $P_{T_0,T_0}(k)$, defined as
\begin{equation}
\langle T_0^{*}(\vec{k}) T_0(\vec{k'}) \rangle \equiv (2\pi)^3 \delta_{\mathrm{D}}(\vec{k}-\vec{k'}) P_{T_0,T_0}(k)
\label{eq:pspec_def}
\end{equation}
where $(2\pi)^3 \delta_{\mathrm{D}}(\vec{k}-\vec{k'}) = \int \mathrm{d}^3x \ e^{-i(\vec{k}-\vec{k'})\cdot\vec{x}}$ is a Dirac delta function. Explicitly, $P_{T_0,T_0}(k)$ is the 3-dimensional average of the Fourier transform of the 2-point correlation function, $\xi(|\vec{x}-\vec{x'}|)=\xi(r)$: $\int \mathrm{d}^3\mathrm{r} \ e^{-i \vec{k}\cdot\vec{r}}\xi(r)$. We note, however, that the power spectrum in Equation~\ref{eq:pspec_def} can include contributions from astrophysical signal at the target and/or other redshifts, as well as instrument noise, characterized by $P_N$. In Sections~\ref{sec:line_ambiguity} and \ref{sec:continuum}, we explained why the main astrophysical source of surface brightness fluctuations in the data cube is expected to be the CO line transitions (from $J=1,2,3,4$). As our estimated noise power is 2--3 orders of magnitude greater than the predicted CO power spectrum signal, the instrument noise introduces a significant bias in the measured power spectrum throughout all $k$ probed by the survey. Thus, in this low signal-to-noise regime, we seek a way to \emph{remove} the noise-bias from the data in order to accurately measure the CO signal; we avoid \emph{subtracting} this noise-bias term from the data based on independent estimates of $P_N$ that may not reflect the true noise amplitude in the data, or exhibit deviations from gaussianity that would, for example, invalidate Equation~\ref{eq:pnoise}, which assumes the measured RMS describes a white-noise random field. 

\citet{Dillon2014} demonstrate that it is possible---and, indeed, preferable when the expected signal-to-noise is sub-unity, as in the current generation of Reionization-era 21~cm intensity mapping experiments discussed in their paper---to remove the noise-bias by computing the cross-power spectrum of two data cubes, e.g., $T_{\mathrm{I}}$ and $T_{\mathrm{II}}$, that are derived as subsets of the original data cube, $T_0$, in a manner that preserves the real- and Fourier-spaces sampled by $T_0$. For ASPECS data, we can split the original CASA measurement set\footnote{Throughout, this paper, we use the tilde ( $\widetilde{}$ ) symbol to denote a visibility dataset $\widetilde{T}$ used to generate an image cube $T$.} $\widetilde{T}_0$ into two subsets, such that the sum of visibilities in $\widetilde{T}_{\mathrm{I}}$ and $\widetilde{T}_{\mathrm{II}}$ gives $\widetilde{T}_0$ , i.e., $\widetilde{T}_{\mathrm{I}} + \widetilde{T}_{\mathrm{II}} = \widetilde{T}_0$, in order to produce corresponding image cubes $T_{\mathrm{I}}$ and $T_{\mathrm{II}}$. The working assumption here is that the cross-power spectrum between $T_{\mathrm{I}}$ and $T_{\mathrm{II}}$, $P_{T_{\mathrm{I}}, T_{\mathrm{II}}}(k)$, given by 
\begin{equation}
\langle T_{\mathrm{I}}^{*}(\vec{k}) T_{\mathrm{II}}(\vec{k'}) \rangle \equiv (2\pi)^3 \delta_{\mathrm{D}}(\vec{k}-\vec{k'}) P_{T_{\mathrm{I}},T_{\mathrm{II}}}(k),
\label{eq:xspec_TITII}
\end{equation}
contains only astrophysical signal and, in principle, any residual correlated noise; random noise present in each cube will be uncorrelated and produce zero mean signal in the cross. Hereafter, we refer to $P_{T_{\mathrm{I}}, T_{\mathrm{II}}}(k)$ as the noise-bias free power spectrum.

Errors on $P_{T_{\mathrm{I}}, T_{\mathrm{II}}}(k)$, $\delta P_{T_{\mathrm{I}}, T_{\mathrm{II}}}(k)$, can be similarly evaluated by first creating additional subsets $\tau$ of two visibility datasets from each parent visibility dataset, $\widetilde{T}_{\mathrm{I}}$ or $\widetilde{T}_{\mathrm{II}}$, such that $\widetilde{\tau}_1 + \widetilde{\tau}_2 = \widetilde{T}_{\mathrm{I}}$ and $\widetilde{\tau}_3 + \widetilde{\tau}_4 = \widetilde{T}_{\mathrm{II}}$, e.g. Then, the cross-power spectrum (performed, again, in the image domain) between the mathematical differences of each pair is computed to yield the error on $P_{T_{\mathrm{I}}, T_{\mathrm{II}}}(k)$, 
\begin{equation}
\delta P_{T_{\mathrm{I}}, T_{\mathrm{II}}}(k) = \gamma_N \langle \left(\tau_1(\mathbf{k}) - \tau_2(\mathbf{k})\right)^* \left(\tau_3(\mathbf{k}) - \tau_4(\mathbf{k})\right) \rangle.
\label{eq:error_on_Pofk}
\end{equation}
In this scheme, the purpose of differencing the two data cubes derived from either $T_{\mathrm{I}}$ or $T_{\mathrm{II}}$ is to remove signal from $T_{\mathrm{I}}$ or $T_{\mathrm{II}}$, respectively, such that the mathematical difference represents noise-only data. Then, one can compute the cross-power spectrum between the pair of differences to remove the noise-bias in the noise-only data, yielding a so-called ``noise-bias free" error on the ``noise-bias free" power spectrum, $P_{T_{\mathrm{I}}, T_{\mathrm{II}}}(k)$. Furthermore, we can create two additional realizations of $\delta P_{T_{\mathrm{I}}, T_{\mathrm{II}}}(k)$ by re-ordering the differences, and obtain a final, average error, $\langle \delta P_{T_{\mathrm{I}}, T_{\mathrm{II}}}(k) \rangle$ as follows: 
\begin{align}
\langle \delta P_{T_{\mathrm{I}}, T_{\mathrm{II}}}(k) \rangle &= \frac{\gamma_N}{3} \left( \langle \left(\tau_1(\mathbf{k}) - \tau_2(\mathbf{k})\right)^* \left(\tau_3(\mathbf{k}) - \tau_4(\mathbf{k})\right) \rangle \right. \nonumber \\
& \qquad {} + \left. \langle \left(\tau_1(\mathbf{k}) - \tau_3(\mathbf{k})\right)^* \left(\tau_2(\mathbf{k}) - \tau_4(\mathbf{k})\right) \rangle \right. \nonumber \\
&  \qquad {} + \left. \langle \left(\tau_1(\mathbf{k}) - \tau_4(\mathbf{k})\right)^* \left(\tau_2(\mathbf{k}) - \tau_3(\mathbf{k})\right) \rangle \right)
\label{eq:final_error}
\end{align}
The pre-factor, $\gamma_N$, is determined by relating the expected (and actual) noise properties of $\tau_1$, $\tau_2$, $\tau_3$, and $\tau_4$, to the parent data cubes $T_{\mathrm{I}}$ and $T_{\mathrm{II}}$ and grandparent data cube $T_0$. Specifically, in the case that only thermal noise is present in the data, then---as long as the number of visibilities in $\widetilde{T}_0$ is divided equally among $\widetilde{T}_{\mathrm{I}}$ and $\widetilde{T}_{\mathrm{II}}$, and the weights (determined by antenna system temperatures) on the visibility data are not dramatically different in one subset compared to the other---the resulting mean RMS in $T_\mathrm{I}$ and $T_{\mathrm{II}}$ will be equal to $\sqrt{2}\langle\sigma_{N,\mathrm{40chn}}\rangle$, where $\langle\sigma_{N,\mathrm{40chn}}\rangle$ refers to the mean RMS in $T_0$, derived earlier in this Section; and, the cross-power spectrum (Equation~\ref{eq:xspec_TITII}) will have a noise covariance equal to $2\langle\sigma_{N,\mathrm{40chn}}\rangle^2$. Similarly, if the same conditions hold for the splitting of CASA measurement sets corresponding to $\widetilde{T}_{\mathrm{I}}$ into $\widetilde{\tau}_1$ and $\widetilde{\tau}_2$, and $\widetilde{T}_{\mathrm{II}}$ into $\widetilde{\tau}_3$ and $\widetilde{\tau}_4$, then each image $\tau_1$ through $\tau_4$ will have a mean RMS $=2\langle \sigma_{N,\mathrm{40chn}} \rangle$.  Figure~\ref{fig:timesplit_properties} (lower righthand panel), which shows that the RMS at all $\nu_{obs}$ in $\tau_1$ through $\tau_4$ (colored curves), matches well the RMS level representing $2\langle \sigma_{N,\mathrm{40chn}} \rangle$ (black, dotted curve), confirms the assumptions that our noise is predominantly thermal, that the visibility datasets have been divided equally among the subsets, and that visibility weights do not differ significantly among the subsets. (We point out that the RMS in $\tau_4$ is slightly higher than the black, dotted curve at some frequency intervals, and have verified that this discrepancy is due to a smaller number of visibilities entering into this subset at those frequencies.) Then, the images representing mathematical differences $(\tau_1 -\tau_2)$ and $(\tau_3 - \tau_4)$ will each have mean RMS = $2\sqrt{2} \langle \sigma_{N,\mathrm{40chn}} \rangle$, and the resulting cross-power spectrum (Equation~\ref{eq:error_on_Pofk}) will have a noise covariance equal to $8 \langle \sigma_{N,\mathrm{40chn}} \rangle^2$. Therefore, we find $\gamma_N=0.25$. The process is visualized as a tree diagram in the upper panel of Figure~\ref{fig:timesplit_properties}. 

In practice, we forego the step of dividing $\widetilde{T}_0$ into subsets $\widetilde{T}_{\mathrm{I}}$ and $\widetilde{T}_{\mathrm{II}}$, and begin by dividing $\widetilde{T}_0$ into quarters $\widetilde{\tau}_1$ through $\widetilde{\tau}_4$. The CASA measurement set represented by $\widetilde{T}_0$ contains visibility data corresponding to 11 frequency ranges, or spectral windows,\footnote{We do not refer to the four 1.875~GHz spectral windows of the ALMA sidebands.} that have been stitched together from all available data so that spectral windows in $T_0$ are ordered from lowest to highest observed frequency, with no overlapping regions. (Please refer to Section~2.4 of \citet{Walter2016_survey} for more details on the construction of $T_0$.) For each spectral window in $T_0$, there are multiple blocks of visibility data corresponding to the various execution blocks scheduled by ALMA for observing. Each block, in turn, is typically comprised of 8--9 scans that repeat 17 times to cover the entire spatial area of the mosaic. Using CASA task \textsf{split}, we select and distribute these scans evenly among four subsets. We repeat these steps for every block of scans and every spectral window, and merge visibilities in the four subsets using CASA task \textsf{concat}---followed by \textsf{statwt}, for a homogenous weighting system in the concatenated data---to produce the subsets $\widetilde{\tau}_1$, $\widetilde{\tau}_2$, $\widetilde{\tau}_3$, and $\widetilde{\tau}_4$. Our choice of dividing $T_0$ this way guards against possible frequency and/or temporal biases, as each subset contains the full range of frequency (84--115~GHz)---required, moreover, so that all image cubes probe the same line-of-sight distance---and time (December 2--December 21, 2016) covered by the observations.  The splittings have also resulted in small (i.e., less than the 0.36~arcsecond cell size in the gridded image cubes) variations in beam sizes between subsets (cf. lower lefthand panel in Figure~\ref{fig:timesplit_properties}), which is necessary when taking the cross-power spectrum, or performing other mathematical operations like subtraction, between any two images.  \footnote{Splitting $\widetilde{T}_0$ must be done in such a way that preserves the real and Fourier-spaces probed by $T_0$. For example, if $\widetilde{T}_0$ were split into two sets $\widetilde{T}_{\mathrm{I}}$ and $\widetilde{T}_{\mathrm{II}}$ that contained visibilities from the first and second half of channels, respectively, in $T_0$, then $\widetilde{T}_{\mathrm{I}} + \widetilde{T}_{\mathrm{II}} = \widetilde{T}_0$ would still hold, but the images $T_{\mathrm{I}}$ and $T_{\mathrm{I}}$ would each probe only half of the volume in $T_0$.} 

\section{Results} \label{sec:results}
 
\subsection{Limits from detected sources}

\subsubsection{Mean surface brightness at 99~GHz} \label{sec:mean_tco}

A direct measurement of the mean CO surface brightness, $\langle T_{\comath} \rangle$, across observed frequencies $\nu_{obs}=84.3$--114.8~GHz provides an empirical point of comparison to model predictions in the context of CO intensity mapping experiments at moderate redshfits, and also places a constraint on foreground emission for cosmic microwave background (CMB) experiments aiming to map spectral distortions at high redshift. We repeat the analysis performed for the ASPECS-Pilot program by \cite{Carilli2016} to place a lower limit on $\langle T_{\comath} \rangle$ at $\nu_{obs}=99$~GHz, based on blindly detected sources in the ASPECS LP 3~mm survey. 

Following \citet{Carilli2016}, we consider the aggregate emission from all observed CO transitions that contribute to the mean sky brightness at 99~GHz. We begin by summing the line fluxes of the sixteen blindly detected CO emission line candidates reported in GL19 to obtain a total CO flux of $6.91\pm0.19$~Jy~km~s$^{-1}$ or, equivalently, $\left(2.28\pm0.06\right)\times10^{6}$~Jy~Hz. Dividing this total flux by $\Delta\nu_{BW}$ yields a total mean CO flux density $\langle S_{\nu} \rangle = \left(7.53\pm0.21\right)\times10^{-5}$~Jy. Finally, to derive a mean surface brightness in units of $\mu$K, we apply the Rayleigh-Jeans approximation: $\langle T_{\comath} \rangle \sim 1360 \langle S_{\nu} \rangle \lambda_{obs}^2 / A_{S}^{\mathrm{blind}} = 0.55\pm0.02$~$\mu$K, where $\lambda_{obs}$ is the observed wavelength (in units of cm) and $A_{S}^{\mathrm{blind}}$ is the survey area (in units of arcsec$^2$) utilized in the line search, corresponding to the region of the mosaic where primary beam attenuation is less than 20\%: $1.69\times10^4$~arcsec$^2$ at 99~GHz. Following the prescription in \citet{Moster2011}, we estimate a 19.5\% relative uncertainty on $\langle T_{\comath} \rangle$ due to cosmic variance in the pencil beam survey by combining the fractional uncertainties\footnote{The relative uncertainty on the mean CO(2-1), CO(3-2), and CO(4-3) surface brightnesses due to cosmic variance is 17\%, 23\%, and 54\%, for minimum stellar masses probed of $6\times10^{9}$~M$_{\odot}$, $2\times10^{10}$~M$_{\odot}$, and $3\times10^{10}$~M$_{\odot}$, respectively.} calculated for each identified line entering in to the above flux sum, given the survey depth in stellar mass, mean redshift, and survey volume probed by the respective $J$ transition. Because \citet{Moster2011} estimates cosmic variance as the product of galaxy bias and the dark matter cosmic variance, their prescription is strictly applicable here in the case where the galaxy bias is identical to the bias $b_{\comath}$ of CO emission with respect to the matter density field.

For ASPECS-Pilot, which consisted of a single $\sim1$~arcmin$^2$ pointing with the same spectral coverage as ASPECS LP, $\langle T_{\comath} \rangle$ was found to be $0.94\pm0.09$~$\mu$K at 99~GHz \citep{Carilli2016}, which is a factor of 1.72 times greater than reported here for ASPECS LP. Since the time of publication of that analysis, however, four of the ten line candidates reported by ASPECS-Pilot (namely, ``3~mm.4," ``3~mm.7", ``3~mm.8", and ``3~mm.9" in Table~2 of \citet{Walter2016_survey}) have been re-classified as ``unconfirmed"---i.e., likely spurious, given their narrow line widths---sources based on the improved line search algorithms developed in GL19, and are excluded from the present analysis. Additionally, two of the ASPECS-Pilot line candidates (``3~mm.6" and ``3~mm.10") are outside the ASPECS LP survey coverage, and are similarly excluded. Thus, when including emission from only the four remaining confirmed sources from the original ten sources listed in \citet{Walter2016_survey}, one finds that the total observed CO flux scales linearly with the decrease in observed survey area, resulting in a revised $\langle T_{\comath} \rangle=0.55\pm0.05$~$\mu$K for ASPECS-Pilot, consistent with our new measurement.

It is important to note that the measurement of $\langle T_{\mathrm{CO}} \rangle$ presented here is considered a lower limit because the blind detections represent only a fraction of the total CO emission in the ASPECS LP survey volume; the fraction recovered by blind detections is determined by the sensitivity limit of the survey and the shape of the relevant CO luminosity functions (LFs). We compute the mean CO surface brightness based on the observed CO(2-1), CO(3-2), and CO(4-3) LFs for ASPECS LP presented in D19 as follows:
\begin{equation}
\label{eq:sbar_co}
\langle T_{\comath(J\mhyphen(J-1))} \rangle = \int \mathrm{dlog}_{10}L_{\comath(J\mhyphen(J-1))} \Phi(L_{\comath(J\mhyphen(J-1))}) \frac{L_{\comath(J\mhyphen(J-1))}}{4\pi D_L^2}yD_{A,co}^2,
\end{equation}
where $D_L$, $y$, and $D_{A,co}$ refer, respectively, to the luminosity distance, the derivative of the co-moving radial distance with respect to the observed frequency (i.e., $y=\mathrm{d}\chi/\mathrm{d}\nu = \lambda_{rest}(1+z)^2/H(z)$), and the co-moving angular diameter distance, and are all evaluated at $z_{cen,\comath(J\mhyphen(J-1))}$. Here, $\Phi(L_{\comath(J\mhyphen(J-1))})$ is originally expressed as a function of the integrated source brightness temperature, $L'_{\comath(J\mhyphen(J-1))}$ (in units of K~km~s$^{-1}$~pc$^2$), $\Phi(L_{\comath(J\mhyphen(J-1))}')$, and is given in the logarithmic Schechter form 
\begin{equation}
\log_{10} \Phi(L_{\comath(J\mhyphen(J-1))}') = \log_{10} \Phi_* + \alpha \log_{10}\left(\frac{L_{\comath(J\mhyphen(J-1))}}{L'_{\comath(J\mhyphen(J-1))*}}\right)\frac{1}{\mathrm{ln}10}\frac{L_{\comath(J\mhyphen(J-1))}'}{L'_{\comath(J\mhyphen(J-1))*}} + \log_{10}\left(\mathrm{ln}10\right).
\label{eq:LF_form}
\end{equation}
We convert from $L_{\comath(J\mhyphen(J-1))}'$ to $L_{\comath(J\mhyphen(J-1))}$ (in units solar luminosity) via 
\begin{equation}
L_{\comath(J\mhyphen(J-1))} = 3\times 10^{-11} \nu_{rest, \comath(J\mhyphen(J-1))}^3 L_{\comath(J\mhyphen(J-1))}'
\end{equation}
from \citet{CarilliWalter2013}. Fits to the luminosity function data have yielded Schechter parameters $\alpha$, $\Phi_*$, and $L'_{\comath(J\mhyphen(J-1))*}$ with uncertainties summarized in Table~\ref{tab:schechter_params}. Note that the faint-end slope, $\alpha$, has been fixed at $\alpha=-0.2$ for all LFs.
 
\begin{table}
\centering
\caption{CO LF Schechter parameters from D19}
\begin{tabular}{c c c c c}
\hline \hline
Line & Redshift & $\alpha$ & $\log_{10} \Phi_*$ & $\log_{10} L'_{*}$ \\
        &                          &               & [$\log_{10}$~(Mpc$^{-3}$~dex$^{-1}$)] & [$\log_{10}$~(K~km~s$^{-1}$~pc$^2$)] \\
  (1) &         (2)              &   (3)        &     (4)                                                    &       (5)         \\
\hline
CO(2-1) & 1.43            & -0.2 (fixed)        &    $-2.79^{+0.09}_{-0.09}$ & $10.09^{+0.10}_{-0.09}$ \\  
CO(3-2) & 2.61            & -0.2  (fixed)       &    $-3.83^{+0.13}_{-0.12}$ & $10.60^{+0.20}_{-0.15}$ \\
CO(4-3) & 3.80            & -0.2  (fixed)       &     $-3.43^{+0.19}_{-0.22}$ & $9.98^{+0.22}_{-0.14}$ \\
\hline
\multicolumn{5}{p{.7\textwidth}}{---\emph{Notes:} (1) Line transition (2) Mean redshift of LF redshift bin (3) Faint-end slope parameter in Eq.~\ref{eq:LF_form} (4) Normalization parameter in Eq.~\ref{eq:LF_form} (5) Characteristic luminosity parameter in Eq.~\ref{eq:LF_form}} 
\end{tabular}
\label{tab:schechter_params}
\end{table}

Integrating the LFs  (Equation~\ref{eq:sbar_co}) from an upper luminosity limit $L'_{upp}=10^{12}$~K~km~s$^{-1}$~pc$^2$ down to the mean 7-$\sigma$ line sensitivity\footnote{For reference, the mean 7-$\sigma$ line sensitivity for ASPECS in CO(2-1), CO(3-2), and CO(4-3) is $2.68\times10^9$~K~km~s$^{-1}$~pc$^2$ ($9.85\times10^{5}$~L$_{\odot}$), $3.70\times10^9$~K~km~s$^{-1}$~pc$^2$ ($4.58\times10^6$~L$_{\odot}$), and $3.93\times10^9$~K~km~s$^{-1}$~pc$^2$ ($1.15\times10^{7}$~L$_{\odot}$), respectively.} $L'_{min,7\mhyphen\sigma}$ in the respective redshift interval covered by each CO transition, which reflects the ASPECS LP detection threshold,\footnote{The signal-to-noise ratio threshold SNR $\ge6.8$ applied to the catalog of all possible line candidates (including candidates down to low SNR) yields the 16 high fidelity detections presented in Table~\ref{tab:COdets}} yields a mean total surface brightness $\langle T_{\comath} \rangle_{\mathrm{LF,7\sigma}} = 0.49$--1.78~$\mu$K, where the quoted range reflects the uncertainty in the LF parameters; please see Table~\ref{tab:sco} for a breakdown of the inferred $\langle T_{\comath} \rangle$ by $J$ transition. Extending the lower limit of integration down to $L'_{min}=10^8$~K~km~s$^{-1}$~pc$^2$ at all redshifts implies a total mean surface brightness of $\langle T_{\comath} \rangle_{\mathrm{LF}} = 0.72$--2.24~$\mu$K. Therefore, we estimate that our blind detections represent $\langle T_{\comath} \rangle_{\mathrm{LF,7\sigma}}/\langle T_{\comath} \rangle_{\mathrm{LF}}=68.1$--79.5\% of the total CO surface brightness at this observed frequency. 

\begin{table}
\centering
\caption{Mean CO surface brightness inferred from Schechter-form LFs}
\begin{tabular}{c c c}
\hline \hline
Line & $\langle T_{\comath(J\mhyphen(J-1))} \rangle$ ($L'_{min}= L'_{min,7\mhyphen\sigma}$) & $\langle T_{\comath(J\mhyphen(J-1))} \rangle$ ($L'_{min} = 10^8$~K~km~s$^{-1}$~pc$^2$)  \\
 & [$\mu$K] & [$\mu$K] \\
 (1) & (2) & (3) \\
 \hline
 CO(2-1) & $0.53^{+0.32}_{-0.21}$ & $0.72^{+0.38}_{-0.25}$ \\
 CO(3-2) & $0.25^{+0.31}_{-0.12}$ & $0.29^{+0.33}_{-0.13}$ \\
 CO(4-3) & $0.12^{+0.25}_{-0.08}$ & $0.20^{+0.32}_{-0.11}$ \\
 \hline
 \multicolumn{3}{p{.7\textwidth}}{---\emph{Notes:} (1) Line transition (2) Mean CO surface brightness calculated by integrating Eq.~\ref{eq:sbar_co} with lower and upper limits of integration $L'_{min} = L'_{min,7\mhyphen\sigma}$ and $L'_{upp} = 10^{12}$~K~km~s$^{-1}$ (3) Same as column (2), except for $L'_{min} = 10^8$~K~km~s$^{-1}$~pc$^2$} 
\end{tabular}
\label{tab:sco}
\end{table}



\subsubsection{CO shot noise power} \label{sec:pshot_co_limit}

As with the limit on mean CO surface brightness, the blindly detected sources in Table~\ref{tab:COdets} can also be used to place a lower limit on the expected CO shot noise power.

The total CO shot noise power from only the detected sources, $\left[P_{\comath,\comath}^{shot}(k_{\cotwomath})\right]_{\mathrm{det}}$, will contain contributions from galaxies emitting in the observed transitions $J = 2, 3,$ and 4:
\begin{align}
\left[P_{\comath,\comath}^{shot}(k_{\cotwomath})\right]_{\mathrm{det}} &= \left[P_{\cotwomath,\cotwomath}^{shot}(k_{\cotwomath})\right]_{\mathrm{det}} \nonumber \\
& \qquad {} + \left[P_{\cothreemath,\cothreemath}^{shot}(k_{\cotwomath})\right]_{\mathrm{det}}  \nonumber \\
& \qquad {} + \left[P_{\cofourmath,\cofourmath}^{shot}(k_{\cotwomath})\right]_{\mathrm{det}},
\label{eq:pshot_det}
\end{align}
where $\left[P_{\cothreemath,\cothreemath}^{shot}(k_{\cotwomath})\right]_{\mathrm{det}}$ and $\left[P_{\cofourmath,\cofourmath}^{shot}(k_{\cotwomath})\right]_{\mathrm{det}}$ have been converted to the CO(2-1) frame using Equation~\ref{eq:pspec_sum_kco21}. Each term on the righthand side of Equation~\ref{eq:pshot_det} can be determined analytically by summing the $N$ individual line fluxes per the expression
\begin{equation}
\sum_{i=1}^{N} \frac{1}{V_{S}} \left(\frac{L_{\comath(J\mhyphen(J-1))}}{4\pi D_L^2} y D_{A,co}^2 \right)^2,
 \label{eq:pshot_discrete}
\end{equation}
where $V_{S}$ refers to the survey volume at $z_{cen,\comath(J\mhyphen(J-1))}$. Note that the above expression for shot noise power has units of surface brightness, squared, times volume ($\mu$K$^2$~(Mpc~h$^{-1}$)$^3$), and is equal to the same value at all $k$, appropriate for a Poisson-sampling of galaxies.

Starting with the CO(2-1), CO(3-2), and CO(4-3) source fluxes from GL19, reported in Table~\ref{tab:COdets}, we find, for the entire ASPECS LP~3~mm survey volume used in the blind search, lower limits on the expected shot noise power of $\left[P_{\cotwomath,\cotwomath}^{shot}(k_{\cotwomath})\right]_{\mathrm{det}}=63.64$~$\mu$K$^2$~(Mpc~h$^{-1}$)$^3$, $\left[P_{\cothreemath,\cothreemath}^{shot}(k_{\cothreemath})\right]_{\mathrm{det}} = 98.49$~$\mu$K$^2$~(Mpc~h$^{-1}$)$^3$, and $\left[P_{\cofourmath,\cofourmath}^{shot}(k_{\cofourmath})\right]_{\mathrm{det}} = 1.05$~$\mu$K$^2$~(Mpc~h$^{-1}$)$^3$, respectively. For the cropped region (black, dotted square in Figure~\ref{fig:noise_maps}) corresponding to the volume used in the power spectrum analysis, we find the CO(2-1), CO(3-2), and CO(4-3) line emitters each give rise to a respective shot noise power of 73.99~$\mu$K$^2$~(Mpc~h$^{-1}$)$^3$, 71.06~$\mu$K$^2$~(Mpc~h$^{-1}$)$^3$, and 1.21~$\mu$K$^2$~(Mpc~h$^{-1}$)$^3$. The slightly higher shot noise power predicted in the cropped region for CO(2-1) and CO(4-3) is due to the decrease in volume after the crop; the CO(3-2) shot noise power decreases due to the fact that two of five detected sources are located outside of the boundary of the cropped region. After converting the CO(3-2) and CO(4-3) shot noise power into the CO(2-1) frame, we sum each contribution to arrive at a total shot noise power at $z_{cen}=1.315$ arising from the blind detections: $\left[P_{\comath,\comath}^{shot}(k_{\cotwomath})\right]_{\mathrm{det}} = 118.45$~$\mu$K$^2$~(Mpc~h$^{-1}$)$^3$ and 113.24~$\mu$K$^2$~(Mpc~h$^{-1}$)$^3$ for the full survey and cropped region, respectively. 

Finally, we estimate the expected shot noise power based on the D19 CO LFs, 
\begin{equation}
P_{\comath(J\mhyphen(J-1)),\comath(J\mhyphen(J-1))}^{shot} = \int \mathrm{dlog_{10}} L_{\comath(J\mhyphen(J-1))} \ \Phi(L_{\comath(J\mhyphen(J-1))}) \left[ \frac{L_{\comath(J\mhyphen(J-1))}}{4\pi D_L^2} y D_{A,co}^2 \right]^2,
\label{eq:pshot_coLF}
\end{equation}
and find that the detected sources (i.e., integrating Equation~\ref{eq:pshot_coLF} down to the relevant 7-$\sigma$ line sensitivity limit) recover 95.2--97.7\% of $P_{\cotwomath,\cotwomath}^{shot}(k_{\cotwomath})$, 98.6--99.7\% of $P_{\cothreemath,\cothreemath}^{shot}(k_{\cothreemath})$, and 84.9--96.0\% of $P_{\cofourmath,\cofourmath}^{shot}(k_{\cofourmath})$. In total, the recovered fraction is 96.0--98.6\% of $P_{\comath,\comath}^{shot}(k_{\cotwomath})$.


\subsection{Measurement of CO auto-power spectrum at $0.001 \lesssim z \lesssim 4.5$} \label{sec:autopower}

The noise-bias free auto-power spectrum, $P_{\comath, \comath}(k_{\cotwomath})$, is presented in Figure~\ref{fig:autopower}. We have averaged the power spectrum in linear bins of width d$k_{\cotwomath} = 2\pi / r_{\perp,max} = 4.1$~h~Mpc$^{-1}$, measuring  CO fluctuations on scales from $k_{\cotwomath}\sim10$~h~Mpc$^{-1}$ to 100~h~Mpc$^{-1}$. Formally, the ASPECS survey volume provides access to 3D modes (i.e, modes containing both $k_{\perp}$ and $k_{\parallel}$ components) down to the fundamental mode $k_{\cotwomath}=4.1$~h~Mpc$^{-1}$ (see Table~\ref{tab:coord_mapping}), though the number of independent modes $N_m$ ($=196$, or one mode per every channel in the cube) in this lowest wavenumber bin is small and the resulting signal-to-noise on the power spectrum is low; it has been discarded in this analysis.

Errors on the power spectrum at each $k_{\cotwomath}$ bin, $\langle \delta P_{\comath,\comath}(k_{\cotwomath})\rangle$, have been calculated using a 6-degree polynomial fit to the raw values calculated per Equation~\ref{eq:final_error}, as we would expect $\langle \delta P_{\comath,\comath}(k_{\cotwomath})\rangle$  to approach a smooth function as the number of realizations of the noise-only cubes---$\left(\tau_1(\mathbf{k}) - \tau_2(\mathbf{k})\right)$, $\left(\tau_3(\mathbf{k}) - \tau_4(\mathbf{k})\right)$, etc.---approaches infinity.

As an independent check on our error estimation, we also compute the noise-bias free power spectrum of noise-only simulated data cubes, $P_{N,N}(k_{\cotwomath})$, created with CASA task \textsf{simobserve}. The output of \textsf{simobserve} are CASA measurement sets, $\widetilde{\tau}_{1,N}$, $\widetilde{\tau}_{2,N}$, $\widetilde{\tau}_{3,N}$, and $\widetilde{\tau}_{4,N}$, that have been generated to mock the ASPECS observational setup, including identical mosaic pointing pattern and antenna configuration, which determine the mosaic power pattern and synthesized beam sizes, respectively. We then produce dirty image data cubes with the same parameters (e.g., 40-channel re-binning in frequency) adopted for the real data, and normalize the flux densities in each cube so that the RMS of each frequency slice (or channel map) at a given $\nu_{obs}$ for a given simulated cube (e.g., $\tau_{1,N}$) is identical to the RMS noise of the corresponding data cube (e.g., $\tau_1$) at the same $\nu_{obs}$ (Figure~\ref{fig:timesplit_properties}). In this way, we have constructed noise-only simulated image cubes, $\tau_{1,N}$--$\tau_{4,N}$, with noise properties similar to the real data cubes $\tau_1$--$\tau_4$. The resulting noise-bias free power spectrum of this simulated, noise-only dataset is shown alongside $P_{\comath, \comath}(k_{\cotwomath})$ in Figure~\ref{fig:autopower}. 

To improve signal-to-noise on $P_{\comath, \comath}(k_{\cotwomath})$, we have averaged the power within individual wavenumber bins into two wider bins containing the first and second halves linearly of the full $k_{\cotwomath}$ range, and a third set containing all $k_{\cotwomath}$ bins in the available range. We then report the inverse-variance weighted mean, and corresponding inverse-variance weighted error, for the bin representing the power spectrum averaged across all $N_{b} = 23$ bins from $k_{\cotwomath} = 9.55$~h~Mpc$^{-1}$ to 100.05~h~Mpc$^{-1}$,
\begin{equation}
\langle P_{\comath,\comath}(k_{\cotwomath})\rangle_{tot} = -45 \pm 77 \ \mu\mathrm{K}^2 \ (\mathrm{Mpc} \ \mathrm{h}^{-1})^3.  \nonumber \\
\end{equation}
We compute similar quantities for the bin containing the lower half (9.55~h~Mpc$^{-1}$ $\leq k_{\cotwomath} \leq 54.98$~h~Mpc$^{-1}$) of modes only, $\langle P_{\comath,\comath}(k_{\cotwomath}) \rangle_{low}$, and the upper half (59.20~h~Mpc$^{-1}$ $\leq k_{\cotwomath} \leq 100.05$~h~Mpc$^{-1}$) of modes only, $\langle P_{\comath,\comath}(k_{\cotwomath}) \rangle_{high}$, finding, overall,
\begin{gather}
\langle P_{\comath,\comath}(k_{\cotwomath}) \rangle_{low} = -260 \pm 170 \ \mu\mathrm{K}^2 \ (\mathrm{Mpc} \ \mathrm{h}^{-1})^3 \nonumber \\
\langle P_{\comath,\comath}(k_{\cotwomath}) \rangle_{high} = +10 \pm 86 \ \mu\mathrm{K}^2 \ (\mathrm{Mpc} \ \mathrm{h}^{-1})^3. \nonumber
\end{gather}
The measurements above are generally consistent with non-detections (i.e., $\langle P_{\comath,\comath}(k_{\cotwomath})\rangle = 0$) at the quoted 1-$\sigma$ or 1.5-$\sigma$ (in the case of $\langle P_{\comath,\comath}(k_{\cotwomath}) \rangle_{low}$) level, and are comparable to the noise-bias free power spectrum measured for the simulated noise-only cubes:
\begin{gather}
\langle P_{N,N}(k_{\cotwomath})\rangle_{tot} = +41 \pm 87 \ \mu\mathrm{K}^2 \ (\mathrm{Mpc} \ \mathrm{h}^{-1})^3  \nonumber \\
\langle P_{N,N}(k_{\cotwomath})\rangle_{low} = -47 \pm 180 \ \mu\mathrm{K}^2 \ (\mathrm{Mpc} \ \mathrm{h}^{-1})^3  \nonumber \\
\langle P_{N,N}(k_{\cotwomath})\rangle_{high} = +70 \pm 100 \ \mu\mathrm{K}^2 \ (\mathrm{Mpc} \ \mathrm{h}^{-1})^3 \nonumber
\end{gather}
Furthermore, the reported power spectra in each ``total," ``low," and ``high" $k_{\cotwomath}$ bin agrees within $\sim1$-$\sigma$ of each other, suggesting that our measurement does not discern any spectral structure. Thus, we adopt $\langle P_{\comath,\comath}(k_{\cotwomath})\rangle_{tot}$ as representative of the measured, flat power spectrum, and use it to place a 3-$\sigma$ upper limit on the noise-bias free CO power spectrum, $P_{\comath,\comath}(k_{\cotwomath}) \le  190$~$\mu$K$^2$~(Mpc~h$^{-1}$)$^3 = \langle P_{\comath,\comath}(k_{\cotwomath})\rangle_{tot} + 3\times\langle \delta P_{\comath,\comath}(k_{\cotwomath}) \rangle_{tot}$. 

The analytically estimated shot noise power based only on ASPECS LP blind detections (cf. Section~\ref{sec:pshot_co_limit}),  $\left[P_{\comath,\comath}^{shot}(k_{\cotwomath})\right]_{\mathrm{det}} = 113.24$~$\mu$K$^2$~(Mpc~h$^{-1}$)$^3$, lies roughly a factor of 2 below our upper limit. (We refer here to $\left[P_{\comath,\comath}^{shot}(k_{\cotwomath})\right]_{\mathrm{det}}$ calculated for the cropped region used in the power spectrum analysis, with CO flux values from GL19.)

We note that the effect of the CMB radiation is likely negligible on the cumulative measurement presented here. Assuming gas kinetic temperatures close to $\sim40$~K, which are appropriate for the CO-emitting sources in ASPECS based on the output dust temperatures from MAGPHYS SED fits \citep{Boogaard2019_3mm}, corrections due to CMB on the CO line fluxes contributing to the emission are expected to be less than 25\% at $z\leq4$, with the possible exception of CO(4-3) at $z\sim4$ in low density gas ($n_{\mathrm{H}}=10^{3.2}$~cm$^{-3}$) and non-LTE conditions (Sections~3.3 and 3.4 in \citet{daCunha2013}). In that case, the intrinsic CO(4-3) flux can be up to a factor of 2 higher than observed. However, since the product of distortion factors converting CO(4-3) power into the CO(2-1) frame is small, we do not expect this to have a significant effect on the aggregate emission measured by the power spectrum, even if such conditions were representative of the ISM.

The two theoretical models we include in our comparison in Figure~\ref{fig:autopower} (``MAGPHYS" and ``Popping") also are factors of $\sim2$ to $\sim4.5$ below our upper limit. The ``Popping" model refers to the semi-analytic model described in \citet{Popping2016}. The ``MAGPHYS model" refers to a model CO catalog for a subset of $\sim1,000$ sources in HUDF with (1) a (photometric, grism, or spectroscopic) redshift that is in the allowable ranged covered by CO transitions in ASPECS LP up to $J=4$, and a reliable fit to the spectral energy distribution (SED) using MAGPHYS from the rest-frame UV/optical wavelengths out to the infrared based on (2) a 1.6~$\mu$m flux density greater than 0.1~$\mu$Jy, and (3) detections in, at least, 5 photometric bands. (For more details on the SED fitting of HUDF sources within the ASPECS field, see D19 and B19.) CO(1-0) luminosities for the sources in this catalog have been estimated by scaling to the output IR luminosity from MAGPHYS according to \citet{CarilliWalter2013}, and then using excitation corrections from \citet{Daddi2015} to predict higher $J$ CO line luminosities. We estimated the total CO shot noise power for the model CO catalog per Equation~\ref{eq:pshot_discrete}, finding that---if we include all sources with stellar masses $M_{*}\ge10^{7}$~M$_{\odot}$---$P_{\comath,\comath}^{shot}(k_{\cotwomath}) = 98.13$~$\mu$K$^2$~(Mpc h$^{-1}$)$^3$, which is consistent with the 3-$\sigma$ upper limit measured from the power spectrum, and is comparable to $\left[P_{\comath,\comath}^{shot}(k_{\cotwomath})\right]_{\mathrm{det}}$.

Expectations for $P_{\comath,\comath}^{shot}(k_{\cotwomath})$---separated into the constituent power from specific $J$ transitions, when possible---based on the above models, empirically derived CO luminosity functions (Section~\ref{sec:coLF_compare}), and higher-redshift CO(1-0) measurements (Section~\ref{sec:co10_compare}), have been gathered, for reference, in Table~\ref{tab:pshot_summary}. Note that predictions based on semi-analytic simulations from \citet{Popping2016} are known to underestimate the number density of bright sources compared to observed CO luminosity functions (D19), and thus may represent an unrealistically low prediction for $P_{\comath,\comath}(k_{\cotwomath})$. (Please see following section, particularly Equation~\ref{eq:pshot_coLF}, for details connecting the shot noise power spectrum measurement to the luminosity function.) The modelling framework in that paper has since been updated in \citet{Popping2019}, now including, e.g., predictions using hydrodynamical simulations,  but the authors find that the models, again, underpredict the bright-end ($L'_{\comath} > 10^{10}$~K~km~s$^{-1}$~pc$^2$) of the luminosity functions by factors of 1--3~dex. (We direct the reader to \citet{Popping2019} for a discussion of the potential origins of the discrepancy between their models and observations.) Also, we point out that the discrepancy between the \citet{Popping2016} predictions for CO(1-0) shot noise power and the expected CO(1-0) shot noise from the ``MAGPHYS" model and the ASPECS observations can be ascribed to the small volume coverage of the ASPECS survey volume at $z\sim0.28$, which limits the number of bright sources in the ``MAGPHYS" model catalog and precludes a Schechter function fit for CO(1-0) LF data in the D19 analysis. (See also discussion on cosmic variance in \citet{Popping2019}.) Specifically, in Table~\ref{tab:pshot_summary}, the expected CO(1-0) shot noise power from the ASPECS LF is calculated by integrating the LF only in luminosity bins where where LF data is available, i.e., in the range from $L'_{\coonemath} = 10^8$--$10^9$~K~km~s$^{-1}$~pc$^{2}$, instead of the full range of $L'_{\comath}$ ($=10^8$--$10^{12}$~K~km~s$^{-1}$~pc$^{2}$) encapsulated by the Schechter function fits provided at higher $z$.

\begin{table}
\centering
\caption{Predictions for CO shot noise power, $P_{\comath(J\mhyphen(J-1)), \comath(J\mhyphen(J-1))}^{shot}(k_{\comath(J\mhyphen(J-1)), z_{cen,\comath(J\mhyphen(J-1))}})$ [$\mu$K$^2$~(Mpc~h$^{-1}$)$^3$]}
\begin{tabular}{c c c c c c}
\hline \hline
Line & Popping et al. 2016 & MAGPHYS model & ASPECS LF & COLDz$^a$ LF & COPSS~II$^b$ \\ 
  (1)         &             (2)                   &            (3)               &     (4)            &       (5)     &  (6)                    \\
\hline
CO(1-0) &         0.55                     &    0.04         &    $0.040^{+0.31}_{-0.03}$   &         --         &    --               \\
CO(2-1) &          8.15                   &       89.23       &   $80^{+71}_{-38}$ & $170^{+509}_{-120}$$^a$ &  $1600\pm700$$^a$ \\
CO(3-2) &       17.82                 &       8.61            &   $130^{+320}_{-82}$ &       --        &        --               \\
CO(4-3) &          8.77                   &       6.33          &   $27^{+90}_{-19}$     &     --          &       --                  \\
Total$^b$ &       43.31                  &       98.13        &  $170^{+290}_{-92}$ &    --          &       --              \\
\hline
\multicolumn{6}{p{0.8\textwidth}}{$^a$ Total area surveyed: $\sim60$~arcmin$^2$}   \\
\multicolumn{6}{p{0.8\textwidth}}{$^b$ Total area surveyed: $\sim0.7$~deg$^2$}  \\
\multicolumn{6}{p{0.8\textwidth}}{$^c$ CO(2-1) power inferred from CO(1-0) LF measurement. See text in Section~\ref{sec:co10_compare} for details regarding the applied conversion.}  \\
\multicolumn{6}{p{0.8\textwidth}}{$^d$ Sum of CO(1-0), CO(2-1), CO(3-2), and CO(4-3) shot noise power, converted to the CO(2-1) frame.}  \\
\multicolumn{6}{p{0.8\textwidth}}{---\emph{Notes:} (1) Line transition (2) \citep{Popping2016} (3) Model based on SED-fitting and $L'_{\comath}$--$L_{IR}$ relation (see text for details) (4) CO LFs from D19. CO LFs refer to Schechter fits, integrated from $L'_{min}=10^8$~K~km~s$^{-1}$~pc$^2$ to $L'_{max}=10^{12}$~K~km~s$^{-1}$~pc$^2$ except for $J=1$, where no Schechter fit was performed. For CO(1-0), we have used tabulated LF data for each luminosity bin from $L'_{min}=10^8$~K~km~s$^{-1}$~pc$^2$ to $L'_{max}=10^9$~K~km~s$^{-1}$~pc$^2$ (5) \citep{Riechers2018} (6) \citep{Keating2016}} 
\end{tabular}
\label{tab:pshot_summary}
\end{table}

\subsubsection{Comparison to CO LFs derived from ASPECS LP} \label{sec:coLF_compare}

Integrating the luminosity functions measured by ASPECS LP (with lower and upper limits of integration equal to $10^8$~K~km~s$^{-1}$~pc$^2$ and $10^{12}$~K~km~s$^{-1}$~pc$^2$) per Equation~\ref{eq:pshot_coLF} yields the following estimates of shot noise power for CO(2-1), CO(3-2), and CO(4-3), respectively: $80^{+71}_{-38}$~$\mu$K$^2$~(Mpc~h$^{-1}$)$^3$, $130^{+320}_{-82}$~$\mu$K$^2$~(Mpc~h$^{-1}$)$^3$, and $27^{+90}_{-19}$~$\mu$K$^2$~(Mpc~h$^{-1}$)$^3$. In order to compare to our measurement of the total CO shot noise power, we convert the individual shot noise powers estimated for each transition into the CO(2-1) frame, and sum, finding  $P_{\comath,\comath}^{shot}(k_{\cotwomath}) = 170^{+290}_{-92}$~$\mu$K$^2$~(Mpc~h$^{-1}$)$^3$. The 3-$\sigma$ upper limit on $P_{\comath,\comath}(k_{\cotwomath}) \le 190$~$\mu$K$^2$~(Mpc~h$^{-1}$)$^3$ determined via the power spectrum places a more stringent constraint on the total CO shot noise power; it rules out a significant fraction of the allowable range, $P_{\comath,\comath}^{shot}(k_{\cotwomath})=78$~$\boldsymbol{\mu}$K$^\mathbf2$~(Mpc~h$^{-1}$)$^3$ to 460~$\mu$K$^2$~(Mpc~h$^{-1}$)$^3$, obtained from the LF fits. The low sensitivity, however, on the power spectrum measurement precludes us from determining the amplitude of individual contributions from the different CO transitions to the aggregate value, so translating our limit on $P_{\comath,\comath}(k_{\cotwomath})$ to constraints on individual LFs is highly speculative---as would be attempting to constrain Schechter parameters of the individual CO LFs. We note, however, that the fixed faint-end slope $\alpha = -0.2$ in D19 implies that the shot noise power spectrum is relatively insensitive to the low-luminosity systems in the CO LF. In this case, the shot noise power is more sensitive to the Schechter parameters $\Phi_{*}$ and $L_{*}$, and our measurement suggests either lower normalizations $\Phi_*$ or a knee in the Schechter function that occurs at lower luminosities $L'_*$. This may be particularly applicable to the observed CO(3-2) and CO(4-3) LFs at $z>2$, where the empirical constraints span a limited range in luminosity compared to the CO(2-1) LF at $z\sim1$, and resulting uncertainties on the Schechter parameters are large.

 
\subsubsection{Comparison to higher redshift CO(1-0) observations} \label{sec:co10_compare}

In this section, we compare our measured $P_{\comath,\comath}(k_{\cotwomath})$ to independent observational constraints on CO(1-0) shot noise power at higher redshift.

First, we consider the CO(1-0) luminosity functions determined at $z=2$--3 by the VLA COLDz program \citep{Riechers2018}, which targeted survey volumes in COSMOS ($\sim9$~arcmin$^2$) and GOODS-N ($\sim51$~arcmin$^2$) across 8~GHz of bandwidth in Ka band ($\nu_{obs}\approx30$--38~GHz) with typical synthesized beam sizes $\sim3$~arcsecond. We compute the probability distribution of  $P_{\coonemath,\coonemath}(k_{\coonemath}, z_{cen}=2.4)$ using Equation~\ref{eq:pshot_coLF} and the Schechter function parameter samples from the posterior distributions obtained with the Approximate Bayesian Computation method for the merged COSMOS and GOODS-N dataset (see Figure 6 in \citet{Riechers2018}). We find that the distribution has median $P_{\coonemath,\coonemath}(k_{\coonemath}, z_{cen}=2.4) = 276.58$~$\mu$K$^2$~(Mpc~h$^{-1}$)$^3$, with a probable range of $P_{\coonemath,\coonemath}(k_{\coonemath}, z_{cen}=2.4) = 74.75$~$\mu$K$^2$~(Mpc~h$^{-1}$)$^3$ (5th percentile) to 1119.35~$\mu$K$^2$~(Mpc~h$^{-1}$)$^3$  (95th percentile).

The COLDz constraints on the CO(1-0) shot noise power spectrum at $z_{cen,\coonemath}=2.4$ can be converted to a constraint on the CO(2-1) luminosity density at $z\sim1$ (and, thus, the CO(2-1) shot noise power spectrum at the same redshift) assuming (1) the CO(2-1) line is thermalized and has the same brightness temperature as CO(1-0) at $z_{cen}=2.4$, and (2) there is no evolution in the CO(2-1) luminosity density from $z_{cen,\coonemath}=2.4$ to $z_{cen,\cotwomath}=1.3$. The first assumption is reasonable for the low $J$ transitions relevant here, and is supported by observations of a variety of high-$z$ systems (see, e.g., Table~2 in \citet{CarilliWalter2013}, which shows $L'_{\cotwomath}/L'_{\coonemath}\sim0.9$ for submillimeter galaxies, color-selected star-forming galaxies, etc.). The latter assumption is likely false in detail, but, given that the cosmic star formation rate density is relatively flat between $z=1$--3, it is a reasonable first approximation; note also that the cosmic molecular gas densities at $z\sim1$ and $z\sim3$ are indistinguishable by current empirical standards. In any case, assuming (2) results in an underestimation of $P_{\cotwomath,\cotwomath}$ if the CO(2-1) luminosity density at $z=2.4$ is, in reality, lower than the luminosity density at $z=1$, and vice versa. So, assuming (1) and (2) are valid,  we have $P_{\coonemath,\coonemath}(k_{\coonemath}, z_{cen}=2.4) = P_{\cotwomath, \cotwomath}(k_{\cotwomath}, z_{cen}=2.4)$, which implies a CO(2-1) luminosity density $\rho_{\cotwomath}(z_{cen}=2.4) = \rho_{\cotwomath}(z_{cen}=1.3) = 5.63\times10^{10} $~L$_{\odot}$~Mpc$^{-3}$, yielding $P_{\cotwomath,\cotwomath}^{shot}(k_{\cotwomath}, z_{cen}=1.3)=167.03$~$\mu$K$^2$~(Mpc~h$^{-1}$)$^3$; the 5th and 95th percentile similarly give lower and upper bounds on $P_{\cotwomath,\cotwomath}^{shot}(k_{\cotwomath}, z_{cen}=1.3)=45.14$~$\mu$K$^2$~(Mpc~h$^{-1}$)$^3$ and 675.98~$\mu$K$^2$~(Mpc~h$^{-1}$)$^3$, respectively. This range is consistent with both our upper limit on the measured $P_{\comath,\comath}(k_{\cotwomath}, z_{cen}=1.3)$, as well as $P_{\cotwomath,\cotwomath}^{shot}(k_{\cotwomath}, z_{cen}=1.3)$ estimated from the CO(2-1) LF fit derived from ASPECS LP data (cf. Section~\ref{sec:coLF_compare}).\footnote{Note that larger uncertainties on shot noise power derived from COLDz are largely due to differences in fitting the measured LFs: the authors in \citet{Riechers2018} treated the faint-end slope of the LF as a free parameter, while \citet{Decarli2019_3mm} kept it fixed during their fitting.} A direct comparison of the CO(1-0) LF inferred\footnote{\citet{Decarli2019_3mm} used the CO(3-2) LF measured at $z_{cen}=2.6$ to infer a CO(1-0) LF after accounting for CO excitation. The fiducial prescription to convert from CO(3-2) to CO(1-0) luminosities was based on \citet{Daddi2015}, but the authors there explored alternative lower and higher excitation scenarios, as well, finding that their results are qualitatively robust to the adopted CO line ratio.} by ASPECS at $z_{cen}=2.6$ and the CO(1-0) LF measured by COLDz at nearby redshift also reveals an excellent agreement across the overlapping CO(1-0) luminosity range from $L'_{\coonemath}=0.1$--$4\times10^{11}$~K~km~s$^{-1}$~pc$^2$, which suggests that our assumptions (1) and (2) are reasonable. Importantly, the apparent agreement at the bright-end of the CO(1-0) LFs also suggests the impact of cosmic variance on the shot noise power spectrum presented in this study is modest, since the shot noise measurement is inherently more sensitive to high luminosity systems (because of the $\propto L^2$ dependence in Equation~\ref{eq:pshot_coLF}).

We also compare the measured CO power spectrum of ASPECS LP data to the power spectrum measurement of CO(1-0) at wavenumbers $k_{\coonemath}\sim$1--10~h~Mpc$^{-1}$ at $z\sim2-3$ from COPSS~II \citep{Keating2016}. COPSS~II was a dedicated intensity mapping experiment carried out with the Sunyaev-Zel'dovich Array to observe non-contiguous fields (totalling a survey area $\sim0.7$~deg$^2$) on the sky at $\sim2$~arcmin spatial resolution.  \citet{Keating2016} present a marginal 2-$\sigma$ detection of $P_{\coonemath,\coonemath}^{shot}(k_{\coonemath}) = 3000\pm1300$~$\mu$K$^2$~(Mpc~h$^{-1}$)$^3$ at $z_{cen}=2.8$, which is at least a factor of 6 larger than the COLDz measurement at similar redshift. Following the same procedure as outlined above, we estimate the CO(2-1) shot noise at $z_{cen}=1.3$ based on the COPSS~II measurement, finding $1600\pm700$~$\mu$K$^2$~(Mpc~h$^{-1}$)$^3$, which 4.8 to 12 times higher than our current 3-$\sigma$ upper limit, though the ASPECS measurement probes CO power at higher wavenumbers ($>10$~h~Mpc$^{-1}$), compared to COPSS~II.

\subsection{CO-galaxy statistics}
\label{sec:crosspower_TITII}

While the noise-bias free auto-power spectrum (Section~\ref{sec:autopower}) provides an unbiased view of the aggregate CO-emitting galaxy population, the detection of the auto-power spectrum is inherently challenging given that the measurement weights noisy voxels\footnote{Voxels refer to 3D resolution elements defined by the beam area and channel width.} equally with voxels containing CO emission. Adjusting these weights by including information from another field that is correlated in spatial distribution with the target field of CO fluctuations could yield a higher fidelity measurement on the CO properties (e.g., shot noise power, mean surface brightness) of the secondary population represented in the additional field, as long as this additional dataset does not contribute significantly to the noise.

Given that we expect the CO fluctuations to originate within the interstellar medium of galaxies, we consider the available galaxy catalogs with spectroscopic redshifts (spec-$z$) in HUDF as potential datasets to perform power spectrum analyses with ASPECS data. Spectroscopic redshifts are required in order to match the spectral precision of the ASPECS LP data cube, where astrophysical emission is expected to be contained within the channel width $\Delta\nu_{40chn}$, corresponding, to a redshift resolution  $\Delta z_{40chn}/(1+z_{cen,\cotwomath}) = 0.0013$. In contrast, photometric redshifts measured via SED-template fitting procedures, e.g. are typically characterized by uncertainties  $\Delta z_{phot} / (1+z_{phot}) \sim 0.05$ \citep[e.g.][]{Coe2006, Brammer2008}, and are unsuitable for use in the power spectrum analysis.


The MUSE Ultra Deep Field (UDF) survey \citep{Bacon2017} has yielded $\sim1500$ spectroscopic redshifts within a 3$'$ by 3$'$ field in HUDF \citep{Inami2017}, which includes the ASPECS LP areal footprint. The redshift range covered by MUSE UDF extends beyond the ASPECS LP Band 3 spectral coverage, but there is overlap at $z < 1.5$ and $z> 3$ due primarily to the identification in MUSE spectra of rest-frame optical and UV emission features, respectively, such as [O~II] and Lyman $\alpha$ emission; from $z\sim2$--3, MUSE spectroscopic redshifts are determined with the presence of various UV emission and/or absorption features.\footnote{Please see Figure~1 of B19 for the redshift distribution---color-coded by the respective spectral feature(s) used for redshift determination---of MUSE galaxies within the ASPECS LP survey volume.} There are, in total, 680 sources identified by MUSE that are available in the ASPECS LP areal and redshift coverage, and 415 of these lie within the cropped region used here for the power spectrum analysis; unless otherwise noted, we include all spec-$z$ reported in the MUSE catalog.\footnote{As discussed in \citet{Inami2017}, spec-$z$ have been assigned confidences ranging from 1 to 3; redshifts with $\mathrm{CONFID} = 2$ or 3 are considered ``secure,'' and those with $\mathrm{CONFID} = 1$ have been determined as a ``possible" redshift from the presence of a spectral line with uncertain identifcation.} Of these 415, 24 MUSE sources fall within the redshift interval and volume probed by CO(1-0) in ASPECS LP, and can be used to search for CO(1-0) emission in the ASPECS LP data; for CO(2-1), there are 128 MUSE sources in the relevant redshift range and area; for CO(3-2), 64 sources; and, for CO(4-3), 199 sources. Note that the relative number of MUSE sources available to correlate with the different CO $J$ transitions is affected by the availability and intrinsic strength of the various, aforementioned spectral features used to determine MUSE spec-$z$ within each redshift interval. 

Since galaxies in the MUSE spec-$z$ catalog have been selected at shorter rest wavelengths than the mm-wave emission observed in ASPECS, it is not \emph{a priori} known how closely they correlate with the CO fluctuations. However, based on the counterpart analysis of the 16 secure ASPECS LP CO detections, which all have optical/near-infrared counterparts, we expect a non-zero positive correlation. In fact, as described in B19, all ten CO(2-1) blind detections in ASPECS LP have a counterpart MUSE spec-$z$, with velocity offsets typically less than 100~km~s$^{-1}$ between the CO and MUSE redshift (cf. Table~1 in B19); two of the five CO(3-2) blind detections in ASPECS LP have MUSE spec-$z$; the single CO(4-3) blind detection does not have a MUSE spec-$z$.

In the following sections, we explore two statistics, namely, a masked auto-power spectrum, $P_{\comath,\comath}^{gal}(k_{\cotwomath})$ (Section~\ref{sec:xpower_nbf}) and the cross-shot noise power spectrum $P_{\comath,gal}(k_{\cotwomath})$ (Section~\ref{sec:xshot}), to assess the contributions to the observed shot noise power and mean CO surface brightness from MUSE-selected galaxies, respectively.

\subsubsection{Masked noise-bias free auto-power spectrum} \label{sec:xpower_nbf}

We obtain $P_{\comath,\comath}^{gal}(k_{\cotwomath})$ by adopting the same methods to measure the noise-bias free auto-power spectrum, $P_{\comath,\comath}(k_{\cotwomath})$, with one key difference: masking was performed on $\tau_1$, $\tau_2$, $\tau_3$, and $\tau_4$ to remove flux densities from voxels beyond a spatial radius of $\sim$1~arcsecond and spectral width of 0.165~GHz (or 1 channel) from the center of known positions of MUSE galaxies listed in \citet{Inami2017}, prior to estimating the power spectrum. The $\sim1$~arcsecond radius was chosen so that any enclosed source flux would be encompassed by a full beamwidth; the true source flux is not recovered in the case of extended emission (cf. Table~\ref{tab:COdets} for sources identified as extended and corresponding fluxes extracted per GL2019). Explicitly, we are evaluating a statistic defined as 
\begin{equation}
\langle T_{\comath}^{gal*}(\vec{k}) T_{\comath}^{gal}(\vec{k'}) \rangle \equiv (2\pi)^3 \delta_{\mathrm{D}}(\vec{k}-\vec{k'}) P^{gal}_{\comath,\comath}(k).
\end{equation}
In the above equation, $T_{\comath}^{gal}$ represents the masked ASPECS data cube, which has been weighted at every $i^{\mathrm{th}}$ voxel according to 
\begin{equation}
(T_{\comath}^{gal})_i = w_i (T_{\comath})_i,
\end{equation}
where $w_i = 1$ for voxels containing a MUSE galaxy, and $w_i=0$ otherwise. Thus, in the same way that the noise-bias free auto-power spectrum measurement at $k=10$--100~h~Mpc$^{-1}$ yielded constraints on the second moments of the CO LFs (per Equation~\ref{eq:pshot_coLF}), the masked noise-bias free auto-power spectrum measurement at the same $k$ range constrains the second moments of the LFs of CO-emitting MUSE galaxies.

Errors on the masked noise-bias free auto-power spectrum, $\delta P_{\comath,\comath}^{gal}(k_{\cotwomath})$, were obtained from Equation~\ref{eq:final_error}, and include an additional contribution estimated from 100 simulations of random MUSE source positions. This Poisson term was deemed necessary to prevent underestimating error based only on Equation~\ref{eq:final_error}; after removing significant source flux from the cubes in the masking step, the pre-factor $\gamma=0.25$ that appears in this Equation might no longer accurately describe the relation of noise properties in $\tau_1$, $\tau_2$, $\tau_3$, and $\tau_4$ to those in $T_\mathrm{I}$ and $T_{\mathrm{II}}$.

Figures~\ref{fig:xspec_muse_allJ} and \ref{fig:xspec_muse_coJ} show the measured masked noise-bias free CO auto-power spectrum, $P_{\comath,\comath}^{\mathrm{MUSE}}(k_{\cotwomath})$, between CO fluctuations in the ASPECS LP data cube and 3D positions from the MUSE spec-$z$ catalog. In these figures, we compare the power measured when including (1) MUSE positions of the CO blind detections (solid red curves), (2) all available MUSE positions (solid black curves), and (3) all available MUSE positions, excluding those corresponding the positions of the CO blind detections (dashed red curves). To facilitate comparison of the relative contributions of each sample described by (1)--(3), we have scaled the $y$-axis in each plot by a factor $1/\langle P^{\mathrm{MUSE}}_{\comath,\comath}(k_{\cotwomath})\rangle^{\mathrm{all}}$, where the denominator represents the total noise bias-free cross-power using all MUSE positions and including potential CO emission from $J=1$--4 transitions. In the case where (3) yields $P^{\mathrm{MUSE}}_{\comath,\comath}=0$, we can be confident that there is no measured ``excess" power from MUSE galaxies with previously undetected CO emission. Note that, although our power spectrum measurements cannot probe fluctuations on scales $k_{\cotwomath} \gtrsim 100$~h~Mpc$^{-1}$, which correspond to the sub-beam size pixel gridding, we include these wavenumbers in the plot to show the effect of the beam size on the measured power, which follows an exponential drop-off as expected; the sensitivity on the auto-power spectrum measurement was insufficient to ``detect" the beam roll-off, so we truncated the spectrum in Figure~\ref{fig:autopower} at $k_{\cotwomath}\approx100$~h~Mpc$^{-1}$.

We find that the noise-bias free cross-power spectrum is detected at high significance with an amplitude suggesting that the majority of CO emission contained within the survey volume is accounted for by the ASPECS blind detections, and that the rest-frame optical/UV galaxies in the same field closely trace the observed CO emission (Figure~\ref{fig:xspec_muse_allJ}). The latter result is unsurprising, given that 12/16 of CO blind detections had known MUSE counterparts. Only a small level of excess power---$\langle P^{\mathrm{MUSE}}_{\comath,\comath}(k_{\cotwomath} \rangle^{\mathrm{blind \ removed}} /\langle P^{\mathrm{MUSE}}_{\comath,\comath}(k_{\cotwomath}\rangle^{\mathrm{all}}= 7.05\pm3.2$\%,  where we have averaged over wavenumber bins from $k_{\cotwomath}= 9.55$ to 54.98~h~Mpc$^{-1}$ to avoid the effects of the exponential beam roll-off\footnote{If we average over all $k_{\cotwomath}$ from 9.55 to 100.05~h~Mpc$^{-1}$, we find the excess power is $\langle P^{\mathrm{MUSE}}_{\comath,\comath}(k_{\cotwomath} \rangle^{\mathrm{blind \ removed}} /\langle P^{\mathrm{MUSE}}_{\comath,\comath}(k_{\cotwomath}\rangle^{\mathrm{all}}= 7.1\pm1.2$\%.} seen at higher $k_{\cotwomath}$---is observed to originate from the positions of MUSE galaxies in the same field that do not have previously detected ASPECS counterparts, amounting to 3.8-10.7\% of the total masked CO auto-power measured when including all MUSE positions with potential CO $J=1,2,3,$ or 4 emission. In other words, the percentage of power recovered by the MUSE sources with previously detected ASPECS counterparts is 89--96\%. 

The high SNR on $P_{\comath,\comath}^{\mathrm{MUSE}}(k_{\cotwomath})$ enables further decomposition of the total masked CO auto-power spectrum into the individual $J$ transitions contributing to the aggregate signal (Figure~\ref{fig:xspec_muse_coJ}). Here, we identify emission from the CO(3-2) transition as the principal source of residual power after removing the ASPECS blind detections in the total masked CO auto-power spectrum in Figure~\ref{fig:xspec_muse_coJ}, while the power derived from all other $J$ transitions is zero. The level of residual power detected in the masked CO(3-2) power spectrum,\footnote{$P^{\mathrm{MUSE}}_{\cothreemath,\cothreemath}(k_{\cotwomath})=1.3\pm0.23$~$\mu$K$^2$~(Mpc h$^{-1}$)$^3$, when averaging over all $k_{\cotwomath}$ up to 100~h~Mpc$^{-1}$, instead of the lower half of modes, as adopted in the main text.} $P^{\mathrm{MUSE}}_{\cothreemath,\cothreemath}(k_{\cotwomath})=1.8\pm0.56$~$\mu$K$^2$~(Mpc h$^{-1}$)$^3$, which represents $5.2\pm1.6$\% of the total masked CO auto-power $\langle P^{\mathrm{MUSE}}_{\comath,\comath}(k_{\cotwomath}\rangle^{\mathrm{all}}$ (including all $J$ transitions), or 14--22\% of the total masked CO(3-2) power spectrum amplitude.  

Next we consider the nature of sources contributing to the excess in CO(3-2) emission relative to the expected power from the CO(3-2) blind detections only. The level of observed power could be attributed to a single bright source, e.g., with CO(3-2) flux of the order $\sim0.10$~Jy~km~s$^{-1}$, or multiple fainter sources. Since a CO(3-2) source flux $\sim0.10$~Jy~km~s$^{-1}$ at $2\leq z \leq3$ implies a line luminosity $4.2\times10^6$~L$_{\odot}$ that is just below the mean sensitivity limit for the survey for that redshift range, it is possible that a single bright source with this flux would have been previously undetected in the ASPECS line search. The scenario where the excess power originates from CO(3-2) emitters below the individual detection threshold is also plausible, however, given that the MUSE catalog contains galaxies with lower stellar mass $M_*$ and star formation rates (SFRs) than probed by the CO blind detections. Specifically, at $2 \leq z \leq 3$, MUSE sources probe down to $M_*\sim10^9$~M$_{\odot}$ and SFR~$\sim0.3$~M$_{\odot}$~yr$^{-1}$, while the ASPECS-detected galaxies have $M_*\geq10^{10}$~M$_{\odot}$ and SFR~$\geq10$~M$_{\odot}$~yr$^{-1}$ (B19). We can test these scenarios by masking the ASPECS data cube down to progressively lower flux thresholds, and determine the flux level where the excess power vanishes. Figure~\ref{fig:xspec_muse_co32_masking} shows the results of this analysis, where voxels with flux densities with $\vert F_{\nu}\vert> 1.0,$ 0.75, 0.50, 0.25, and 0.1~mJy have been blanked so that the remaining emission in the datacubes is due to sources (or noise) fainter than the masking threshold; note that we consider the absolute value of the flux densities because we are working with dirty image cubes that contain $F_{\nu}<0$. From Figure~\ref{fig:xspec_muse_co32_masking} it is clear that roughly 75\% of the excess power originates from voxels with flux densities greater than 0.25--0.50~mJy, which imply fluxes $F_{\nu}\Delta\nu_{40chn} > 0.12$--0.24~Jy~km~s$^{-1}$, suggesting that a single relatively bright emitter is responsible for the majority of the observed power from galaxies with previously undetected CO sources. We note that this flux level is fainter than any of the previous blind CO(3-2) detections in Table~\ref{tab:COdets}, and is comparable to a 0.17~Jy~km~s$^{-1}$ ASPECS CO(3-2) detection identified with a MUSE spec-$z$ prior at $z=2.028$; this source cannot be responsible for the excess power discussed here, however, as it lies outside the region of sky used in the power spectrum analysis. Nonzero power due to voxels with $F_{\nu}<0.25$~mJy implies that very faint sources with fluxes less than $0.10$~Jy~km~s$^{-1}$ may exist in the data cube and contribute to the observed excess in the CO-galaxy power spectrum, but their overall contribution is small (i.e., $<25$\%).

Since the masking analysis suggests the observed excess CO(3-2) power could be due to a single source, we can try to identify this source by masking (i.e., setting to zero) one-by-one each MUSE source position; note that only one MUSE position is masked at a time. If the excess power is due to a single source, then the power will remain unchanged (within measured uncertainties) until the MUSE position corresponding to the CO(3-2) emission is masked and the power goes to zero. Following this procedure, we observe the power drop to zero (magenta curve in Figure~\ref{fig:xspec_muse_coJ}) upon masking source MUSE ID 24. Masking of all other MUSE sources resulted in negligible changes to the power spectrum. Examination of the data cube $T_0$ reveals no significant flux at the source position corresponding to MUSE ID 24 (RA$=53.160088$~deg, declination$=-27.776356$~deg, $\nu_{obs}=97.57$~GHz). However, this location is within a beam's width of a known blind detected source, ASPECS-LP.3mm.01 or MUSE ID 35 (at RA$=53.160587$~deg, declination$=-27.776120$, $\nu_{obs}=97.58$, and so the $\sim1$~arcsecond radius used to extract emission from MUSE positions when computing the masked auto-power spectrum encompasses flux from the blindly detected source. Thus, we conclude that the observed excess in CO(3-2) power is not due to any MUSE sources with previously undetected CO(3-2) emission. 

We thus recompute the masked auto-power spectrum using all MUSE positions with potential CO $J=1,2,3,$ or 4 emission, now excluding MUSE ID 24, in order to revise the estimate of the fraction of total masked auto-power recovered by MUSE positions corresponding to ASPECS blind detections. We find that the MUSE positions corresponding to ASPECS blind detections recover $106\pm6.5$\% of the total CO shot noise power. 

\subsubsection{Cross-shot noise power spectrum} \label{sec:xshot}

Additionally, we measure the cross-power spectrum between the ASPECS data and MUSE position field, and refer to this quantity as the cross-shot noise power spectrum, $P_{\comath,gal}(k_{\cotwomath})$. In this case, we work directly with the ASPECS data cube $T_0$, which has the lowest RMS compared to the subsets $T_{\mathrm{I}}$, $T_{\mathrm{II}}$, etc.  Since the noise in $T_0$ has not been quantified, the error on $P_{\comath,gal}(k_{\cotwomath})$ is derived via simulation of random MUSE positions only, and is a lower estimate of the true error.

We normalize the grid of MUSE galaxy positions to have units corresponding to the dimensionless density fluctuation field $\delta_{gal}(\vec{x}_i) = \left( n_{gal}(\vec{x}_i) - \langle n_{gal} \rangle \right) / \langle n_{gal} \rangle $, where $n_{gal}(\vec{x}_i)$ refers to the number density of galaxies at position $\vec{x}_i$ in the cube, and $\langle n_{gal} \rangle$ is the mean number density of galaxies in the full volume. The cross-shot noise power spectrum between $T_0$ and the dimensionless density fluctuation cube, $G$, is then
\begin{equation}
\langle T_0^{*}(\vec{k}) G(\vec{k'}) \rangle \equiv (2\pi)^3 \delta_{\mathrm{D}}(\vec{k}-\vec{k'}) P_{\comath,gal}(k),
\label{eq:cross_shot_cogal}
\end{equation}
with units of $\mu$K~(Mpc~h$^{-1}$)$^{3}$.

As derived in, e.g., \citet{Breysse2019} and \citet{Wolz2017}, $P_{\comath,gal}(k)$ is proportional to the mean CO surface brightness of CO-emitting MUSE galaxies, $\langle T_{{\comath}, gal} \rangle$,
\begin{equation}
\label{eq:pcross_tco_ngal}
P_{CO,gal}(k) = \frac{\langle T_{{\comath}, gal} \rangle}{\langle n_{gal} \rangle}.
\end{equation}
Since the factor $1/\langle n_{gal}\rangle$ is equal to the amplitude of the shot noise term of the galaxy auto-power spectrum, $P^{shot}_{gal,gal}$ (units of (Mpc~h$^{-1}$)$^3$), we rearrange Equation~\ref{eq:pcross_tco_ngal} to write 
\begin{equation}
\label{eq:tcogal}
\langle T_{{\comath}, gal} \rangle = \frac{P_{\comath,gal}(k)}{P^{shot}_{gal,gal}},
\end{equation}
which has units of $\mu$K. 

Constraints on $\langle T_{\mathrm{CO,gal}}\rangle$ are obtained from the cross-shot noise power spectrum, $P_{\comath,gal}(k_{\cotwomath})$, according to Equation~\ref{eq:tcogal}. As illustrated in Figure~\ref{fig:pcross_tcogal}, we strongly detect the cross-shot noise power spectrum in both considered $J=3$ and $J=2$ transitions, finding no significant contribution to the observed mean CO surface brightness from MUSE sources with previously undetected CO(2-1) or CO(3-2) emission. It is interesting to note, however, that the average value of $\langle T_{\cotwomath,gal}\rangle$ is slightly nonzero after removing the blindly detected sources from the sample (red dashed curve, upper panel). After we discard the 18 MUSE positions with potential CO(2-1) emission that have poorly characterized MUSE spectra (or CONFID = 1, i.e., a non-secure redshift based on a singly unidentified line), then there is marginal (2.5$\sigma$) detection of excess CO(2-1) surface brightness from MUSE emitters without previous ASPECS CO detections (orange dashed curve) which can contribute $0.07\pm0.02$~$\mu$K, or $19\pm7.8$\% of the total observed CO(2-1) emission.\footnote{When averaging all bins from $k=10$-100~h~Mpc$^{-1}$, we find the significance of this ``excess" improves, and the previously undetected sources contribute $23\pm5.8$\% of the total observed CO(2-1) emission.} Not shown in Figure~\ref{fig:pcross_tcogal}, we measure non-detections of CO(1-0) and CO(4-3) surface brightnesses ($-0.0020\pm0.0082$~$\mu$K and $-0.022\pm0.028$~$\mu$K, respectively) in MUSE galaxies.

\paragraph{Implications for CO LFs} If there is a one-to-one correlation between CO-emitting galaxies and MUSE-selected galaxies, i.e., then  $\langle T_{\comath,gal}\rangle = \langle T_{\comath}\rangle$ (and $P^{\mathrm{NBF}}_{\comath,gal}(k_{\cotwomath}) = P_{\comath,\comath}(k_{\cotwomath})$), and one can use the above measurements on $\langle T_{\comath},gal \rangle$ (and $P^{\mathrm{NBF}}_{\comath,gal}(k_{\cotwomath})$) to constrain the CO luminosity function via Equation~\ref{eq:sbar_co} (and Equation~\ref{eq:pshot_coLF}). For example, if the observed $\sim20$\% excess surface brightness in CO(2-1) from galaxies without ASPECS detections is real, and the MUSE galaxies represent the complete population of total CO(2-1) emitters, then one can deduce that the ASPECS survey recovers $\gtrsim80$\% of the CO(2-1) surface brightness at its sensitivity threshold. Keeping other Schechter parameters in Table~\ref{tab:schechter_params} fixed, this suggests a relatively flat faint-end slope $\alpha \leq -0.1$ for the CO(2-1) LF at $z\sim1$.\footnote{This constraint on $\alpha$ is robust to changes in $\Phi_*$ and $L'_*$ within the quoted uncertainties in Table~\ref{tab:schechter_params}.} Of course, if there exist CO(2-1) emitters that do not have MUSE spec-$z$, then $\langle T_{\comath,gal}\rangle \neq \langle T_{\comath}\rangle$, and we cannot reliably infer constraints on the CO(2-1) LF. Given the high percentage (100\%) of ASPECS CO(2-1) detections with MUSE spec-$z$ counterparts, it is possible that $\langle T_{\cotwomath,gal}\rangle$ is not dramatically different from $\langle T_{\cotwomath}\rangle$. For CO(3-2), however, the percentage is much lower (40\%), so we do not attempt to draw conclusions about the CO(3-2) LF based on $\langle T_{\cothreemath,gal} \rangle$.

\section{Beyond ASPECS: Detecting the CO power spectrum} \label{sec:lim_vs_gs}

Based on the analytic estimate $[P^{shot}_{\comath,\comath}(k_{\cotwomath})]_{\mathrm{det}} = 113.24$~$\mu$K$^2$~(Mpc~h$^{-1}$)$^3$ from blindly detected sources (Section~\ref{sec:pshot_co_limit}), and our 3$\sigma$ upper limit $P^{shot}_{\comath,\comath}(k_{\cotwomath}) \leq 187.29$~$\mu$K$^2$~(Mpc~h$^{-1}$)$^3$, we consider the question: what would be needed in an ALMA line survey to obtain a significant ($\geq5\sigma$) detection on the CO shot noise power spectrum?

The empirically determined uncertainty (Equation~\ref{eq:final_error}) on the measured shot noise power $\langle P_{\comath,\comath}(k_{\cotwomath}) \rangle_{tot} = -45$~$\mu$K$^2$~(Mpc~h$^{-1}$)$^3$ was found to be $\langle \delta P_{\comath,\comath}(k_{\cotwomath}) \rangle_{tot} = 77$~$\mu$K$^2$~(Mpc~h$^{-1}$)$^3$. If the ASPECS blind detections can account for the bulk of the observed power, i.e., if $[P^{shot}_{\comath,\comath}(k_{\cotwomath})]_{\mathrm{det}}$ approximates closely $P_{\comath,\comath}(k_{\cotwomath})$ in our measured $k$ range, which is supported by the cross-power spectrum analysis in Section~\ref{sec:xpower_nbf}, then $\langle \delta P_{\comath,\comath}(k_{\cotwomath}) \rangle$ must be reduced by a factor of 3.4 in order to obtain a $5\sigma$ detection on the total CO shot noise power, or $\sigma_{N,\mathrm{40chn}}$ must be reduced be a factor of $\sqrt{3.4}=1.8$, since $\langle \delta P_{\comath,\comath}(k_{\cotwomath}) \rangle \propto \sigma_{N,\mathrm{40chn}}^2$. Since $\langle \delta P_{\comath,\comath}(k_{\cotwomath}) \rangle \propto \sigma_{N,\mathrm{40chn}}^2$, this can be achieved by increasing the integration time $t_{int}$ by a factor of 3.4, as $\sigma_{N,\mathrm{40chn}} \propto t_{int}^{-1/2}$.

Alternatively, the signal-to-noise SNR on $P^{shot}_{\comath,\comath}(k_{\cotwomath})$ can be improved by increasing the number of independent $k$ modes in the survey. Because SNR $= P_N / (N_m)^{1/2}$, the total number of modes would need to be increased by a factor $(3.4)^2 = 12$. In order to cover the same redshift range in CO, this would require enlarging the survey area to $12\Delta\theta_S^2 = 12(1.8 \ \mathrm{arcmin})^2 = 39$~arcmin$^2$ while scanning in frequency across the same 30~GHz bandwidth, which is substantially more expensive than the $\sim3$-fold increase in $t_{int}$ estimated above. However, increased areal coverage would enable observations of more massive galaxies that are not captured in ASPECS LP.

The community is considering the Next Generation VLA (ngVLA) project as an order of magnitude improvement in observational capabilities in the 1--115~GHz regime over existing facilities, such as ALMA and the JVLA \citep{Selina2018, Murphy2018}. The ngVLA core array will have $100 \times 18$~m diameter antennas within $\sim 1$~km diameter, and a minimum bandwidth of 20 GHz. While the primary beam, and thus the instantaneous field of view, is a factor 2.25 smaller than that of ALMA at a given frequency, the collecting area is about 4.5 times larger, and the bandwidth is at least 2.5 times larger. The implied time to cover the same cosmic volume to the same sensitivity is then a factor of about 20 shorter than for the current ASPECS program.\footnote{Note that we have assumed an observed wavelength range centered at 3~mm, identical to ASPECS LP, for this comparison, and ngVLA will also offer powerful cm-wave capabilities.}

 \subsection{Power spectrum vs. Individual galaxy detection}

Given that the ASPECS blind detections recover $\sim100$\% of the observed noise-bias free cross-power between CO and MUSE galaxies, it is difficult to motivate longer integration times to obtain a detection on the power spectrum. To explain why the ASPECS LP 3mm survey is more efficient at recovering the CO shot noise power by detecting individual galaxies, we consider the relationship between the significance of the individual detections and the expected uncertainty on the shot noise power spectrum.

Suppose the survey can detect galaxies down to some minimum luminosity $L_{min, det}$. Explicitly, the relation between the mean surface brightness sensitivity $\sigma_N$ (in units of Jy~sr$^{-1}$, for example) and $L_{min,det}$ is
\begin{equation}
\frac{L_{min,det}}{4\pi D_L^2 \Delta\nu_{chn}\Delta\theta_b^2} = \epsilon \sigma_N,
\end{equation}
where $\Delta\nu_{chn}$ and $\Delta\theta_b^2$ are in units of Hz and steradian, respectively, and $\epsilon$ is the required significance for a detection. For example, in Section~\ref{sec:mean_tco}, we set the ASPECS blind detection threshold as $\epsilon=7$. Then, we can rewrite $P_N=\sigma_N^2 V_{vox}$ (Equation~\ref{eq:pnoise}) in terms of $L_{min,det}$ and $\epsilon$:
\begin{align}
P_N &= \frac{1}{\epsilon^2} \left(\frac{L_{min,det}}{4\pi D_L^2 \Delta\nu_{chn}\Delta\theta_b^2}\right)^2 yD_A^2 \Delta\nu_{chn}\Delta\theta_b^2 \nonumber \\
        &= \frac{A^2 L_{min,det}^2}{\epsilon^2 V_{vox}},
\end{align}
where $A = y D_A^2 = c D_L^2 / (H(z) \nu_{rest})$. 

For the simple toy case, where all galaxies have identical luminosity $L = L_{min,det}$ and number density $n$, the shot noise power (cf. Equation~\ref{eq:pshot_coLF}) can be written as $P_{shot} = A^2 L_{min,det}^2 n$, and the SNR on $P_{shot}$ as
\begin{align}
\mathrm{SNR} &= \frac{P_{shot}}{P_N}\sqrt{N_m}\nonumber \\
                        &= \epsilon^2 \sqrt{N_m} n V_{vox} \nonumber \\
                        &= \frac{\epsilon^2 N_{gal}}{\sqrt{N_m}}.
                        \label{eq:SNR_lmindet_sameL}
\end{align}
Note that we have used $n V_{vox} = n V_{survey} \times (V_{vox}/V_{survey}) = N_{gal} / N_m$, where $N_{gal}$ is the total number of galaxies in the survey, to arrive at Equation~\ref{eq:SNR_lmindet_sameL}.

For the more realistic case where all galaxies follow a Schechter-form luminosity function, $P_{shot}~=~A^2~L_*^2~\Phi_* \Gamma(3~+~\alpha)$ \citep[e.g.,][]{LidzTaylor2016}, and we obtain
\begin{align}
\mathrm{SNR} &= \frac{\epsilon^2}{\sqrt{N_m}} \frac{L_*^2\Gamma(3+\alpha)}{L_{min,det}^2}\Phi_{*} V_{survey} \nonumber \\
                        &= \frac{\epsilon^2 \mathcal{N}_{gal}}{\sqrt{N_m}},
\end{align}
where $\mathcal{N}_{gal} = L_*^2\Gamma(3+\alpha)/L_{min,det}^2\Phi_{*} V_{survey}$ is the effective number of detectable galaxies with in the survey volume.

For the ASPECS survey and relevant Schechter function LFs, $\epsilon=7$ and $\mathcal{N}_{gal} = 15$ CO(2-1) $+\ 7$ CO(3-2) $+\ 5$ CO(4-3) $ = 27$~galaxies. $N_m$ is $1.18\times10^7$ modes, or the fraction of independent modes that falls within $k=10$--100~h~Mpc$^{-1}$. Note that, for mode-counting, we refer to the $T_{0,\mathrm{2chn}}$ cube defined by a grid of roughly 64 beams by 64 beams by 3935 channels, because this is the cube where the threshold for individual detection was defined. Then, SNR $= 0.39$, and the CO power spectrum is not expected to be detected at high significance;\footnote{Note that $\mathcal{N}_{gal}$ and, thus, the derived SNR contains contributions from different $J$ transitions. If we project the SNR into the CO(2-1) frame, as we have done for the power spectrum analysis, we must scale the relative contributions to the total SNR by the appropriate distortion factors, resulting in SNR $= 0.29$.} the shot noise signal from individual galaxies is ``diluted" by the factor $1/\sqrt{N_m}$. In this regime of small $\sigma_N$ and survey volume, the ``traditional galaxy survey analysis," where one identifies individual sources above the flux limit, provides a better SNR on the shot-noise term than the power spectrum analysis; the number and formulae here demonstrate this.

\section{Conclusions} \label{sec:conclusion}

\begin{figure}
\center
\includegraphics[width=0.75\textwidth]{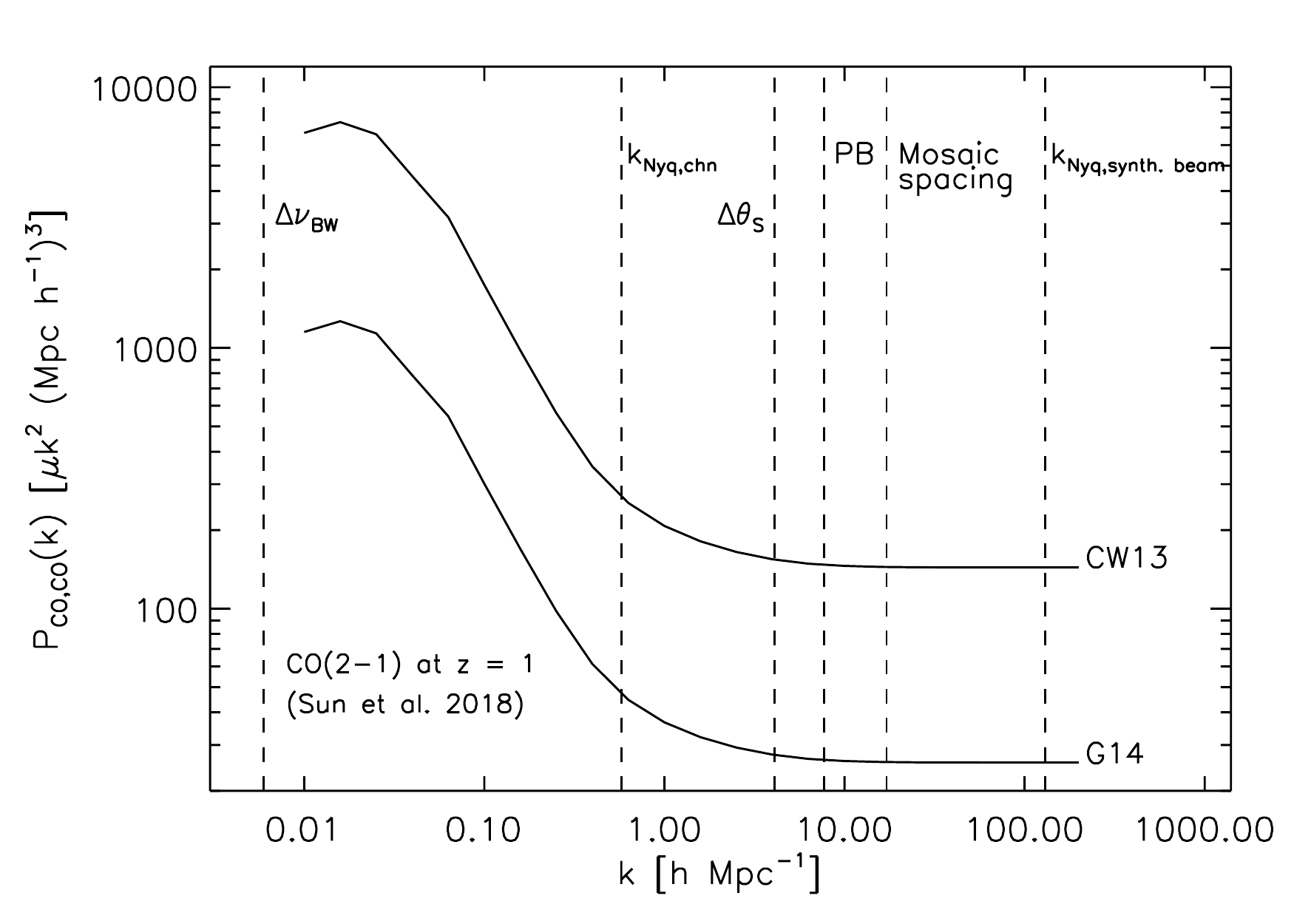}
\caption{Total predicted CO(2-1) power from \citet{Sun2018}. ``CW13'' and ``G14'' refer to different prescriptions used for relating CO luminosities to estimated IR luminosities in the model, based on the compilations in \citet{CarilliWalter2013} and \citet{Greve2014}. Power from clustering and shot noise dominate at $k<1$~h~Mpc$^{-1}$ and $k>1$~h~Mpc$^{-1}$, respectively. Vertical dashed line indicate $k$ values corresponding to ASPECS survey parameters, namely, survey bandwidth $\Delta\nu_{BW}$, channel width, survey width $\Delta\theta_S$ (1.84~arcmin), the antenna primary beam (0.98~arcmin), mosaic spacing (25.4~arcsec), and synthesized beam size $\Delta\theta_b$ ($1.8 \times 1.5$~arcsec).}
\label{fig:pco21_model_kscales}
\end{figure}

We presented a power spectrum analysis of the ASPECS~LP~3mm dataset. Key results from this analysis are:

\begin{enumerate}[label=(\roman*)]
\item We derive a lower limit on the mean total CO surface brightness $\langle T_{\comath} \rangle$ at 99~GHz by summing the observed fluxes from the 16 ASPECS blind detections: $\langle T_{\comath} \rangle \geq 0.55\pm0.02$~$\mu$K. As the CO clustering power is proportional to  $\langle T_{\comath} \rangle^2$, this information from individually detected sources provides information on the CO clustering amplitude at large physical scales ($k < 1.0$~h~Mpc$^{-1}$), as well as an indication of foreground contamination in CMB spectral distortion mapping experiments.
\item We derive an upper limit (3$\sigma$) on the 3D CO autopower spectrum at $10 \lesssim k \lesssim 100$~h~Mpc$^{-1}$:  $P_{\comath, \comath}(k_{\cotwomath}) \leq 187.29$~$\mu$K$^2$~(Mpc~h$^{-1}$)$^3$, which is broadly consistent with ASPECS observed CO LFs presented in D19. The upper limit measured in this study places a significantly tighter constraint on total CO power than predictions based on the Schechter-form LFs, due to the large uncertainties in parameters $\Phi_*$ and $L_*$ for the individual LFs at $z>2$. Extrapolating the observed CO(1-0) LF at $z=2$--3 in the COLD$z$ survey to a CO(2-1) power in the ASPECS redshift range is consistent with our result, while the same procedure applied to the COPSS~II measured shot noise power yields a CO(2-1) power that is $>5$ times greater than our upper limit.
\item We report detections of the masked noise-bias free auto-power spectrum, $P^{gal}_{\comath,\comath}(k_{\cotwomath})$ and cross-shot noise power spectrum, $P_{\comath,gal}(k_{\cotwomath})$, between CO and rest-frame optical/UV tracers from the MUSE spectroscopic redshift catalog yield constraints on the second and first moments, respectively, of the CO-emitting MUSE galaxy luminosity functions. We found that $106\pm6.5$\% of the measured power in total CO shot noise is comprised of surface brightness fluctuations from MUSE galaxies with previously detected CO from the ASPECS blind line search. We also constrained the contribution of ASPECS blind detections to the observed CO mean intensity of MUSE emitters, finding that up to $\sim20$\% of $ \langle T_{\cotwomath, \mathrm{MUSE}} \rangle$ is attributed to emission from galaxies below the threshold for individual detection. With the assumption that all observed CO(2-1) emission orginates from MUSE emitters, $P_{\comath,gal}(k_{\cotwomath})$ can be used to place a direct constraint on the mean CO(2-1) surface brightness and, thus, faint-end slope of the luminosity function, suggesting $\alpha\leq-0.1$ at $z\sim1$.

BDU would like to thank Riccardo Pavesi for providing data to aid in the comparison of the ASPECS shot noise power spectrum measurement with COLD$z$ luminosity functions, and Guochao Sun for providing model data to use in Figure~\ref{fig:pco21_model_kscales}. We thank Patrick Breysse for useful discussions pertaining to the interpretation of the CO-galaxy cross-power spectrum. FW acknowledges support from ERC Grant `Cosmic Gas' (740246). DR acknowledges support from the National Science Foundation under grant number AST-1614213 and from the Alexander von Humboldt Foundation through a Humboldt Research Fellowship for Experienced Researchers. The authors jointly thank the anonymous referee for a constructive report that helped to improve this paper.

\emph{Facility}: ALMA. \emph{data}: 2016.1.00324.L. ALMA is a partnership of ESO (representing its member states), NSF (USA) and NINS (Japan), together with NRC (Canada), NSC and ASIAA (Taiwan), and KASI (Re- public of Korea), in cooperation with the Republic of Chile. The Joint ALMA Observatory is operated by ESO, AUI/NRAO and NAOJ.

\end{enumerate}

\newpage

\clearpage

\begin{figure}
\centering
\pgfkeys{/pgf/inner sep=0.5em}
\begin{forest}
for tree={
   draw,
    minimum height=2.2cm,
    anchor=north,
    align=center,
    child anchor=north
},
[\hspace{-1.2cm}{\textbf{\large $\mathbf{\widetilde{T}}_{\mathbf{0}}$}} \\ \hspace{-1.2cm}{$\sigma=\sigma_{N,40\mathrm{chn}}$}, align=center,name=Tnot
    [\hspace{-1.2cm}{\textbf{\large $\mathbf{\widetilde{T}}_{\mathbf{I}}$}} \\ \hspace{-1.2cm}{$\sigma=\sqrt{2}\times\sigma_{N,40\mathrm{chn}}$}, name=Tone
        [\hspace{-1.2cm}\large$\boldsymbol{\widetilde{\tau}}_{\mathbf{1}}$ \\ \hspace{-1.2cm}{$\sigma=2\times\sigma_{N,40\mathrm{chn}}$}]
        [\hspace{-1.2cm}\large$\boldsymbol{\widetilde{\tau}}_{\mathbf{2}}$ \\ \hspace{-1.2cm}{$\sigma=2\times\sigma_{N,40\mathrm{chn}}$}]
     ]
     [\hspace{-1.1cm}{\textbf{\large $\mathbf{\widetilde{T}}_{\mathbf{II}}$}} \\ \hspace{-1.2cm}{$\sigma=\sqrt{2}\times\sigma_{N,40\mathrm{chn}}$}, name=Ttwo
         [\hspace{-1.2cm}\large$\boldsymbol{\widetilde{\tau}}_{\mathbf{3}}$ \\ \hspace{-1.2cm}{$\sigma=2\times\sigma_{N,40\mathrm{chn}}$}]
         [\hspace{-1.2cm}\large$\boldsymbol{\widetilde{\tau}}_{\mathbf{4}}$ \\ \hspace{-1.2cm}{$\sigma=2\times\sigma_{N,40\mathrm{chn}}$}]
     ]
]
\end{forest} \\
\end{figure}
\begin{figure}[h]
\center
\begin{tabular}{c c}
\includegraphics[width=0.45\textwidth]{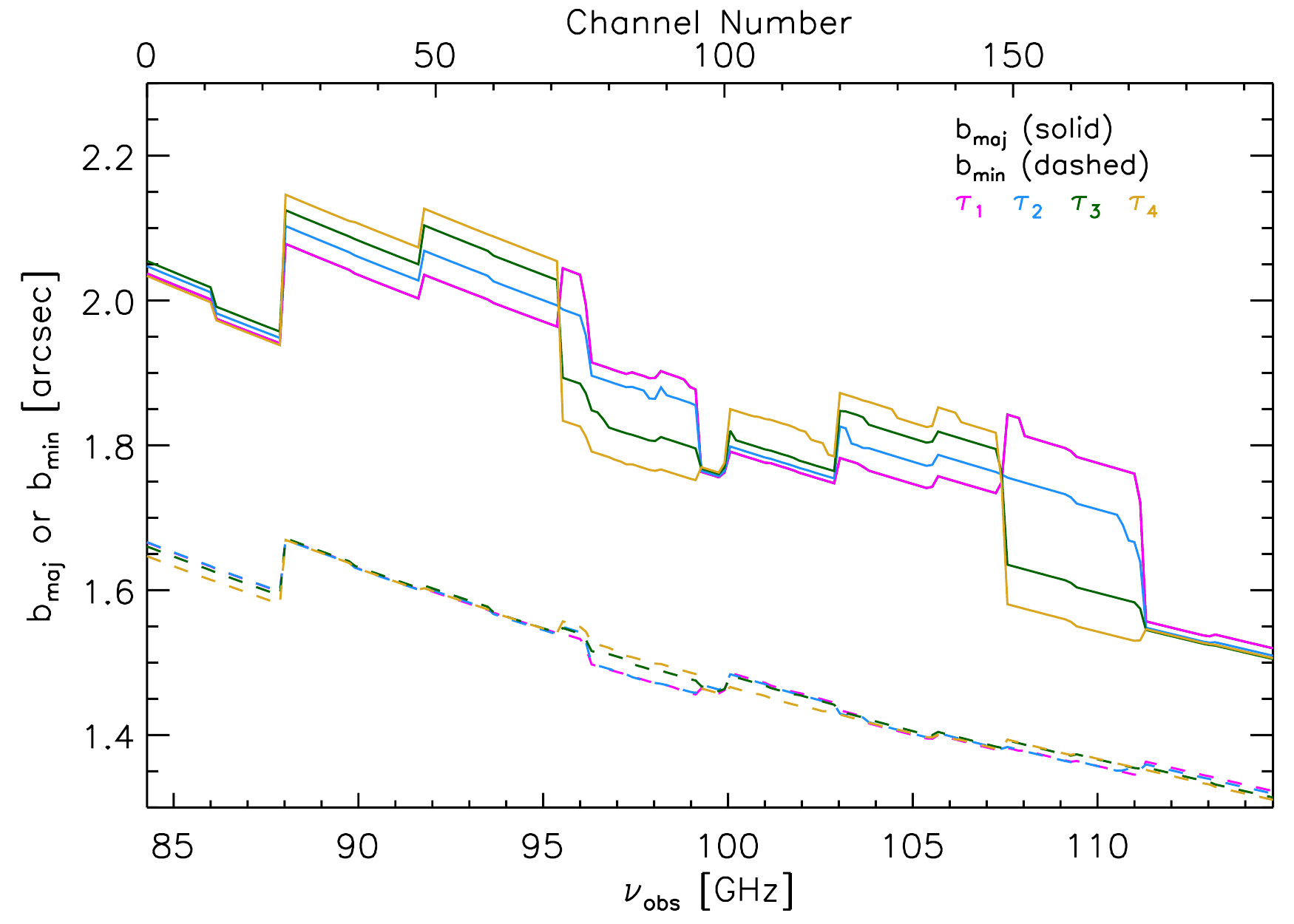} &
\includegraphics[width=0.45\textwidth]{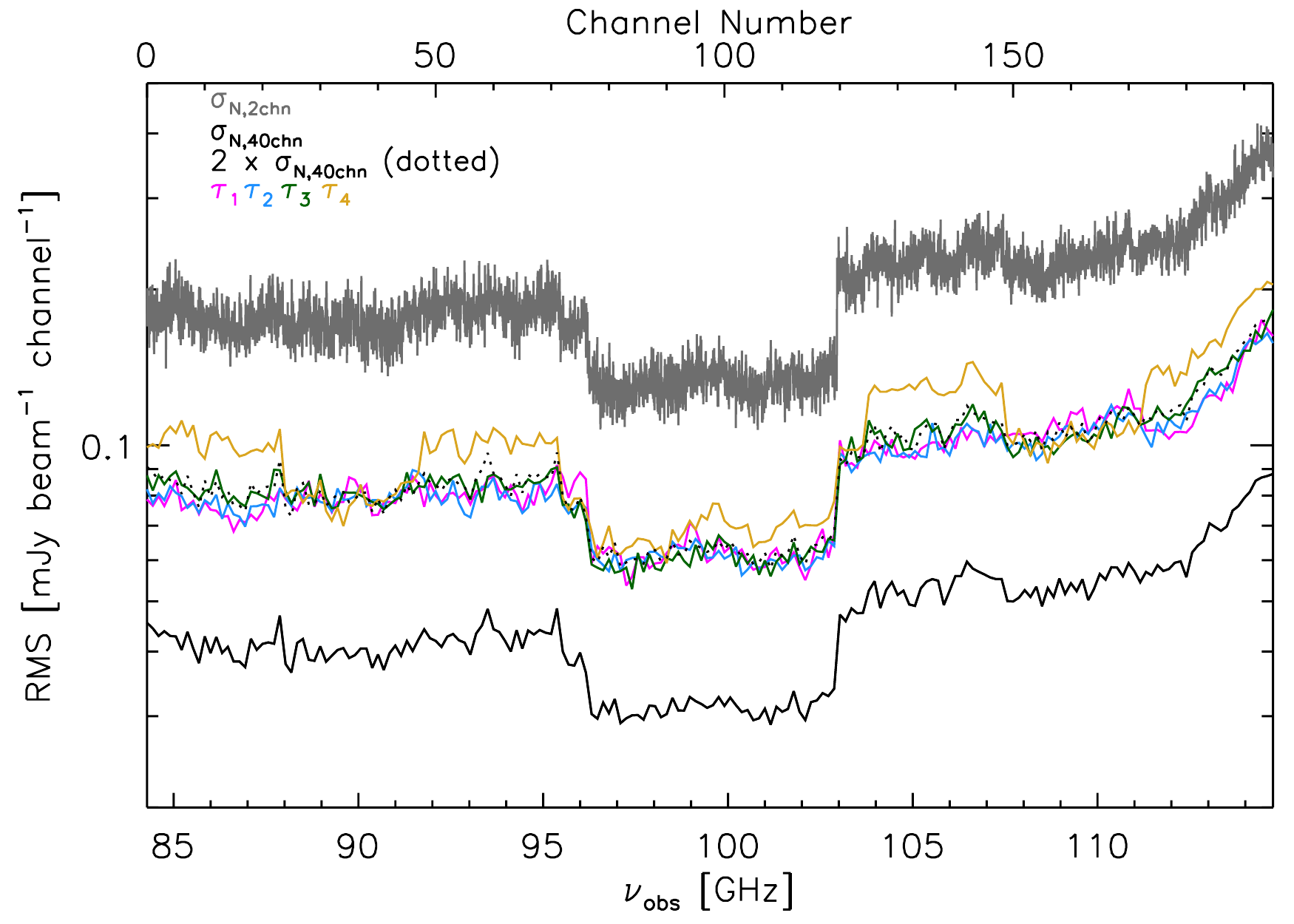} 
\end{tabular}
\caption{\emph{Top:} Tree diagram illustrating derivation of visibility data subsets $\widetilde{T}_\mathrm{I}$, $\widetilde{T}_\mathrm{II}$, $\widetilde{\tau}_1$,  $\widetilde{\tau}_2$,  $\widetilde{\tau}_3$, and $\widetilde{\tau}_4$ from the original visibility dataset $\widetilde{T}_0$, used to determine the noise-bias free power spectrum, $P_{T_{\mathrm{I}}, T_{\mathrm{II}}}(k)$, and corresponding error $\delta P_{T_{\mathrm{I}}, T_{\mathrm{II}}}(k)$. The expected mean RMS of the resulting image generated from each visibility dataset is also labeled. \emph{Bottom left:} Beam major (solid curves) and minor (dashed curves) axes as a function of observed frequency for image subsets $\tau_1$, $\tau_2$, $\tau_3$, $\tau_4$. Numbers in the upper $x$-axis refer to channel number of the data cube, which has been imaged with factor of 40 re-binning in frequency. \emph{Bottom right:} RMS as a function of observed frequency. Upper $x$-axis is the same as in left-adjacent panel.}
\label{fig:timesplit_properties}
\end{figure}

\newpage{}

\begin{figure}[t]
\center
\includegraphics[width=0.8\textwidth]{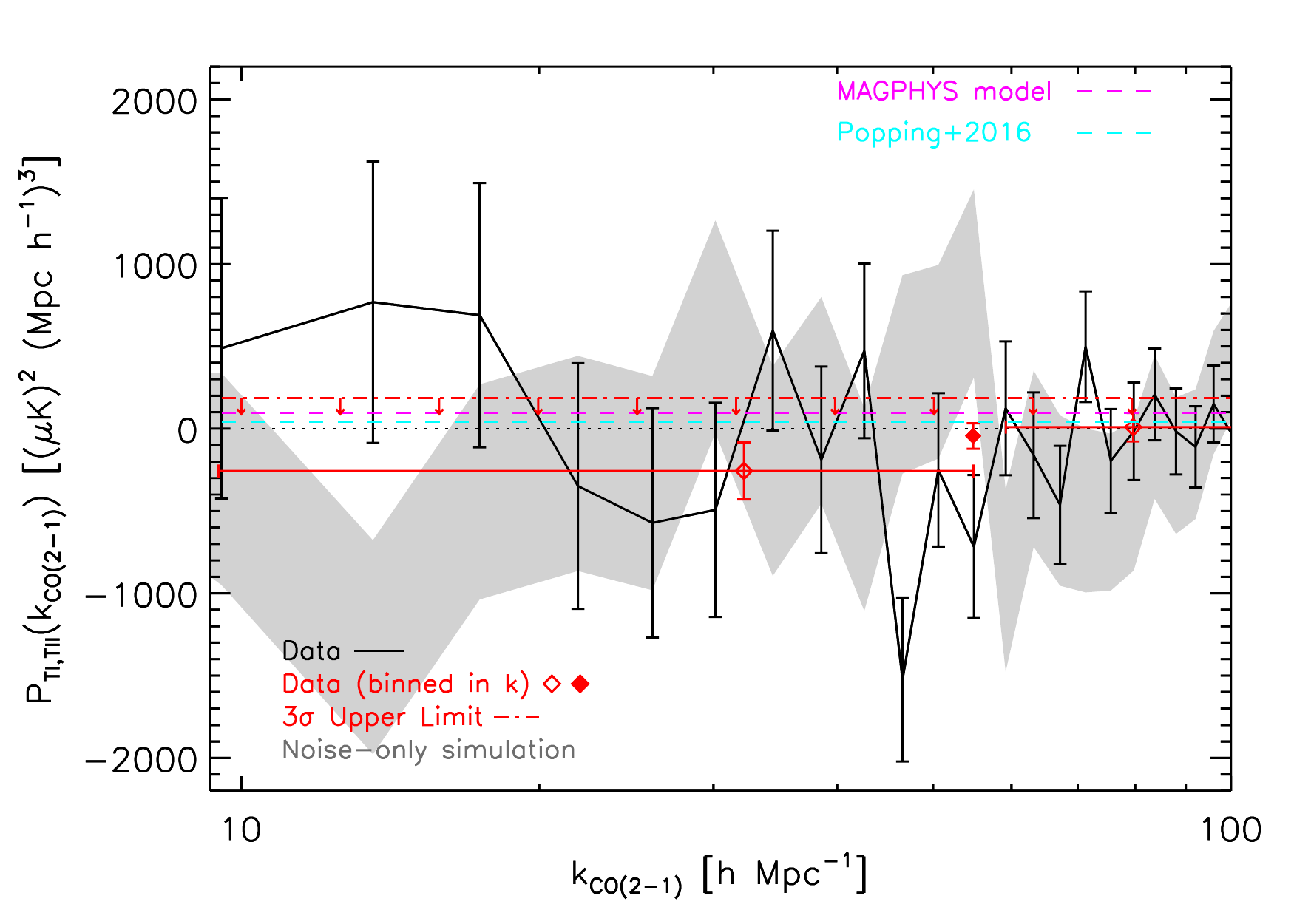}
\caption{Measurement of the noise-bias free CO auto-power spectrum (solid black curve), $P_{\comath,\comath}(k_{\cotwomath})$, in the ASPECS-LP Band 3 survey. Error bars on $P_{\comath,\comath}(k_{\cotwomath})$ represent values from a polynomial fit to raw errors, $\langle \delta P_{\comath,\comath}(k_{\cotwomath})\rangle$, calculated from Equation~\ref{eq:final_error}. The inverse-variance weighted mean CO power is plotted for two bins averaging modes in the upper and lower halves (unfilled red diamonds) of probed $k_{\cotwomath}$, as well as for a bin (filled red diamond) containing the inverse-variance weighted mean CO power for all $k_{\cotwomath}\sim10$-100~h~Mpc$^{-1}$. The 3-$\sigma$ upper limit, calculated using the uncertainty on the latter binned power spectrum, is plotted as the red dot-dashed line. For comparison, the grey swath bounds the 1-$\sigma$ confidence region for the noise-bias free power spectrum of a single realization of a noise-only simulated data cube from CASA task \textsf{simobserve}, with corresponding error bars calculated in the same way as for the real data. Theoretical predictions for $P_{\comath,\comath}^{shot}(k_{\cotwomath})$ from \citet{Popping2016} (cyan dashed line) and a model based on SED-fitting of known sources in the ASPECS survey field (``MAGPHYS model"; magenta dashed line) are also plotted. A dotted black line, which illustrates where the measured power is zero, is drawn for reference.} 
\label{fig:autopower}
\end{figure}

\newpage

\begin{figure}[h]
\center
\includegraphics[width=0.75\textwidth]{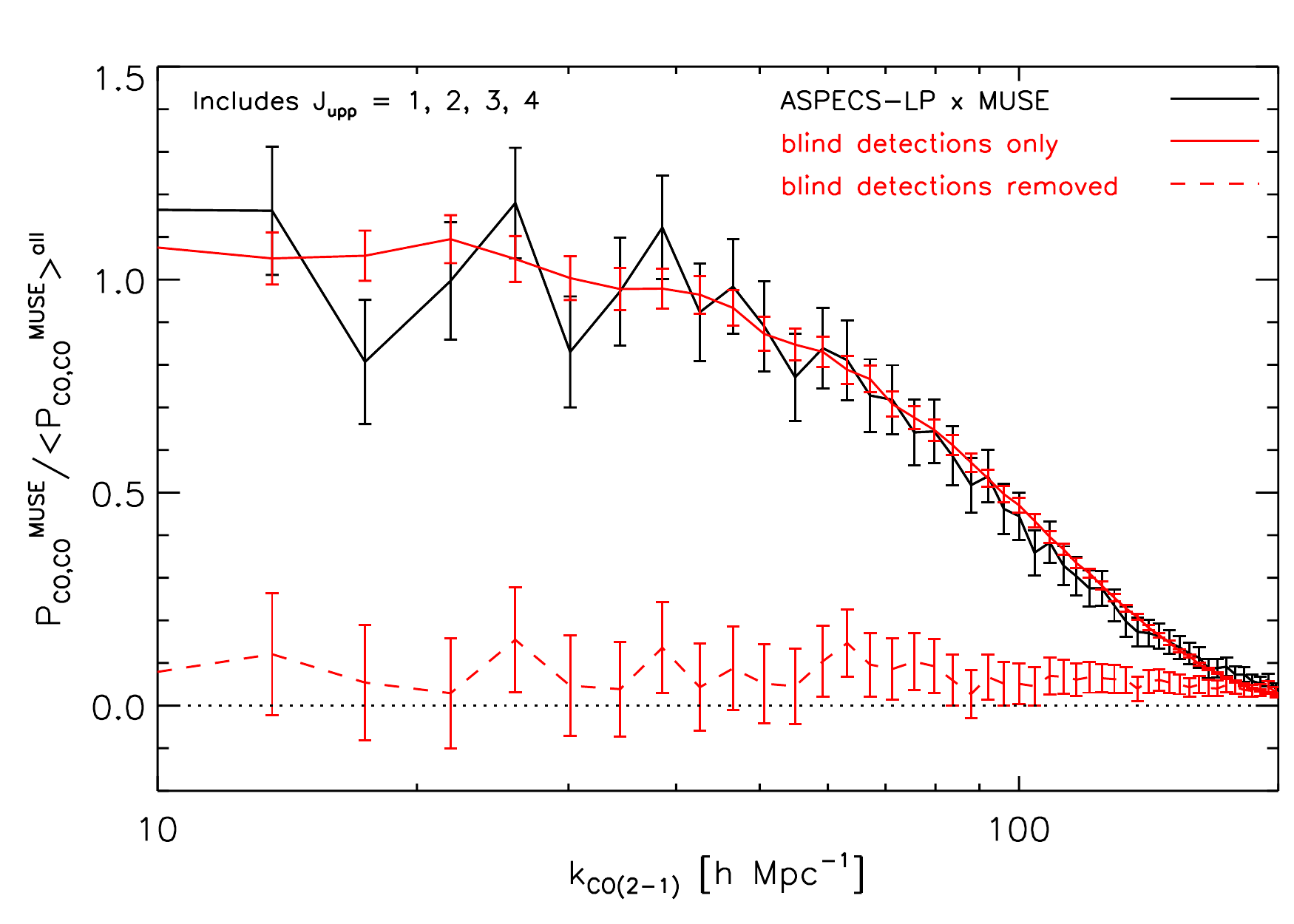}
\caption{Masked noise-bias free auto-power spectrum, $P^{\mathrm{MUSE}}_{\comath,\comath}(k_{\cotwomath})$, of ASPECS LP datacubes with MUSE 3D source positions from \citet{Inami2017}. All 415 MUSE sources with spec-$z$ that fall within redshift ranges observable by ASPECS in CO(1-0), CO(2-1), CO(3-2), and CO(4-3) are included. The solid black curve represents the total power measured using ASPECS LP data and all MUSE sources with potential CO emission up to $J=4$. The solid and dashed red curves show the power using all MUSE sources with and without, respectively, a previously detected CO counterpart from the ASPECS line search. The $y$-axis has been scaled by a factor $1/\langle P^{\mathrm{MUSE}}_{\comath,\comath}(k_{\cotwomath})\rangle^{\mathrm{all}}$, so that each curve represents the contribution of the respective subset of galaxies to the total power measured when using all MUSE galaxies.}
\label{fig:xspec_muse_allJ}
\end{figure}

\begin{figure}[h]
\center
\begin{tabular}{c c}
\includegraphics[width=0.45\textwidth]{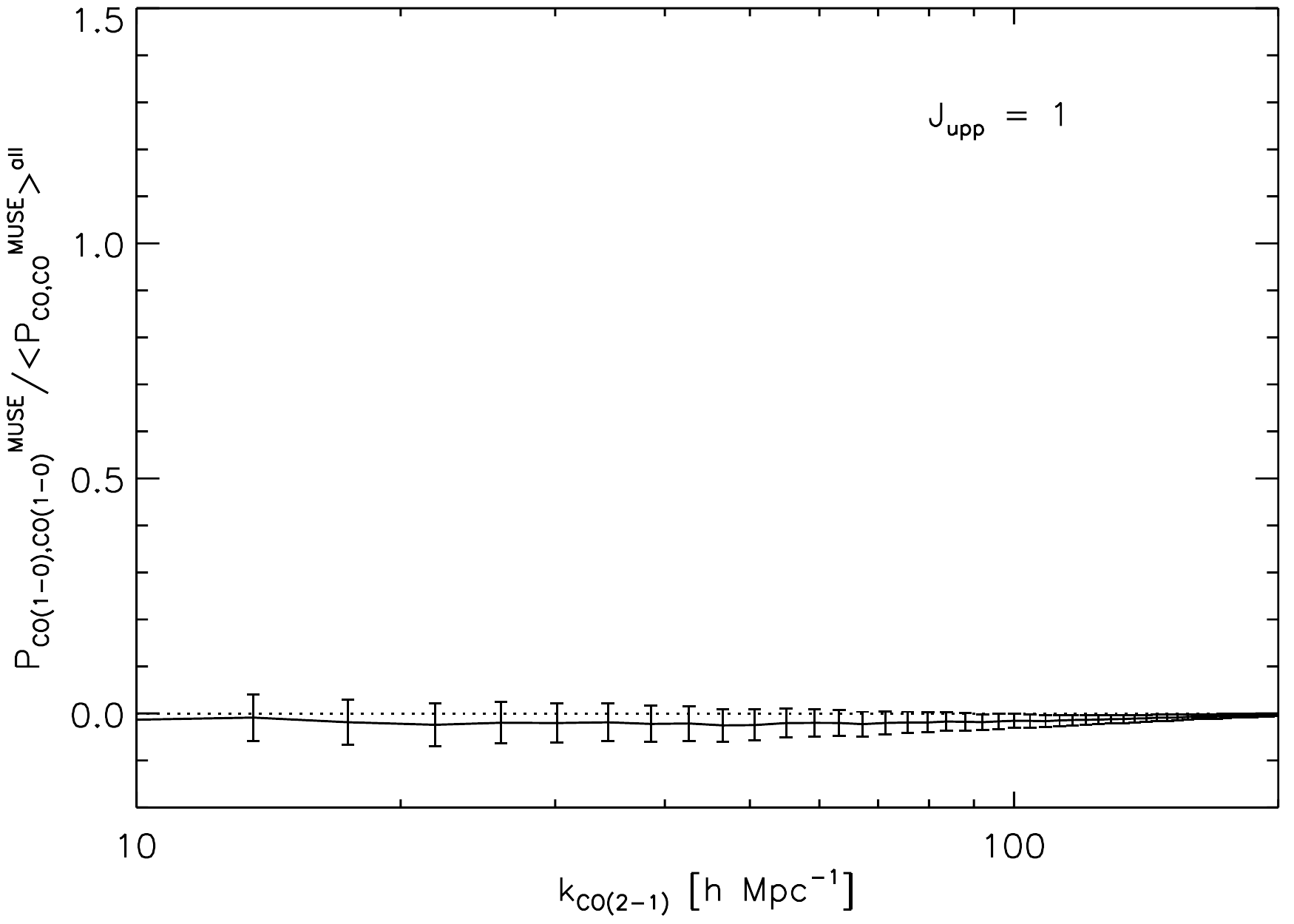} &
\hspace{0.1in}
\includegraphics[width=0.45\textwidth]{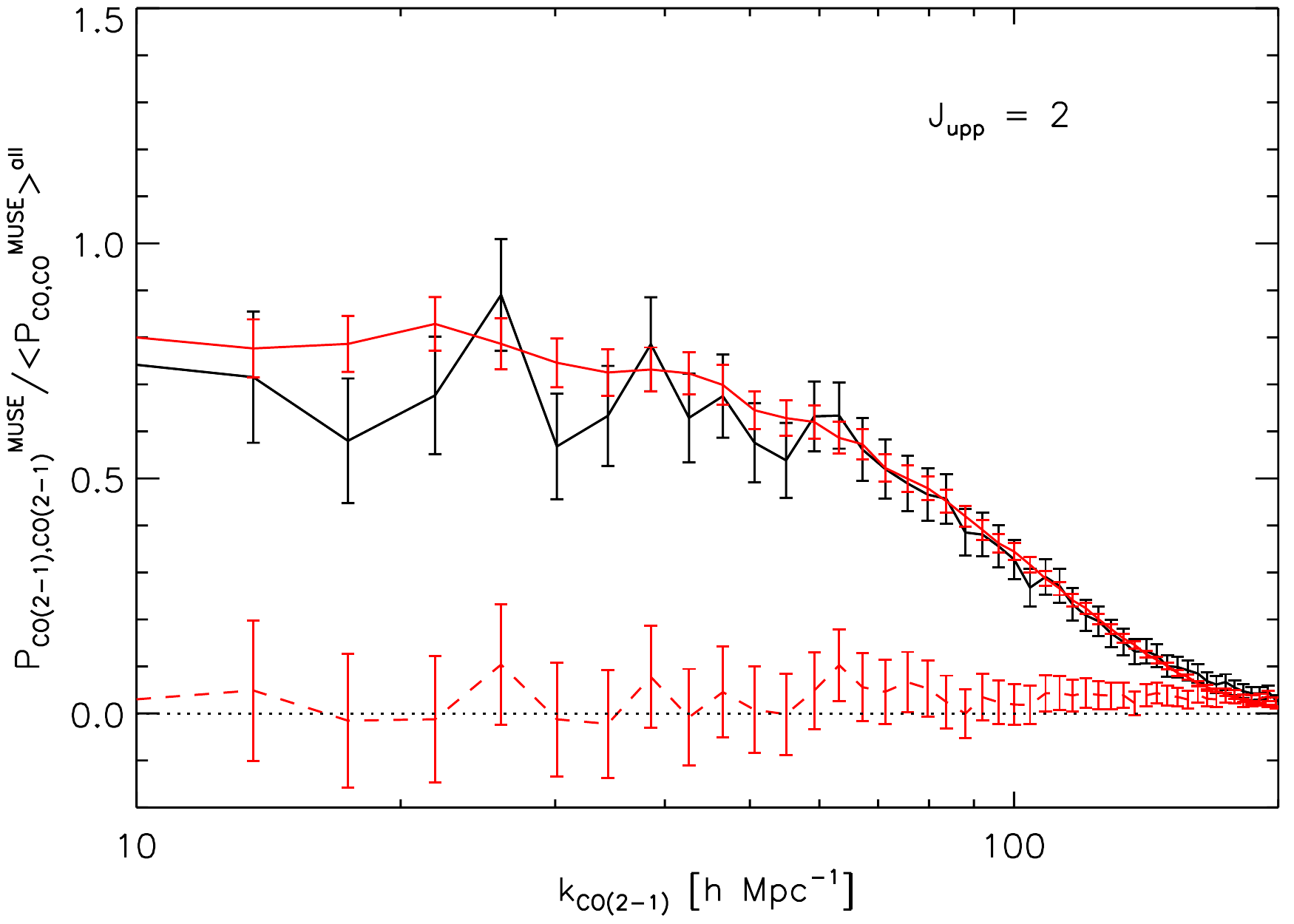} \\
\includegraphics[width=0.45\textwidth]{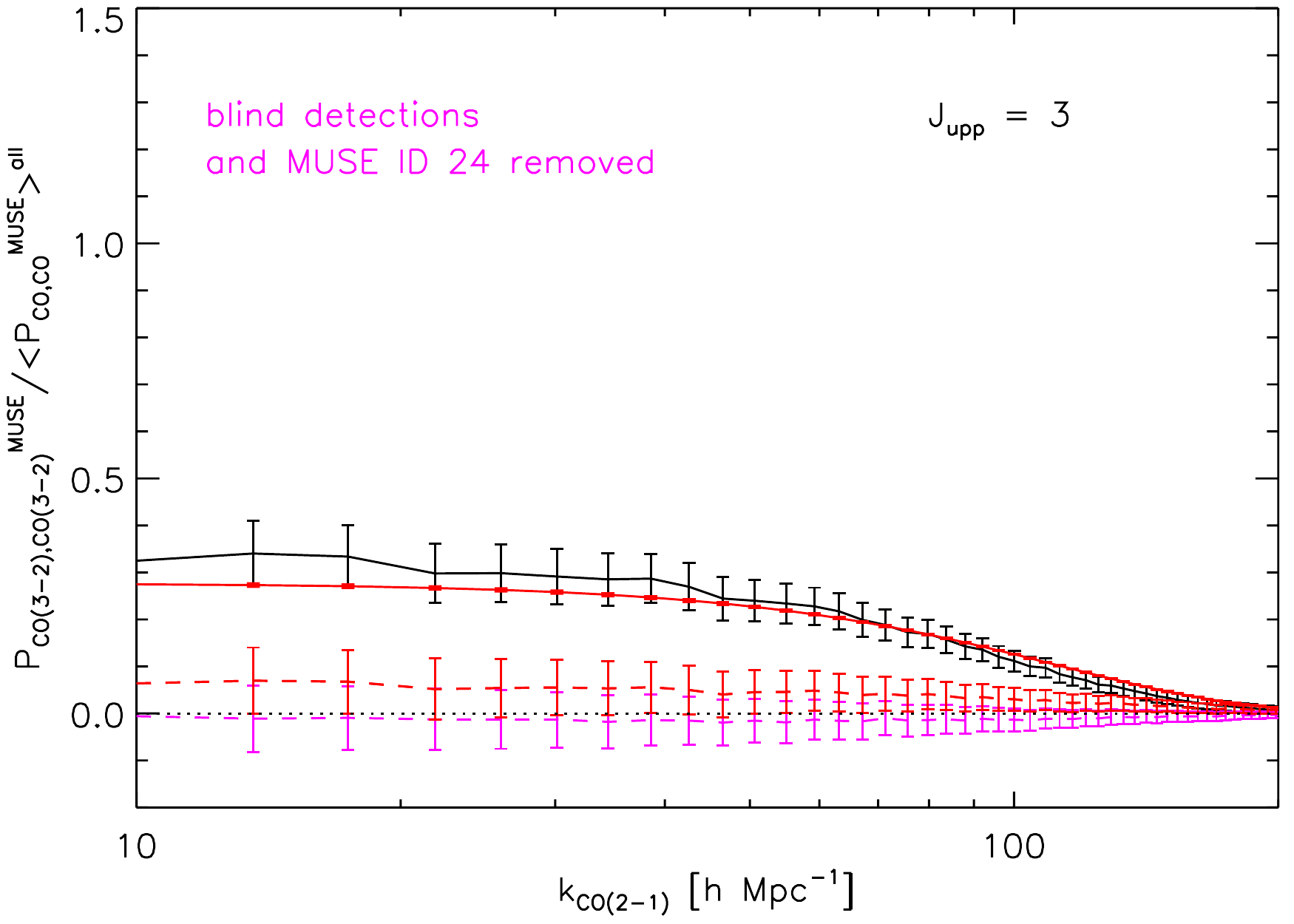} &
\hspace{0.1in}
\includegraphics[width=0.45\textwidth]{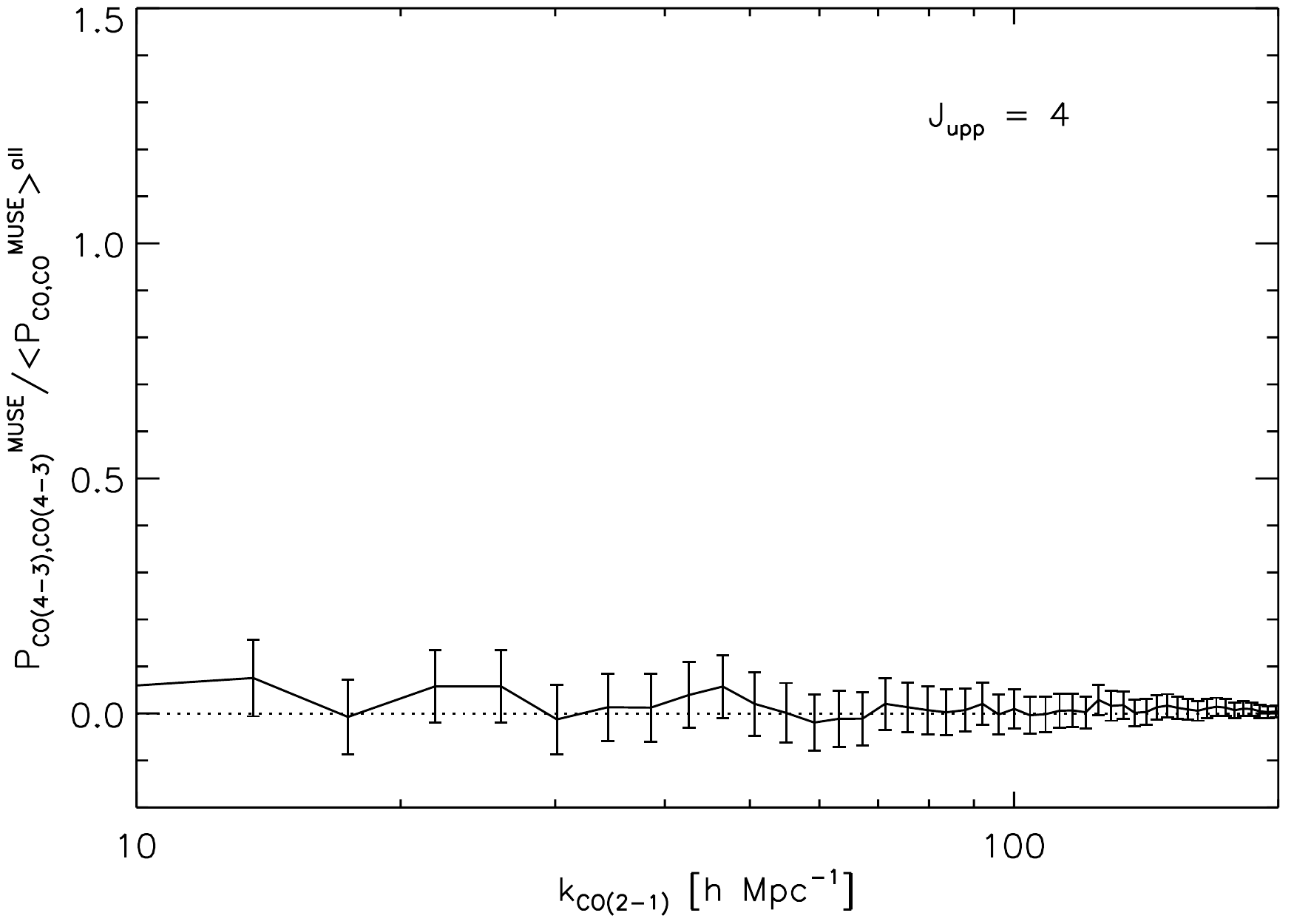} \\
\end{tabular}
\caption{Cross-power spectrum of ASPECS LP datacubes with MUSE 3D positions, separated into contributions by CO $J$ transition. Color-coding is the same as in Figure~\ref{fig:xspec_muse_allJ}. Magenta dashed curve in bottom left panel denotes masked CO(3-2) autopower measured upon removing MUSE ID 24, in addition to MUSE IDs corresponding to ASPECS blind detections.}
\label{fig:xspec_muse_coJ}
\end{figure}

\begin{figure}[h]
\center
\includegraphics[width = 0.75\textwidth]{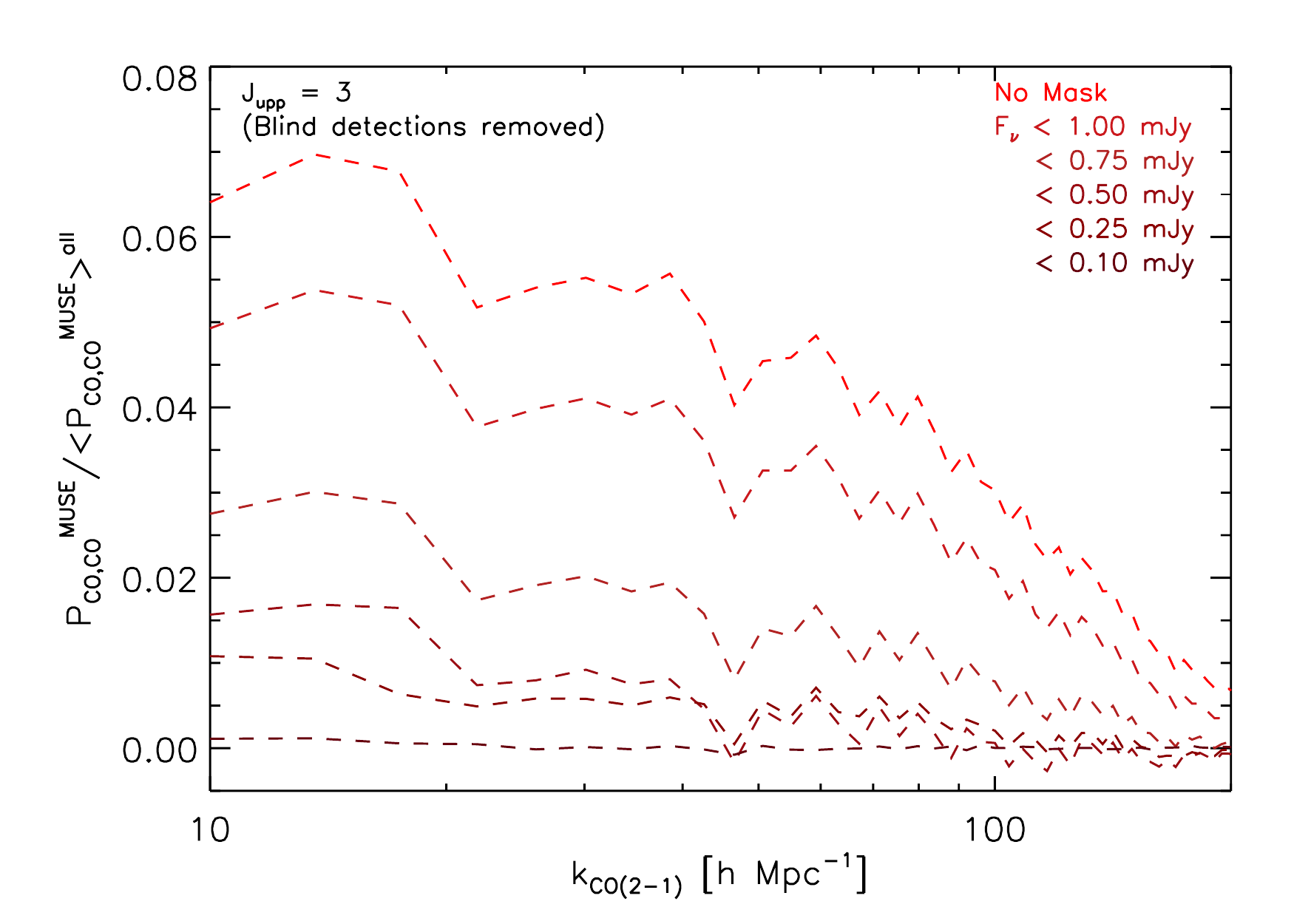}
\caption{Contribution of voxels with different flux density thresholds ($\vert F_{\nu}\vert < 1.00$, 0.75, 0.50, 0.25, and 0.1~mJy) to the masked noise-bias free CO(3-2) auto-power spectrum using MUSE source positions that lack a previous ASPECS blind CO(3-2) detection. The uppermost red dashed curve is identical to the red dashed curve in the bottom left panel in Figure~\ref{fig:xspec_muse_coJ}; please refer to that Figure for error bars on the measured $P_{\cothreemath,gal}(k_{\cotwomath})$.}
\label{fig:xspec_muse_co32_masking}
\end{figure}

\clearpage

\begin{figure}[h]
\center
\includegraphics[width = 0.75\textwidth]{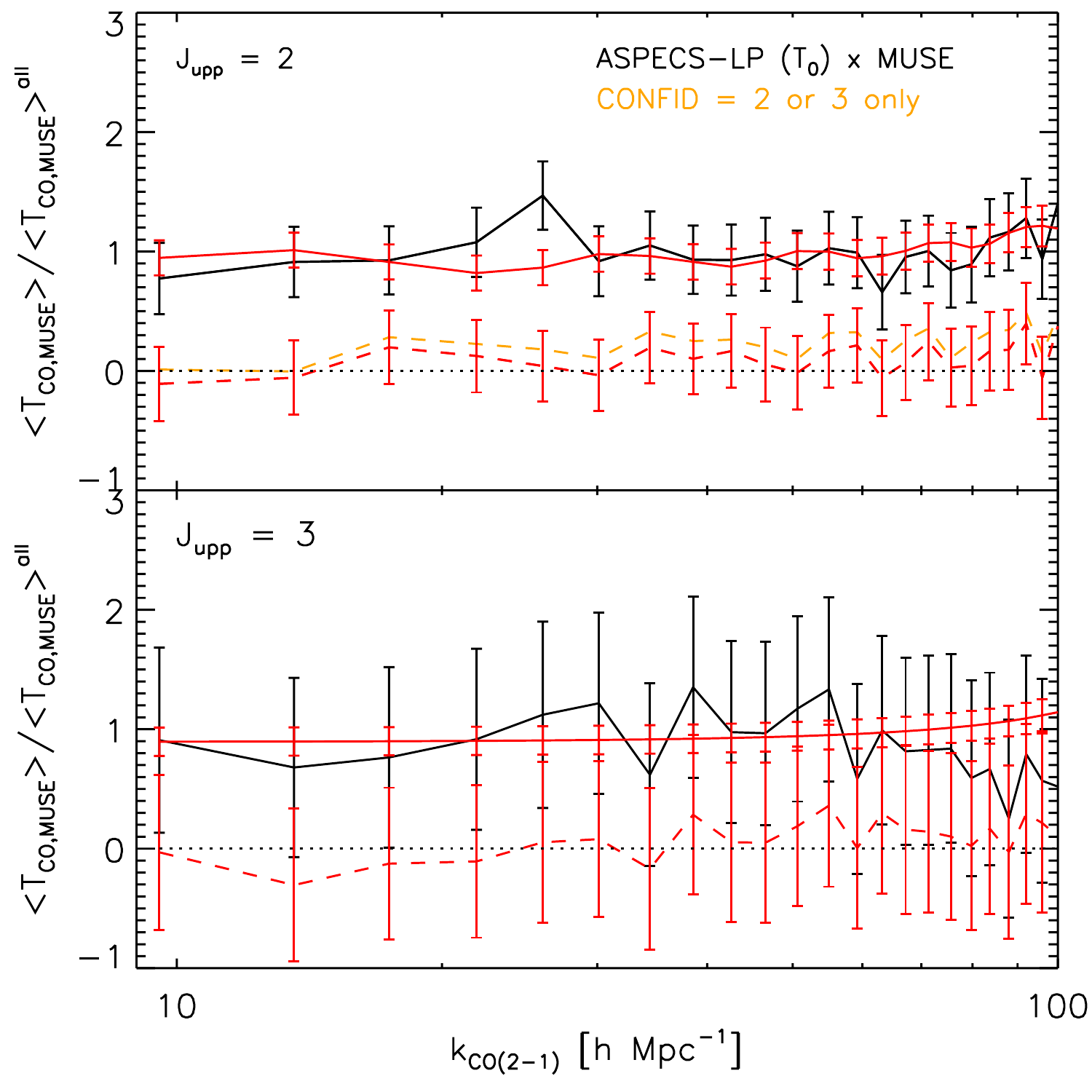}
\caption{Mean CO(2-1) (top panel) and CO(3-2) (bottom panel) surface brightness of MUSE galaxies. Color-coding is the same as in Figure~\ref{fig:xspec_muse_allJ}. Values on the $y$-axis in each panel have been scaled by the total mean CO(2-1) and CO(3-2) surface brightnesses, $\langle T_{\cotwomath, \mathrm{MUSE}} \rangle_{\mathrm{all}}$ and $\langle T_{\cothreemath, \mathrm{MUSE}} \rangle_{\mathrm{all}}$, so that curves plotted represent the relative contribution of MUSE galaxies with (red solid curve) and without (red and orange dashed curves) ASPECS blind detections. In the upper panel for CO(2-1), the orange curve represents MUSE galaxies with potential CO(2-1) emission, excluding spectra with low confidence (CONFID = 1).}
\label{fig:pcross_tcogal}
\end{figure}


\begin{thebibliography}{}
\expandafter\ifx\csname natexlab\endcsname\relax\def\natexlab#1{#1}\fi
\providecommand{\url}[1]{\href{#1}{#1}}
\providecommand{\dodoi}[1]{doi:~\href{http://doi.org/#1}{\nolinkurl{#1}}}
\providecommand{\doeprint}[1]{\href{http://ascl.net/#1}{\nolinkurl{http://ascl.net/#1}}}
\providecommand{\doarXiv}[1]{\href{https://arxiv.org/abs/#1}{\nolinkurl{https://arxiv.org/abs/#1}}}

\bibitem[{{Aravena} {et~al.}(2019){Aravena}, {Decarli},
  {G{\'o}nzalez-L{\'o}pez}, {Boogaard}, {Walter}, {Carilli}, {Popping},
  {Weiss}, {Assef}, {Bacon}, {Bauer}, {Bertoldi}, {Bouwens}, {Contini},
  {Cortes}, {Cox}, {da Cunha}, {Daddi}, {D{\'\i}az-Santos}, {Elbaz}, {Hodge},
  {Inami}, {Ivison}, {Le F{\`e}vre}, {Magnelli}, {Oesch}, {Riechers}, {Smail},
  {Somerville}, {Swinbank}, {Uzgil}, {van der Werf}, {Wagg}, \&
  {Wisotzki}}]{Aravena2019_3mm}
{Aravena}, M., {Decarli}, R., {G{\'o}nzalez-L{\'o}pez}, J., {et~al.} 2019,
  \apj, 882, 136, \dodoi{10.3847/1538-4357/ab30df}

\bibitem[{{Bacon} {et~al.}(2017){Bacon}, {Conseil}, {Mary}, {Brinchmann},
  {Shepherd}, {Akhlaghi}, {Weilbacher}, {Piqueras}, {Wisotzki}, {Lagattuta},
  {Epinat}, {Guerou}, {Inami}, {Cantalupo}, {Courbot}, {Contini}, {Richard},
  {Maseda}, {Bouwens}, {Bouch{\'e}}, {Kollatschny}, {Schaye}, {Marino},
  {Pello}, {Herenz}, {Guiderdoni}, \& {Carollo}}]{Bacon2017}
{Bacon}, R., {Conseil}, S., {Mary}, D., {et~al.} 2017, \aap, 608, A1,
  \dodoi{10.1051/0004-6361/201730833}

\bibitem[{{Beckwith} {et~al.}(2006){Beckwith}, {Stiavelli}, {Koekemoer},
  {Caldwell}, {Ferguson}, {Hook}, {Lucas}, {Bergeron}, {Corbin}, {Jogee},
  {Panagia}, {Robberto}, {Royle}, {Somerville}, \& {Sosey}}]{Beckwith2006}
{Beckwith}, S.~V.~W., {Stiavelli}, M., {Koekemoer}, A.~M., {et~al.} 2006, \aj,
  132, 1729, \dodoi{10.1086/507302}

\bibitem[{{Boogaard} {et~al.}(2019){Boogaard}, {Decarli},
  {Gonz{\'a}lez-L{\'o}pez}, {van der Werf}, {Walter}, {Bouwens}, {Aravena},
  {Carilli}, {Bauer}, {Brinchmann}, {Contini}, {Cox}, {da Cunha}, {Daddi},
  {D{\'\i}az-Santos}, {Hodge}, {Inami}, {Ivison}, {Maseda}, {Matthee}, {Oesch},
  {Popping}, {Riechers}, {Schaye}, {Schouws}, {Smail}, {Weiss}, {Wisotzki},
  {Bacon}, {Cortes}, {Rix}, {Somerville}, {Swinbank}, \&
  {Wagg}}]{Boogaard2019_3mm}
{Boogaard}, L.~A., {Decarli}, R., {Gonz{\'a}lez-L{\'o}pez}, J., {et~al.} 2019,
  \apj, 882, 140, \dodoi{10.3847/1538-4357/ab3102}

\bibitem[{{Brammer} {et~al.}(2008){Brammer}, {van Dokkum}, \&
  {Coppi}}]{Brammer2008}
{Brammer}, G.~B., {van Dokkum}, P.~G., \& {Coppi}, P. 2008, \apj, 686, 1503,
  \dodoi{10.1086/591786}

\bibitem[{{Breysse} \& {Alexandroff}(2019)}]{Breysse2019}
{Breysse}, P.~C., \& {Alexandroff}, R.~M. 2019, arXiv e-prints.
\newblock \doarXiv{1904.03197}

\bibitem[{{Carilli} \& {Walter}(2013)}]{CarilliWalter2013}
{Carilli}, C.~L., \& {Walter}, F. 2013, \araa, 51, 105,
  \dodoi{10.1146/annurev-astro-082812-140953}

\bibitem[{{Carilli} {et~al.}(2016){Carilli}, {Chluba}, {Decarli}, {Walter},
  {Aravena}, {Wagg}, {Popping}, {Cortes}, {Hodge}, {Weiss}, {Bertoldi}, \&
  {Riechers}}]{Carilli2016}
{Carilli}, C.~L., {Chluba}, J., {Decarli}, R., {et~al.} 2016, \apj, 833, 73,
  \dodoi{10.3847/1538-4357/833/1/73}

\bibitem[{{Chang} {et~al.}(2010){Chang}, {Pen}, {Bandura}, \&
  {Peterson}}]{Chang2010}
{Chang}, T.-C., {Pen}, U.-L., {Bandura}, K., \& {Peterson}, J.~B. 2010, \nat,
  466, 463, \dodoi{10.1038/nature09187}

\bibitem[{{Coe} {et~al.}(2006){Coe}, {Ben{\'{\i}}tez}, {S{\'a}nchez}, {Jee},
  {Bouwens}, \& {Ford}}]{Coe2006}
{Coe}, D., {Ben{\'{\i}}tez}, N., {S{\'a}nchez}, S.~F., {et~al.} 2006, \aj, 132,
  926, \dodoi{10.1086/505530}

\bibitem[{{da Cunha} {et~al.}(2013){da Cunha}, {Groves}, {Walter}, {Decarli},
  {Weiss}, {Bertoldi}, {Carilli}, {Daddi}, {Elbaz}, {Ivison}, {Maiolino},
  {Riechers}, {Rix}, {Sargent}, \& {Smail}}]{daCunha2013}
{da Cunha}, E., {Groves}, B., {Walter}, F., {et~al.} 2013, \apj, 766, 13,
  \dodoi{10.1088/0004-637X/766/1/13}

\bibitem[{{Daddi} {et~al.}(2015){Daddi}, {Dannerbauer}, {Liu}, {Aravena},
  {Bournaud}, {Walter}, {Riechers}, {Magdis}, {Sargent}, {B{\'e}thermin},
  {Carilli}, {Cibinel}, {Dickinson}, {Elbaz}, {Gao}, {Gobat}, {Hodge}, \&
  {Krips}}]{Daddi2015}
{Daddi}, E., {Dannerbauer}, H., {Liu}, D., {et~al.} 2015, \aap, 577, A46,
  \dodoi{10.1051/0004-6361/201425043}

\bibitem[{{Decarli} {et~al.}(2019){Decarli}, {Walter},
  {G{\'o}nzalez-L{\'o}pez}, {Aravena}, {Boogaard}, {Carilli}, {Cox}, {Daddi},
  {Popping}, {Riechers}, {Uzgil}, {Weiss}, {Assef}, {Bacon}, {Bauer},
  {Bertoldi}, {Bouwens}, {Contini}, {Cortes}, {da Cunha}, {D{\'\i}az-Santos},
  {Elbaz}, {Inami}, {Hodge}, {Ivison}, {Le F{\`e}vre}, {Magnelli}, {Novak},
  {Oesch}, {Rix}, {Sargent}, {Smail}, {Swinbank}, {Somerville}, {van der Werf},
  {Wagg}, \& {Wisotzki}}]{Decarli2019_3mm}
{Decarli}, R., {Walter}, F., {G{\'o}nzalez-L{\'o}pez}, J., {et~al.} 2019, \apj,
  882, 138, \dodoi{10.3847/1538-4357/ab30fe}

\bibitem[{{Dillon} {et~al.}(2014){Dillon}, {Liu}, {Williams}, {Hewitt},
  {Tegmark}, {Morgan}, {Levine}, {Morales}, {Tingay}, {Bernardi}, {Bowman},
  {Briggs}, {Cappallo}, {Emrich}, {Mitchell}, {Oberoi}, {Prabu}, {Wayth}, \&
  {Webster}}]{Dillon2014}
{Dillon}, J.~S., {Liu}, A., {Williams}, C.~L., {et~al.} 2014, \prd, 89, 023002,
  \dodoi{10.1103/PhysRevD.89.023002}

\bibitem[{{Gonz{\'a}lez-L{\'o}pez} {et~al.}(2019){Gonz{\'a}lez-L{\'o}pez},
  {Decarli}, {Pavesi}, {Walter}, {Aravena}, {Carilli}, {Boogaard}, {Popping},
  {Weiss}, {Assef}, {Bauer}, {Bertoldi}, {Bouwens}, {Contini}, {Cortes}, {Cox},
  {da Cunha}, {Daddi}, {D{\'\i}az-Santos}, {Inami}, {Hodge}, {Ivison}, {Le
  F{\`e}vre}, {Magnelli}, {Oesch}, {Riechers}, {Rix}, {Smail}, {Swinbank},
  {Somerville}, {Uzgil}, \& {van der Werf}}]{GL2019_3mm}
{Gonz{\'a}lez-L{\'o}pez}, J., {Decarli}, R., {Pavesi}, R., {et~al.} 2019, \apj,
  882, 139, \dodoi{10.3847/1538-4357/ab3105}

\bibitem[{{Greve} {et~al.}(2014){Greve}, {Leonidaki}, {Xilouris}, {Wei{\ss}},
  {Zhang}, {van der Werf}, {Aalto}, {Armus}, {D{\'{\i}}az-Santos}, {Evans},
  {Fischer}, {Gao}, {Gonz{\'a}lez-Alfonso}, {Harris}, {Henkel}, {Meijerink},
  {Naylor}, {Smith}, {Spaans}, {Stacey}, {Veilleux}, \& {Walter}}]{Greve2014}
{Greve}, T.~R., {Leonidaki}, I., {Xilouris}, E.~M., {et~al.} 2014, \apj, 794,
  142, \dodoi{10.1088/0004-637X/794/2/142}

\bibitem[{{Illingworth} {et~al.}(2013){Illingworth}, {Magee}, {Oesch},
  {Bouwens}, {Labb{\'e}}, {Stiavelli}, {van Dokkum}, {Franx}, {Trenti},
  {Carollo}, \& {Gonzalez}}]{Illingworth2013}
{Illingworth}, G.~D., {Magee}, D., {Oesch}, P.~A., {et~al.} 2013, \apjs, 209,
  6, \dodoi{10.1088/0067-0049/209/1/6}

\bibitem[{{Inami} {et~al.}(2017){Inami}, {Bacon}, {Brinchmann}, {Richard},
  {Contini}, {Conseil}, {Hamer}, {Akhlaghi}, {Bouch{\'e}}, {Cl{\'e}ment},
  {Desprez}, {Drake}, {Hashimoto}, {Leclercq}, {Maseda}, {Michel-Dansac},
  {Paalvast}, {Tresse}, {Ventou}, {Kollatschny}, {Boogaard}, {Finley},
  {Marino}, {Schaye}, \& {Wisotzki}}]{Inami2017}
{Inami}, H., {Bacon}, R., {Brinchmann}, J., {et~al.} 2017, \aap, 608, A2,
  \dodoi{10.1051/0004-6361/201731195}

\bibitem[{{Keating} {et~al.}(2016){Keating}, {Marrone}, {Bower}, {Leitch},
  {Carlstrom}, \& {DeBoer}}]{Keating2016}
{Keating}, G.~K., {Marrone}, D.~P., {Bower}, G.~C., {et~al.} 2016, \apj, 830,
  34, \dodoi{10.3847/0004-637X/830/1/34}

\bibitem[{{Koekemoer} {et~al.}(2013){Koekemoer}, {Ellis}, {McLure}, {Dunlop},
  {Robertson}, {Ono}, {Schenker}, {Ouchi}, {Bowler}, {Rogers}, {Curtis-Lake},
  {Schneider}, {Charlot}, {Stark}, {Furlanetto}, {Cirasuolo}, {Wild}, \&
  {Targett}}]{Koekemoer2013}
{Koekemoer}, A.~M., {Ellis}, R.~S., {McLure}, R.~J., {et~al.} 2013, \apjs, 209,
  3, \dodoi{10.1088/0067-0049/209/1/3}

\bibitem[{{Kovetz} {et~al.}(2017){Kovetz}, {Viero}, {Lidz}, {Newburgh},
  {Rahman}, {Switzer}, {Kamionkowski}, {Aguirre}, {Alvarez}, {Bock}, {Bond},
  {Bower}, {Bradford}, {Breysse}, {Bull}, {Chang}, {Cheng}, {Chung}, {Cleary},
  {Corray}, {Crites}, {Croft}, {Dor{\'e}}, {Eastwood}, {Ferrara}, {Fonseca},
  {Jacobs}, {Keating}, {Lagache}, {Lakhlani}, {Liu}, {Moodley}, {Murray},
  {P{\'e}nin}, {Popping}, {Pullen}, {Reichers}, {Saito}, {Saliwanchik},
  {Santos}, {Somerville}, {Stacey}, {Stein}, {Villaescusa-Navarro}, {Visbal},
  {Weltman}, {Wolz}, \& {Zemcov}}]{Kovetz2017}
{Kovetz}, E.~D., {Viero}, M.~P., {Lidz}, A., {et~al.} 2017, arXiv e-prints.
\newblock \doarXiv{1709.09066}

\bibitem[{{Li} {et~al.}(2016){Li}, {Wechsler}, {Devaraj}, \& {Church}}]{Li2016}
{Li}, T.~Y., {Wechsler}, R.~H., {Devaraj}, K., \& {Church}, S.~E. 2016, \apj,
  817, 169, \dodoi{10.3847/0004-637X/817/2/169}

\bibitem[{{Lidz} \& {Taylor}(2016)}]{LidzTaylor2016}
{Lidz}, A., \& {Taylor}, J. 2016, \apj, 825, 143,
  \dodoi{10.3847/0004-637X/825/2/143}

\bibitem[{{Madau} \& {Dickinson}(2014)}]{MD2014}
{Madau}, P., \& {Dickinson}, M. 2014, \araa, 52, 415,
  \dodoi{10.1146/annurev-astro-081811-125615}

\bibitem[{{Moster} {et~al.}(2011){Moster}, {Somerville}, {Newman}, \&
  {Rix}}]{Moster2011}
{Moster}, B.~P., {Somerville}, R.~S., {Newman}, J.~A., \& {Rix}, H.-W. 2011,
  \apj, 731, 113, \dodoi{10.1088/0004-637X/731/2/113}

\bibitem[{{Murphy} {et~al.}(2018){Murphy}, {Bolatto}, {Chatterjee}, {Casey},
  {Chomiuk}, {Dale}, {de Pater}, {Dickinson}, {Francesco}, {Hallinan},
  {Isella}, {Kohno}, {Kulkarni}, {Lang}, {Lazio}, {Leroy}, {Loinard},
  {Maccarone}, {Matthews}, {Osten}, {Reid}, {Riechers}, {Sakai}, {Walter}, \&
  {Wilner}}]{Murphy2018}
{Murphy}, E.~J., {Bolatto}, A., {Chatterjee}, S., {et~al.} 2018, in
  Astronomical Society of the Pacific Conference Series, Vol. 517, Science with
  a Next Generation Very Large Array, ed. E.~{Murphy}, 3

\bibitem[{{Neeleman} {et~al.}(2016){Neeleman}, {Prochaska}, {Ribaudo},
  {Lehner}, {Howk}, {Rafelski}, \& {Kanekar}}]{Neeleman2016}
{Neeleman}, M., {Prochaska}, J.~X., {Ribaudo}, J., {et~al.} 2016, \apj, 818,
  113, \dodoi{10.3847/0004-637X/818/2/113}

\bibitem[{{Pavesi} {et~al.}(2018){Pavesi}, {Sharon}, {Riechers}, {Hodge},
  {Decarli}, {Walter}, {Carilli}, {Daddi}, {Smail}, {Dickinson}, {Ivison},
  {Sargent}, {da Cunha}, {Aravena}, {Darling}, {Smol{\v c}i{\'c}}, {Scoville},
  {Capak}, \& {Wagg}}]{Pavesi2018}
{Pavesi}, R., {Sharon}, C.~E., {Riechers}, D.~A., {et~al.} 2018, \apj, 864, 49,
  \dodoi{10.3847/1538-4357/aacb79}

\bibitem[{{Popping} {et~al.}(2016){Popping}, {van Kampen}, {Decarli}, {Spaans},
  {Somerville}, \& {Trager}}]{Popping2016}
{Popping}, G., {van Kampen}, E., {Decarli}, R., {et~al.} 2016, \mnras, 461, 93,
  \dodoi{10.1093/mnras/stw1323}

\bibitem[{{Popping} {et~al.}(2019){Popping}, {Pillepich}, {Somerville},
  {Decarli}, {Walter}, {Aravena}, {Carilli}, {Cox}, {Nelson}, {Riechers},
  {Weiss}, {Boogaard}, {Bouwens}, {Contini}, {Cortes}, {da Cunha}, {Daddi},
  {D{\'\i}az-Santos}, {Diemer}, {Gonz{\'a}lez-L{\'o}pez}, {Hernquist},
  {Ivison}, {Le F{\`e}vre}, {Marinacci}, {Rix}, {Swinbank}, {Vogelsberger},
  {van der Werf}, {Wagg}, \& {Yung}}]{Popping2019}
{Popping}, G., {Pillepich}, A., {Somerville}, R.~S., {et~al.} 2019, \apj, 882,
  137, \dodoi{10.3847/1538-4357/ab30f2}

\bibitem[{{Pullen} {et~al.}(2013){Pullen}, {Chang}, {Dor{\'e}}, \&
  {Lidz}}]{Pullen2013}
{Pullen}, A.~R., {Chang}, T.-C., {Dor{\'e}}, O., \& {Lidz}, A. 2013, \apj, 768,
  15, \dodoi{10.1088/0004-637X/768/1/15}

\bibitem[{{Riechers} {et~al.}(2019){Riechers}, {Pavesi}, {Sharon}, {Hodge},
  {Decarli}, {Walter}, {Carilli}, {Aravena}, {da Cunha}, {Daddi}, {Dickinson},
  {Smail}, {Capak}, {Ivison}, {Sargent}, {Scoville}, \& {Wagg}}]{Riechers2018}
{Riechers}, D.~A., {Pavesi}, R., {Sharon}, C.~E., {et~al.} 2019, \apj, 872, 7,
  \dodoi{10.3847/1538-4357/aafc27}

\bibitem[{{Selina} {et~al.}(2018){Selina}, {Murphy}, {McKinnon}, {Beasley},
  {Butler}, {Carilli}, {Clark}, {Durand}, {Erickson}, {Grammer}, {Hiriart},
  {Jackson}, {Kent}, {Mason}, {Morgan}, {Ojeda}, {Rosero}, {Shillue},
  {Sturgis}, \& {Urbain}}]{Selina2018}
{Selina}, R.~J., {Murphy}, E.~J., {McKinnon}, M., {et~al.} 2018, in
  Astronomical Society of the Pacific Conference Series, Vol. 517, Science with
  a Next Generation Very Large Array, ed. E.~{Murphy}, 15

\bibitem[{{Sun} {et~al.}(2018){Sun}, {Moncelsi}, {Viero}, {Silva}, {Bock},
  {Bradford}, {Chang}, {Cheng}, {Cooray}, {Crites}, {Hailey-Dunsheath},
  {Uzgil}, {Hunacek}, \& {Zemcov}}]{Sun2018}
{Sun}, G., {Moncelsi}, L., {Viero}, M.~P., {et~al.} 2018, \apj, 856, 107,
  \dodoi{10.3847/1538-4357/aab3e3}

\bibitem[{{Switzer} {et~al.}(2013){Switzer}, {Masui}, {Bandura}, {Calin},
  {Chang}, {Chen}, {Li}, {Liao}, {Natarajan}, {Pen}, {Peterson}, {Shaw}, \&
  {Voytek}}]{Switzer2013}
{Switzer}, E.~R., {Masui}, K.~W., {Bandura}, K., {et~al.} 2013, \mnras, 434,
  L46, \dodoi{10.1093/mnrasl/slt074}

\bibitem[{{Taylor} {et~al.}(1999){Taylor}, {Carilli}, \& {Perley}}]{TCP1999}
{Taylor}, G.~B., {Carilli}, C.~L., \& {Perley}, R.~A., eds. 1999, Astronomical
  Society of the Pacific Conference Series, Vol. 180, {Synthesis Imaging in
  Radio Astronomy II}

\bibitem[{{Walter} {et~al.}(2014){Walter}, {Decarli}, {Sargent}, {Carilli},
  {Dickinson}, {Riechers}, {Ellis}, {Stark}, {Weiner}, {Aravena}, {Bell},
  {Bertoldi}, {Cox}, {Da Cunha}, {Daddi}, {Downes}, {Lentati}, {Maiolino},
  {Menten}, {Neri}, {Rix}, \& {Weiss}}]{Walter2014}
{Walter}, F., {Decarli}, R., {Sargent}, M., {et~al.} 2014, \apj, 782, 79,
  \dodoi{10.1088/0004-637X/782/2/79}

\bibitem[{{Walter} {et~al.}(2016){Walter}, {Decarli}, {Aravena}, {Carilli},
  {Bouwens}, {da Cunha}, {Daddi}, {Ivison}, {Riechers}, {Smail}, {Swinbank},
  {Weiss}, {Anguita}, {Assef}, {Bacon}, {Bauer}, {Bell}, {Bertoldi}, {Chapman},
  {Colina}, {Cortes}, {Cox}, {Dickinson}, {Elbaz}, {G{\'o}nzalez-L{\'o}pez},
  {Ibar}, {Inami}, {Infante}, {Hodge}, {Karim}, {Le Fevre}, {Magnelli}, {Neri},
  {Oesch}, {Ota}, {Popping}, {Rix}, {Sargent}, {Sheth}, {van der Wel}, {van der
  Werf}, \& {Wagg}}]{Walter2016_survey}
{Walter}, F., {Decarli}, R., {Aravena}, M., {et~al.} 2016, \apj, 833, 67,
  \dodoi{10.3847/1538-4357/833/1/67}

\bibitem[{{Wolfe} {et~al.}(2005){Wolfe}, {Gawiser}, \& {Prochaska}}]{Wolfe2005}
{Wolfe}, A.~M., {Gawiser}, E., \& {Prochaska}, J.~X. 2005, \araa, 43, 861,
  \dodoi{10.1146/annurev.astro.42.053102.133950}

\bibitem[{{Wolz} {et~al.}(2017){Wolz}, {Blake}, \& {Wyithe}}]{Wolz2017}
{Wolz}, L., {Blake}, C., \& {Wyithe}, J.~S.~B. 2017, \mnras, 470, 3220,
  \dodoi{10.1093/mnras/stx1388}

\end{thebibliography}
\end{document}